\documentclass[useAMS,usenatbib]{mn2e}
\usepackage{graphicx}
\usepackage{aas_macros}
\usepackage[footnotesize]{subfigure}

\def\eg{{e.g.}}
\def\zo{$z\approx1$}
\def\zt{$z\approx2$}

\newcommand\gtsim{\mathrel{\lower0.6ex\hbox{$\buildrel {\textstyle >}
   \over {\scriptstyle \sim}$}}}
\newcommand\ltsim{\mathrel{\lower0.6ex\hbox{$\buildrel {\textstyle <}
   \over {\scriptstyle \sim}$}}}
\newcommand\etal{\mbox{{et al.}}}

\title[Quasar host galaxies at $z=1-2$]
{Star formation in luminous quasar host galaxies at $\mathbf{z=1-2}$\thanks{Based on observations with the NASA/ESA Hubble Space Telescope, obtained at the Space Telescope Science Institute, which is operated by the Association of Universities for Research in Astronomy, Inc. (AURA), under NASA contract NAS5-26555}}
\author[D. J. E. Floyd \etal]{David J. E. Floyd$^{1}$\thanks{Email: david.floyd@monash.edu}
James S. Dunlop$^{2}$, Marek J. Kukula$^{3}$, Michael J. I. Brown$^{1}$,
\newauthor
Ross J. McLure$^{2}$, Stefi A. Baum$^{4,5}$ and Christopher P. O'Dea$^{6,7}$\\
\\
$^{1}$MoCA\thanks{Monash Centre for Astrophysics}, School of Physics, Monash University, VIC 3800, Australia\\
$^{2}$SUPA\thanks{Scottish Universities Physics Alliance}, Institute for Astronomy, School of Physics, University of Edinburgh, Royal Observatory, Edinburgh EH9 3HJ\\
$^{3}$Royal Observatory Greenwich, Royal Museums Greenwich, London SE10 9NF\\
$^{4}$Chester F. Carlson Center for Imaging Science, 54 Lomb Memorial Drive, Rochester, NY 14623, USA\\
$^{5}$Radcliffe Institute for Advanced Study, 10 Garden St. Cambridge, MA 02138, USA\\
$^{6}$Department of Physics, 84 Lomb Memorial Drive, Rochester, NY 14623, USA\\
$^{7}$Harvard Smithsonian Center for Astrophysics, 60 Garden St. Cambridge, MA 02138, USA\\
}
\begin{document}

\pagerange{\pageref{firstpage}--\pageref{lastpage}} \pubyear{2012} \volume{}

\maketitle

\begin{abstract}
We present deep {\em HST/WFPC2}, rest-frame $U$ images of 17 $\sim L^\star$ quasars at \zo\ and \zt\ ($V$ and $I$ bands respectively), designed to explore the host galaxies. We fit the images with simple axisymmetric galaxy models, including a point-source, in order to separate nuclear and host-galaxy emission. We successfully model all of the host galaxies, with luminosities stable to within 0.3~mag. Combining with our earlier {\em NICMOS} rest-frame optical study of the same sample, we provide the first rest-frame $U-V$ colours for a sample of quasar host galaxies. While the optical luminosities of their host galaxies indicate that they are drawn purely from the most massive ($\gtsim L^\star$) early-type galaxy population, their colours are systematically bluer than those of comparably massive galaxies at the same redshift. The host galaxies of the radio-loud quasars (RLQ) in our sample are more luminous than their radio-quiet quasar (RQQ) counterparts at each epoch, but have indistinguishable colours, confirming that the RLQ's are drawn from only the most massive galaxies ($10^{11}-10^{12}$~M$_\odot$ even at \zt), while the RQQ's are slightly less massive ($\sim10^{11}$~M$_\odot$). 
This is consistent with the well-known anti-correlation between radio-loudness and accretion rate.
Using simple stellar population ``frosting'' models 
we estimate mean star formation rates of $\sim350$~M$_\odot$~yr$^{-1}$ for the RLQ's and $\sim100$~M$_\odot$~yr$^{-1}$ for the RQQ's at \zt. By \zo, these rates have fallen to $\sim150$~M$_\odot$~yr$^{-1}$ for the RLQ's and $\sim50$~M$_\odot$~yr$^{-1}$ for the RQQ's. We conclude that while the host galaxies are extremely massive, they remain actively star-forming at, or close to, the epoch of the quasar. 
\end{abstract}


\begin{table*}
  \begin{minipage}{175mm}
    \begin{center}
      \begin{small}
        \caption{\label{tab-obs} Summary of the {\em WFPC2} observations:
        objects at \zo\ and \zt\ were observed using the F606W and F814W 
        filters respectively (corresponding to approximately rest-frame $U$ in each case).
        Each quasar is classified as either radio-loud (RLQ) or radio-quiet (RQQ) -- see 
        Table~\ref{tab-samp} and section~\ref{sec-samp} of the text. 
        A single white dwarf star was observed in each filter for PSF calibration.
        We show each target name used in the HST archive, together with its B1950 IAU name. 
        J2000 co-ordinates were obtained from the Digitised Sky Survey plates 
        maintained by the Space Telescope Science Institute. Redshifts are derived from NED.
        Note that the image of SGP5:46 was contaminated by diffraction spikes from a 
        nearby bright star and had to be excluded from this study.}
        \begin{tabular}{lrccrrrr}
        \hline
        \hline
        Object &IAU Name& Type &$z$ &\multicolumn{2}{c}{J2000 Position} & Integration & Filter \\
                      &(B1950)&       &    & RA ({\it h m s})& Dec ($^{\circ}$ $'$ $''$) & time (s) & \\
        \hline
        {\it SGP5:46}	& {\it 0049$-$277}	&{\it RQQ}	&{\it 0.955$^{(a)}$}&{\it 00:52:22.8}&{\it $-$27:30:03}&{\it 3600} & {\it F606W}\\
        BVF225		& 1301$+$358	&RQQ	&0.910$^{(b)}$	&13:04:10.5	&$+$35:36:51	&3600 & F606W\\
        BVF247		& 1302$+$361	&RQQ	&0.890$^{(b)}$	&13:05:05.0	&$+$35:51:21	&3600 & F606W\\
        BVF262		& 1303$+$360	&RQQ	&0.970$^{(b)}$	&13:05:30.9	&$+$35:17:14	&3600 & F606W\\
        PSF-STAR		& 1313$+$705	&Star	& ...  			&13:38:59.5	&$+$70:16:40	&  320 & F606W\\
        PKS0440-00	& 0440$-$003	&RLQ	&0.844$^{(b)}$	&04:42:38.6	&$-$00:17:43	&3600 & F606W\\
        PKS0938$+$18	& 0938$+$185	&RLQ	&0.940$^{(c)}$	&09:41:23.2	&$+$18:21:06	&3600 & F606W\\
        3C422		& 2044$-$027	&RLQ	&0.942$^{(b)}$	&20:47:10.4	&$-$02:36:23	&3600 & F606W\\
        MC2112$+$172& 2112$+$172	&RLQ	&0.878$^{(b)}$	&21:14:56.7	&$+$17:29:23	&3600 & F606W\\
        4C02.54		& 2207$+$020	&RLQ	&0.976$^{(d)}$	&22:09:32.8	&$+$02:18:41	&3600 & F606W\\
	\\
        SGP2:36		& 0048$-$293	&RQQ	&1.773$^{(e)}$	&00:51:14.3	&$-$29:05:20	& 7800& F814W\\
        SGP2:25		& 0049$-$295	&RQQ	&1.869$^{(e)}$	&00:52:07.6	&$-$29:17:50	& 7800& F814W\\
        SGP2:11		& 0050$-$291	&RQQ	&1.976$^{(a)}$	&00:52:38.5	&$-$28:51:13	& 7800& F814W\\
        SGP3:39		& 0053$-$286	&RQQ	&1.959$^{(e)}$	&00:55:43.4	&$-$28:24:10	& 7800& F814W\\
        SGP4:39		& 0056$-$281	&RQQ	&1.721$^{(e)}$	&00:59:08.9	&$-$27:51:25	& 7800& F814W\\
        PSF-STAR		& 1313$+$705	&Star	& ... 			&13:38:59.5	&$+$70:16:40	&  320&  F814W\\
        PKS1524$-$13	& 1524$-$136	&RLQ	&1.687$^{(b)}$	&15:26:59.4	&$-$13:51:01	& 7800& F814W\\
        B2~2156$+$29	& 2156$+$297	&RLQ	&1.753$^{(b)}$	&21:58:42.0 	&$+$29:59:08	& 7800& F814W\\
        PKS2204$-$20	& 2204$-$205	&RLQ	&1.923$^{(f)}$	&22:07:33.9	&$-$20:38:35	& 7800& F814W\\
        4C45.51		& 2351$+$456	&RLQ	&1.992$^{(g)}$	&23:54:22.3	&$+$45:53:05	& 7800& F814W\\
        \hline
        \end{tabular}

        $^{\mathrm (a)}$\cite{Boyle:UVQSO};
        $^{\mathrm (b)}$\cite{HB_QSO_89};
        $^{\mathrm (c)}$\cite{SDSS_QSO_DR7};
        $^{\mathrm (d)}$\cite{VCV93};
        $^{\mathrm (e)}$\cite{Croom:2dfQSO};
        $^{\mathrm (f)}$\cite{Dunlop:PKS_QSO};
        $^{\mathrm (g)}$\cite{Stickel:radio}.
      \end{small}
    \end{center}
  \end{minipage}
\end{table*}

\section{Introduction}
Comprehensive studies of low redshift ($z<0.4$) quasar host galaxies indicate that the most powerful nuclear activity in present-day galaxies is usually associated with massive, bulge-dominated host galaxies~\citep[\eg][]{disney+95,bahcall+97,hooper+97,mcleod+99,mclure+99,ridgway+01,hamilton+02,dunlop+03,floyd+04,guyon+06,canalizo+07,bennert+08}. This ties in well with the discovery of the black hole -- bulge mass relationship in quiescent galaxies: the host galaxies are exactly the type of systems in which one would expect to find the very massive black holes that are required to power luminous quasars.
However, this conclusion is based largely on their luminosities and morphologies in single-band Hubble Space Telescope (HST) imaging (especially at the highest quasar luminosities--~\citealt{floyd+04}). The stellar populations and masses of quasar host galaxies remain poorly constrained, although ground-based $K$-band imaging~\citep{taylor+96} and off-nuclear optical spectroscopy~\citep{nolan+01} suggest the presence (albeit at low level) of an intermediate age stellar population in some.
It is difficult to disentangle any young stellar population in the host galaxy from the glare of the active galactic nucleus (AGN), and much of our picture of quasar host galaxies is inferred from the properties of their inactive, giant red elliptical, counterparts.

Meanwhile, lower luminosity AGN in the local Universe have clear evidence for younger stellar populations~\citep[\eg][]{borosonoke82}. 
Seyfert galaxies are frequently found with circumnuclear starbursts~\citep{heckman+97,gu+01,gonzalezdelgado+01,cidfernandes+04,davies+07,riffel+09}. The presence of intermediate age stellar populations has also been extensively shown in type 2 (obscured) and lower luminosity type 1 quasars~\citep{kotilainenward94,ronnback+96,brotherton+99,kauffmann+03,jahnke+04a,jahnke+04b,sanchez+04,vandenberk+06,jahnke+07,jahnke+09}.
A decrease in mean stellar age with increasing AGN luminosity is suggested (e.g.~\citealt{kauffmann+03,vandenberk+06}), indicating that we may be missing young stellar populations in the most luminous nearby quasars, due to the glare of the nucleus. 

At higher redshift we see greater levels of star formation, as expected. \citet{mainieri+11} examined $\sim 1800$ obscured (X-ray selected) quasars in the XMM-COSMOS survey, with bolometric luminosities $>8\times10^{45}$~erg~s$^{-1}$. They confirm that all dwell in massive ($>10^{10}$~M$_\odot$) galaxies, with a monotonically increasing AGN fraction with stellar mass. They also find strong star formation in the majority, with the star forming fraction increasing strongly with redshift (62\% star forming at $z\sim1$, 71\% at $z\sim 2$, 100\% at $z\sim3$). 
\citet{trichas+12}, however, show a decreasing fraction of measurable starburst contributions with increasing AGN luminosity.


A key observable is the level of star formation present in quasar host galaxies when compared to inactive galaxies of the same mass at the same epoch.
In this paper we aim to explore the level of star formation in our \zo\ and \zt\ quasar samples, originally imaged in the rest-frame optical using HST's Near Infrared Camera and Multi-Object Spectrometer (NICMOS -- see~\citealt{kukula+01} -- hereafter K01). That paper showed the quasars to be hosted by massive ($\sim10^{11}-10^{12}$~M$_\odot$) elliptical galaxies, with masses consistent with those found at lower redshift (\citealt{mclure+99, dunlop+03, floyd+04} -- hereafter M99; D03; F04 respectively) assuming passively evolving stellar populations. Here we present the follow-up rest-frame $U$-band imaging taken with WFPC2, explore the colours and morphologies of the host galaxies, and constrain their masses and young stellar populations. In particular, we explore their colours in comparison to the colours of massive galaxies at the same epoch.

The layout of the remainder of the paper is as follows. The sample design and observations are discussed in section 2. We describe the data reduction and analysis in section 3. Our results are presented in section 4, and their implications discussed in section 5. Our main conclusions are summarised in section 6. We assume throughout a cosmology with $\Omega_{m}=0.3$, $\Lambda= 0.7$ and $H_{0}=70$~km s$^{-1}$ Mpc$^{-1}$, and convert previous work to this standard where required.


\begin{figure*}
\centering
{\includegraphics[width=85mm,angle=0]{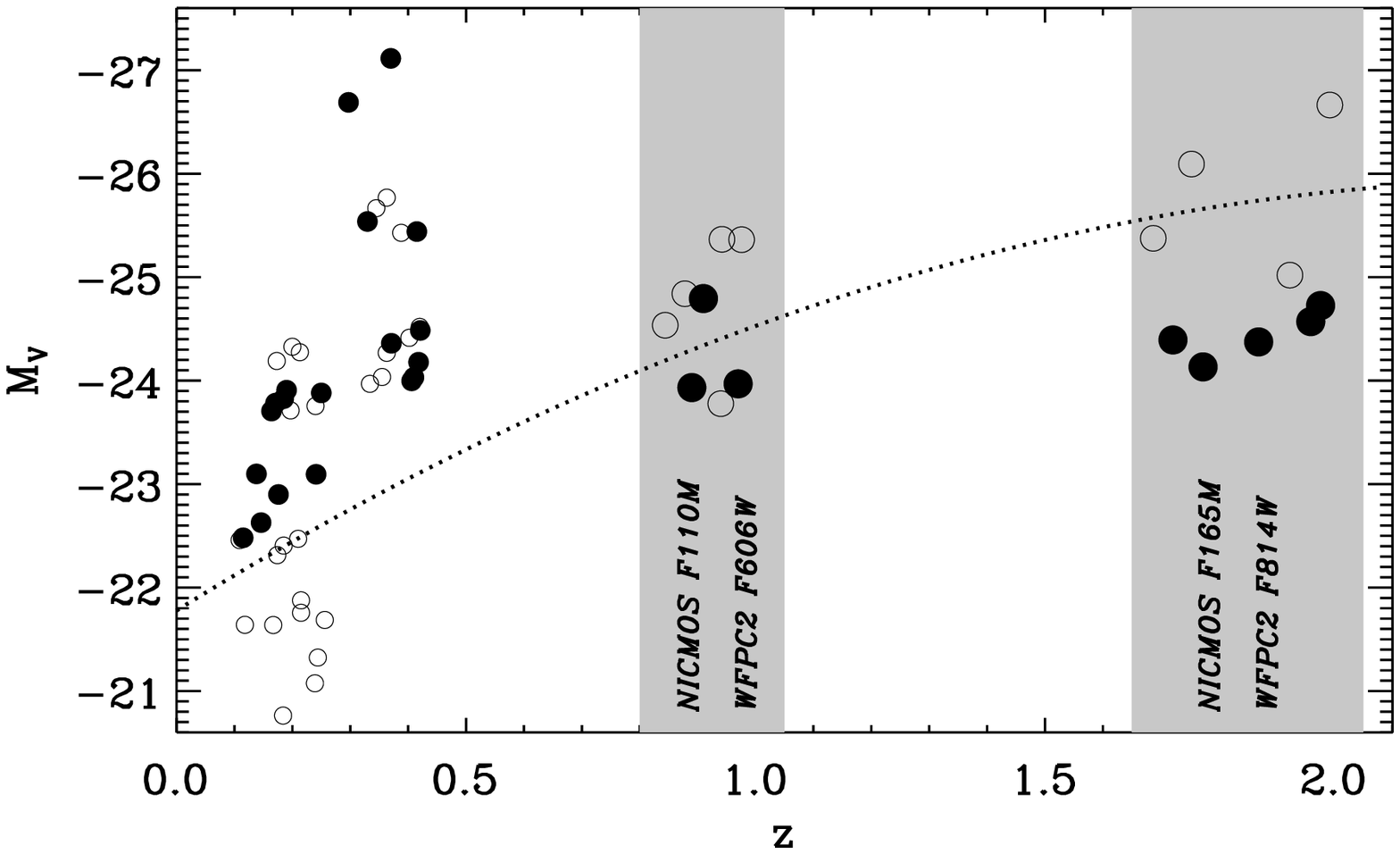}}
\hspace{5mm}
{\includegraphics[width=85mm,angle=0]{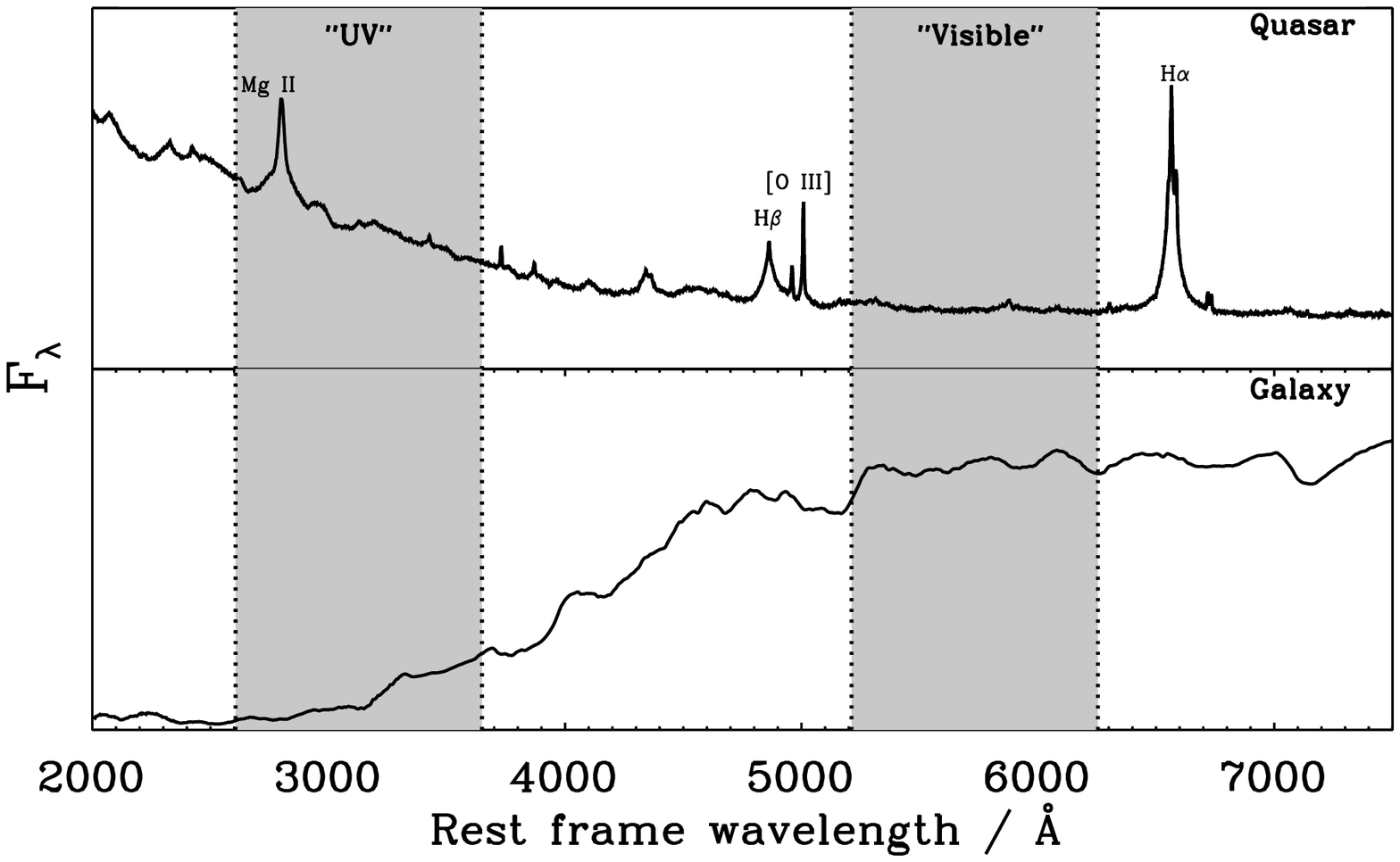}}
\caption{\label{fig-samp}{\bf Left:} Absolute $V$ magnitude versus redshift for the quasars in our host galaxy studies out to $z\approx2$.
Filled circles represent radio-quiet quasars, while open circles represent radio-loud quasars. 
Objects in the present WFPC2 study at \zo\ \& 2 are shown using large symbols, compared to our lower redshift objects (F04, D03, M99).
Our sample spans the knee of the quasar luminosity function at each redshift: the dotted line indicates $M_V^\star$ for the 2QZ quasar luminosity function at each redshift~\citep{croom+04}. 
See section~\ref{sec-samp} for notes on the sample selection.
{\bf Right:} Illustration showing generic spectra for a quasar nucleus (upper panel) and an early-type galaxy (lower panel), with our approximate rest-frame bandpasses marked. We have tailored our filter selection to target the rest-frame $U$ and $V$-band, thus sampling the SED of the host on either side of the break feature at 4000\AA.
While the Mg~II quasar emission line is admitted, we avoid prominent galaxy emission lines (section~\ref{sec-filt}).}
\end{figure*}

\begin{table}
  \begin{center}
    \begin{small}
      \caption{\label{tab-samp} Optical and radio properties of the quasars in the current study.
      We show the target names used in the HST archive. 
      Redshifts, $V$-band magnitudes, colour excesses $E(B-V)$~\citep{schlegel+98} are derived from NED. 
      Absolute magnitudes are confined to the range $-23.9 \leq M_{V} \leq -25.1$ (in rest-frame) to ensure that the quasars
      in each redshift bin are comparable with each other (as well as with the low-redshift quasars
      in our earlier studies, F04, D03, M99).
      RQQs have $L_\mathrm{5GHz}  < 10^{24.5}$~W~Hz$^{-1}$~sr$^{-1}$.
      RLQs have $L_\mathrm{5GHz} > 10^{24.5}$~W~Hz$^{-1}$~sr$^{-1}$ and steep radio spectra.
      For the estimation of luminosities at both radio and optical wavelengths we have assumed a quasar spectrum of the form $F_\nu\propto\nu^{-0.5}$.}
      See section~\ref{sec-samp} for full details of sample selection.
\centering
\begin{tabular}{lccrrr}
  \hline
  \hline
  Object 	& Type 	&$z$	&$V$ 	&$E(B-V)$	&$S_\mathrm{1.4}$\\
       				&      		&     	& /mag	&  / mag		&/mJy\\
\hline
\multicolumn{6}{c}{\bf \zo\ sample (F606W `Wide $V$')}\\
BVF225        		& RQQ 	& 0.910 & 19.6 & 0.012&$<0.15$ \\	
BVF247        		& RQQ 	& 0.890 & 19.2 & 0.014&$<0.15$ \\	
BVF262        		& RQQ 	& 0.970 & 19.6 & 0.008&$<0.15$ \\	
PKS0440$-$00		& RLQ 	& 0.844 & 19.2 & 0.053& 1443\\	
PKS0938$+$18	& RLQ 	& 0.940 & 19.1 & 0.030& 432\\	
3C422         		& RLQ 	& 0.942 & 19.5 & 0.055& 2117\\	
MC2112$+$172	& RLQ 	& 0.878 & 18.7 & 0.134& 425\\	
4C02.54			& RLQ 	& 0.976 & 19.0 & 0.045& 790\\	
\multicolumn{6}{c}{\bf \zt\ sample (F814W `Wide $I$')}\\	       
SGP2:36			& RQQ 	& 1.773 & 20.8 & 0.014&$<0.45$ \\	
SGP2:25			& RQQ 	& 1.869 & 20.8 & 0.016&$<0.45$ \\	
SGP2:11			& RQQ 	& 1.976 & 21.1 & 0.017&$<0.45$ \\	
SGP3:39			& RQQ 	& 1.959 & 20.9 & 0.028&$<0.45$ \\	
SGP4:39			& RQQ 	& 1.721 & 20.9 & 0.022&$<0.45$ \\	
PKS1524$-$13		& RLQ 	& 1.687 & 21.0 & 0.121& 2400\\	
B2~2156+29		& RLQ 	& 1.753 & 17.5 & 0.104& 1223\\	
PKS2204$-$20		& RLQ 	& 1.923 & 20.3 & 0.049& 700\\	
4C45.51			& RLQ 	& 1.992 & 20.6 & 0.122& 1836\\ 	
\hline
\end{tabular}
\end{small}
\end{center}
\end{table}

\section{Observations}
\label{sec-obs}
Our observing strategy was designed to maximise the detection of host galaxy light, as in our previous HST studies of quasar host galaxies (M99; K01; D03; F04). The precautions described in those papers (and reiterated below) were even more critical to the success of the current observations since they target shorter wavelengths than our earlier studies. In the rest-frame UV, the quasar spectral energy distribution (SED) climbs steeply with respect to the host galaxy SED, increasing the nuclear-to-host ratio and making accurate separation of quasar and galaxy light more difficult. The observations were performed in HST Cycle 7 with WFPC2, as summarised in Table~\ref{tab-obs}. We adopted the Wide Field (WF) section of the detector. Despite the higher sampling afforded by the PC's smaller (0\farcs045) pixels, the 0\farcs1 pixels of the WF detectors offered the decisive advantage of better sensitivity to low surface brightness emission.  The WF3 chip was selected for our programme because of its marginally better overall performance relative to the WF2 \& WF4 chips during the twelve months prior to the observations.

\subsection{Sample design}
\label{sec-samp}
Our sample, described in detail by K01, originally consisted of 20 quasars divided into two subsamples in the redshift ranges $0.83\leq z \leq 1.0$ and $1.67 \leq z \leq 2.01$ (referred to hereafter as the ``\zo'' and ``\zt'' samples respectively). Each subsample contained five radio-loud and five radio-quiet objects, and each was matched in terms of its optical luminosity-redshift distribution. The sample was defined to fall in the luminosity range $-24 \leq M_{V} \leq -25$, allowing them to be compared directly to each other and to the quasars in our lower-redshift samples (M99; D03; F04). However, the results presented by K01 were at odds with this criterion, with one $z\sim 2$ quasar shining close to $M_V=-27$, and the $z\approx2$ RQQ's being significantly lower in total optical luminosity than their RLQ counterparts. This appears to be partly due to errors in the (highly heterogeneous) photometric measurements used to define the sample (which pre-dated SDSS and 2QZ), as well as uncertain $K$-corrections from  observed to rest-frame $V$ at significant redshifts. 

The RQQs were derived from the~\cite{Boyle:UVQSO} UVX survey (``SGP'' objects) or~\citet{marshall+84_QSOsamp} survey (``BVF'' objects), and confirmed as radio quiet ($L_{5GHz} < 10^{24.5}$W Hz$^{-1}$sr$^{-1}$) by the Very Large Array (VLA) FIRST\footnote{http://sundog.stsci.edu/} and NVSS\footnote{http://www.cv.nrao.edu/nvss/} surveys. RLQs were selected from the V\'{e}ron-Cetty \& V\'{e}ron (1993) quasar catalogue, with reported 5-GHz radio luminosities $L_{5GHz} > 10^{25.5}$W~Hz$^{-1}$ sr$^{-1}$ and steep radio spectra to ensure that their intrinsic radio luminosities are not boosted by relativistic beaming. The NICMOS observations of two quasars in the original K01 sample (the RQQ ``SGP2:47'' at $z=0.830$ and the RLQ ``1148$+$56W1'' at $z=1.782$) were affected by technical problems, and so these were not re-observed with WFPC2. We also note that the WFPC2 image of SGP5:46 was contaminated by diffraction spikes from a nearby bright star and had to be excluded from the present study.

\subsubsection{Present Sample}
\label{sec-presamp}
The current sample therefore consists of eight objects at \zo\ and nine at \zt, for which the radio and optical properties are listed in Table~\ref{tab-samp}.
The updated redshift versus rest-frame optical luminosity distribution for our sample is shown on the left of Fig.~\ref{fig-samp}, using the HST rest-frame $V$ photometry presented in K01. The QSO's span the knee of the quasar luminosity function at each redshift~\citep{croom+04}, and the radio quiet subsamples are still reasonably well matched to the optical luminosities of our $z\approx0.2$ and 0.4 samples (M99, D03, F04). The shrinking of, and biases inherent within, our sample affect our original aim of exploring the differences between the RLQ and RQQ populations. We focus instead on providing a benchmark colour for all quasar host galaxies at these redshifts. We note that the conclusion drawn by K01 that RQQ hosts are less luminous than RLQ hosts, is potentially biased at $z\approx2$ by the {\em a posteori} luminosity distribution of the quasars themselves (see Fig.~\ref{fig-samp}).
This is discussed further in sections~\ref{sec-kc} and~\ref{sec-RL}.

\subsection{Choice of filters}
\label{sec-filt}
We chose filters that maximise sensitivity to the underlying starlight in the wavelength range of interest, while avoiding strong galaxy emission lines that might positively bias our measurement of the host galaxy luminosity. Our NICMOS observations sampled the rest-frame $V$-band continuum at each redshift, avoiding the strong emission lines [O{\sc iii}]$\lambda5007$ and H$\alpha$ using the F110M ($\approx J$) filter at $z\approx 1$ and F165M ($\approx H$) at $z\approx 2$. With WFPC2 we targetted the objects' rest-frame near-ultraviolet (UV) continuum, in the region $\approx 2300-3400$\AA, corresponding closely to $U$ band. We use WFPC2 F300W as our reference rest-frame $U$ band throughout this work. 

For the $z\approx 1$ sample we have used the F606W (Wide-$V$) filter, and for $z\approx 2$ the F814W (Wide $I$) filter in order to maximise throughput. We have avoided strong galaxy emission lines such as [O~{\sc ii}]$\lambda3727$. The absence of strong galaxy emission lines in the observed bandpass is important as they can render the images sensitive to any large scale regions of emission-line gas which might be present. These are directly associated with the nuclear activity and can mask or confuse our picture of the underlying starlight. However, the quasar broad emission line Mg~{\sc ii}$\lambda$~2796 is included by using such wide filters (see Fig.~\ref{fig-samp}, right panel). The equivalent width of Mg~{\sc ii} in the SDSS composite quasar spectrum~\citep{VandenBerk:2001p3354} is $\approx 32$\AA, and will contribute a slight (up to 0.03~mag) positive bias to our nuclear luminosities. However, since Mg~{\sc ii} arises very close to the nucleus and such high ionization lines are only found in absorption in galaxies, our host galaxy luminosities are unaffected.

\subsection{Extinction and $K$-corrections}
\label{sec-ckc}
For the estimation of luminosities at both radio and optical wavelengths throughout this work, we have assumed a quasar spectrum of the form $F_\nu\sim\nu^{-0.5}$. $K$-corrections are applied to convert to rest-frame magnitudes (see section~\ref{sec-kc}), where we revisit this assumption. We have converted the original counts from the quasar images of K01 to AB mags for comparison in this paper. For galactic extinction corrections we use colour excess values from the map of~\citet{schlegel+98} and a diffuse interstellar dust model ($R_V=3.1$). All magnitudes are AB (so that a flat energy distribution in $F_\nu$ corresponds to zero colour). 

\subsection{Saturation and integration times}
Integration times were calculated to allow the WFPC2 images to reach a similar depth, and to detect the host galaxies out to a similar radius, as in the previous NICMOS observations. To do this we made the null hypothesis that the host galaxies are passively evolving. Any star formation would thus serve to boost the galaxies' luminosities, making them easier to detect. We therefore observed each of the $z \approx 1$ quasars for two orbits through the F606W filter, and each of the $z \approx 2$ quasars for four orbits through the F814W filter.  Each HST orbit was sufficient to include three long exposures (600 or 700~s), plus several snapshots. The latter provided a good characterisation of the bright nuclear point source in the event that this region of the image was saturated in the longer integrations. This resulted in a total on-source integration time of 3600~s for the $z\approx 1$ objects, and 7800~s for those at $z\approx 2$ (see Table~\ref{tab-obs}).

\subsection{Point Spread Function}
\label{sec-psf}
Experience from our previous WFPC2 \& NICMOS imaging studies of quasar host galaxies (M99; K01; D03, F04) has demonstrated that it is essential to obtain a deep, high-dynamic range stellar Point Spread Function (PSF) through each filter in order to accurately characterise the critical outer regions of the quasar. Synthetic TINYTIM\footnote{http://www.stsci.edu/software/tinytim/tinytim.html} PSFs, whilst an excellent match to the inner $\approx 1$~arcsec, fail to adequately reproduce the complex and variable outer structure of the PSF wings. Archive PSFs are rarely precisely centred on the same region of the detector as the target quasars (thus changing the structure of the outer region of the PSF), and do not always have the requisite dynamic range. In any case the form of the WFPC2 PSF changes systematically with time and so measurements made as close to the epoch of the observations as possible are highly desirable. For this study we therefore adopted the same strategy as used in our previous HST programs. An entire orbit was devoted to obtaining high-dynamic range stellar PSFs through each of our two chosen filters. The star selected (GRW$+$70D5824) is a white dwarf, with a flat spectral shape over the wavelengths of interest which is very similar to that of a typical quasar, thus ensuring the best possible match to the nuclear PSF of each target source. Into the centre of this deep, empirical stellar PSF we splice a TINYTIM realisation of the theoretical PSF at the location of the centre of the quasar on the detector. The quality of fit is strongly dependent on the focus term in the TINYTIM model, which affects the amount of light in the central peak. Focus changes significantly during an orbit due to thermal ``breathing'' of the telescope structure\footnote{http://www.stsci.edu/hst/observatory/focus/}. We tested each observation with a range of focus values by performing a $\chi^2$ minimisation on a fit of the PSF to the central region of each quasar (best fits acceptable at $3\sigma$ level). In each case the adopted PSF had a focus close to the value determined for the telescope at the time of observation. 

\begin{table*}
\begin{minipage}{175mm}
\caption{\label{tab-res1} 
Best fit models for the quasars at \zo\ (all Elliptical). Columns are as follows: object name; 
reduced-$\chi^2$ value for the best fit model; $\Delta \chi^2$ between the elliptical and disk-morphology model; number of degrees of freedom in fit, $\nu$; effective radius, $R_{e}$, of best fitting galaxy model in kpc; surface brightness of the host at the effective radius, $\mu_{e}$, in units of $V_\mathrm{AB}$ mag.arcsec$^{-2}$; integrated apparent magnitudes of the nucleus and the host galaxy in $V_\mathrm{AB}$-band; the ratio of integrated nuclear and host galaxy luminosities; position angle of the host (in degrees East of North); the axial ratio of the host. All fluxes are presented as apparent AB magnitudes, uncorrected for galactic extinction. }
\begin{center}
\begin{tabular}{lrrrllllrrr}
\hline
\hline
Object&$\chi^{2}_{red}$&$\Delta\chi^{2}$&$\nu$&$R_{e}$&$\mu_{e}$&$V_\mathrm{AB}^\mathrm{nuc}$&$V_\mathrm{AB}^\mathrm{host}$&$L_{N}/L_{H}$&PA&$a/b$\\
	& & & & /kpc & /mag.arcsec$^{-2}$ & /mag & /mag & &/$^{\circ}$& \\
\hline
\multicolumn{11}{c}{\bf Radio-Quiet Quasars}\\
BVF225 	& 1.122 & 134 & 1526 & $5.5\pm0.2$ & $23.9\pm0.3$ & $18.5\pm0.1$ & $21.2\pm0.1$ & $12.6$ & $176$ & $1.24$ \\
BVF247 	& 1.143 & 97 & 1395 & $4.5\pm0.2$ & $24.5\pm0.5$ & $20.4\pm0.1$ & $22.3\pm0.1$ & $6.1$ & $155$ & $1.25$ \\
BVF262 	& 1.077 & 82 & 1512 & $3.0\pm0.1$ & $23.2\pm0.5$ & $20.0\pm0.1$ & $22.0\pm0.1$ & $5.9$ & $4$ & $1.33$ \\
\multicolumn{11}{c}{\bf Radio-Loud Quasars}\\
PKS0440 	& 1.097 & 55 & 1797 & $11.1\pm0.5$ & $25.5\pm0.4$ & $18.9\pm0.1$ & $21.3\pm0.1$ & $9.2$ & $47$ & $1.12$ \\
PKS0938 	& 1.106 & 45 & 2195 & $3.8\pm0.2$ & $23.3\pm0.6$ & $19.3\pm0.1$ & $21.5\pm0.1$ & $7.4$ & $154$ & $1.20$ \\
3C422 	& 1.073 & 24 & 2812 & $2.7\pm0.1$ & $22.2\pm0.4$ & $19.2\pm0.1$ & $21.2\pm0.1$ & $6.2$ & $87$ & $1.57$ \\
MC2112 	& 1.040 & 3 & 1806 & $1.8\pm0.1$ & $20.5\pm0.1$ & $19.9\pm0.1$ & $20.2\pm0.1$ & $1.4$ & $102$ & $1.74$ \\
4C02.54 	& 1.070 & 31 & 2360 & $6.3\pm0.1$ & $24.2\pm0.1$ & $18.8\pm0.1$ & $21.4\pm0.1$ & $10.2$ & $111$ & $1.42$ \\
\hline
\end{tabular}
\end{center}
\end{minipage}
\end{table*}
\begin{table*}
\begin{minipage}{175mm} 
\caption{\label{tab-res2} 
Best fit models for the quasars at \zt\ (all Elliptical). Columns are as follows: object name; 
reduced-$\chi^2$ value for the best fit model; $\Delta \chi^2$ between the elliptical and disk-morphology model; number of degrees of freedom in fit, $\nu$; effective radius, $R_{e}$, of best fitting galaxy model in kpc; surface brightness of the host at the effective radius, $\mu_{e}$, in units of $I_\mathrm{AB}$ mag.arcsec$^{-2}$; integrated apparent magnitudes of the nucleus and the host galaxy in $I_\mathrm{AB}$-band; the ratio of integrated nuclear and host-galaxy luminosities; position angle of the host (in degrees East of North); the axial ratio of the host. 
All fluxes are presented as apparent AB magnitudes, uncorrected for galactic extinction. 
}
\begin{center}
\begin{tabular}{lrrrllllrrr}
\hline
\hline
Object&$\chi^{2}_{red}$&$\Delta\chi^{2}$&$\nu$&$R_{e}$&$\mu_{e}$&$I_\mathrm{AB}^\mathrm{nuc}$&$I_\mathrm{AB}^\mathrm{host}$&$L_{N}/L_{H}$&PA&$a/b$\\
	& & & & /kpc & /mag.arcsec$^{-2}$ & /mag & /mag & &/$^{\circ}$& \\
\hline
\multicolumn{11}{c}{\bf Radio-Quiet Quasars}\\
SGP2:36 	& 1.244 & 115 & 1164 & $1.6\pm0.1$ & $22.3\pm0.4$ & $21.5\pm0.1$ & $22.5\pm0.1$ & $2.5$ & $136$ & $1.29$ \\
SGP2:25 	& 1.295 & 61 & 1106 & $1.8\pm0.1$ & $23.2\pm0.1$ & $20.8\pm0.1$ & $23.1\pm0.2$ & $8.3$ & $98$ & $1.21$ \\
SGP2:11 	& 1.169 & 5 & 1101 & $4.0\pm0.8$ & $25.2\pm2.0$ & $20.6\pm0.1$ & $23.4\pm0.3$ & $12.3$ & $168$ & $1.15$ \\
SGP3:39 	& 1.325 & 51 & 1189 & $2.2\pm0.2$ & $22.8\pm0.3$ & $20.6\pm0.1$ & $22.5\pm0.1$ & $5.7$ & $107$ & $1.13$ \\
SGP4:39 	& 1.031 & 2 & 1041 & $1.5\pm1.9$ & $22.1\pm2.1$ & $20.7\pm0.1$ & $22.5\pm0.2$ & $5.1$ & $81$ & $1.09$ \\
\multicolumn{11}{c}{\bf Radio-Loud Quasars}\\
PKS1524 	& 1.198 & 39 & 1121 & $2.2\pm0.2$ & $22.0\pm0.6$ & $19.6\pm0.1$ & $21.5\pm0.1$ & $5.5$ & $163$ & $1.31$ \\
B2~2156 	& 1.335 & 45 & 1110 & $3.6\pm0.1$ & $22.3\pm0.1$ & $19.4\pm0.1$ & $20.7\pm0.1$ & $3.4$ & $39$ & $1.10$ \\
PKS2204 	& 1.300 & 25 & 981 & $1.6\pm0.1$ & $21.3\pm0.2$ & $20.8\pm0.1$ & $21.6\pm0.1$ & $2.0$ & $137$ & $1.48$ \\
4C45.51 	& 1.767 & 50 & 1119 & $2.2\pm0.1$ & $22.0\pm0.2$ & $20.6\pm0.1$ & $21.5\pm0.1$ & $2.2$ & $78$ & $1.13$ \\
\hline
\end{tabular}
\end{center}  
\end{minipage}
\end{table*}
\section{Data reduction and analysis}
\label{sec-dr}
\subsection{Data reduction}
Data reduction was carried out using the HST/WFPC2 pipeline and standard procedures in {\sc iraf}. An additional step was required for the images of the PSF star and those quasars for which the nucleus had saturated in the longer exposures (specifically, all nine quasars in the sample at $z\approx 1$ plus SGP2:25 from the $z\approx 2$ sample). The saturated regions of these images were replaced by the corresponding regions (suitably scaled) from the unsaturated snapshot image of the same object.

\begin{figure*}
\centering 
{\includegraphics[width=55mm]{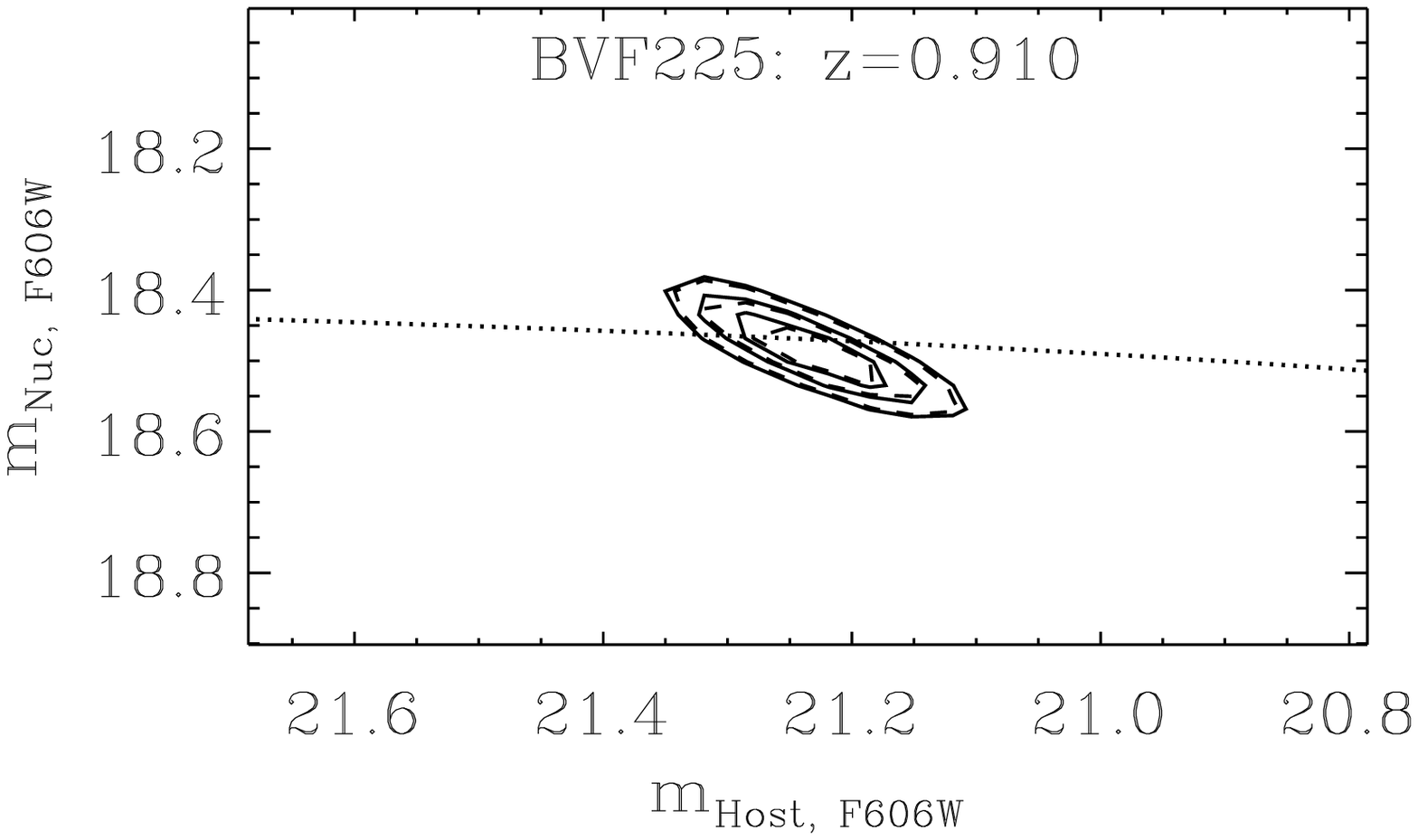}}
{\includegraphics[width=55mm]{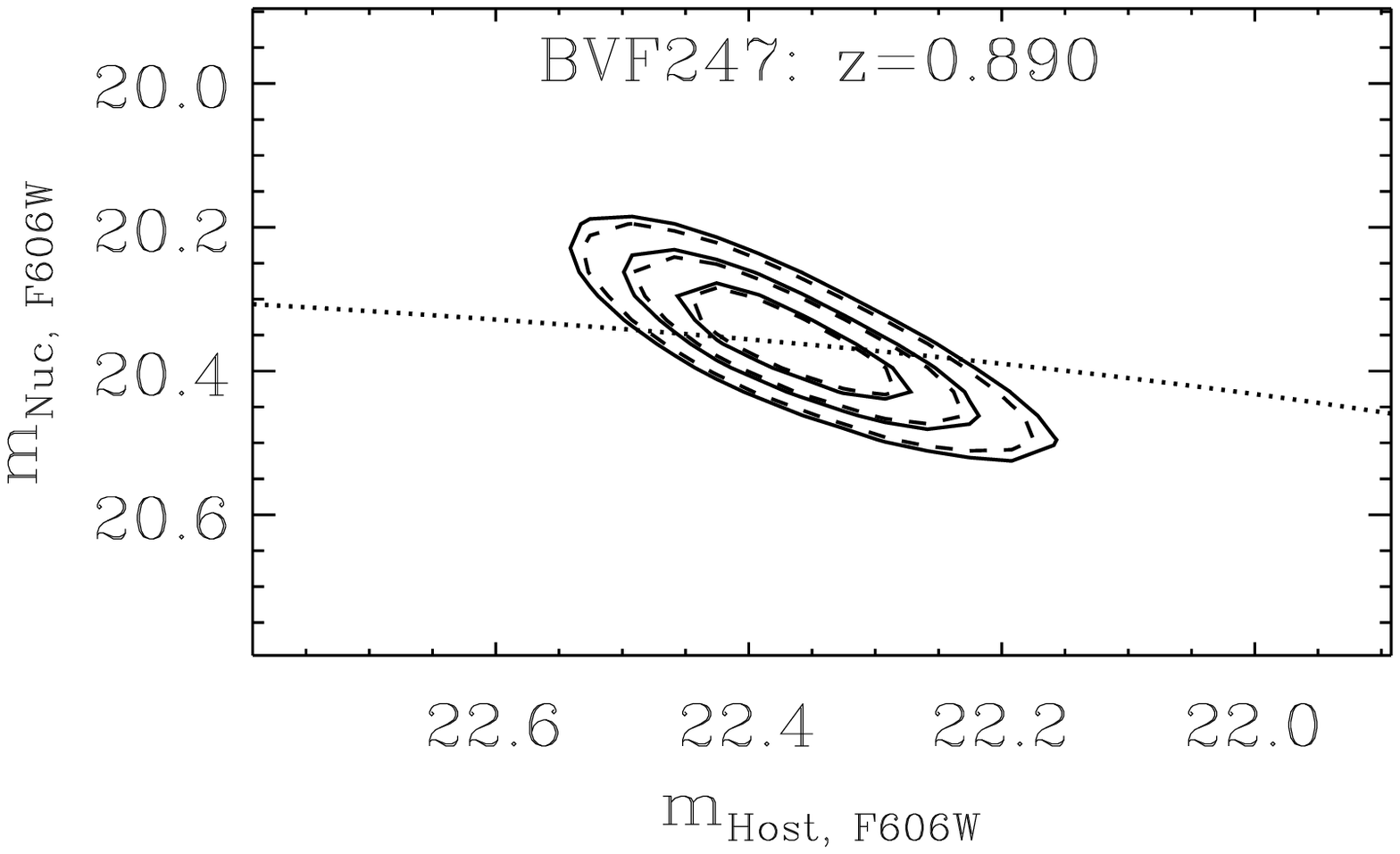}}
{\includegraphics[width=55mm]{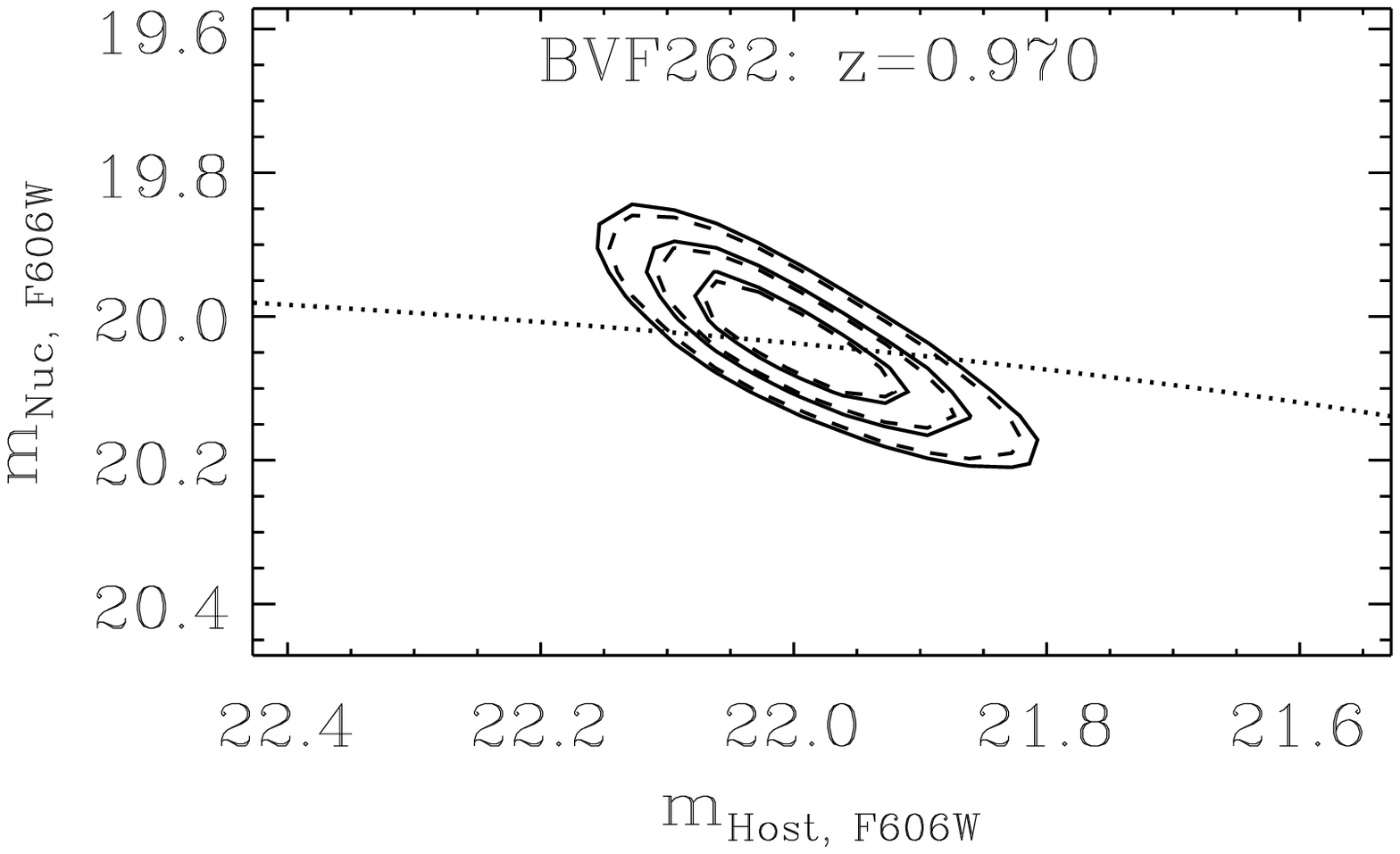}}
\vspace{3mm}

{\includegraphics[width=55mm]{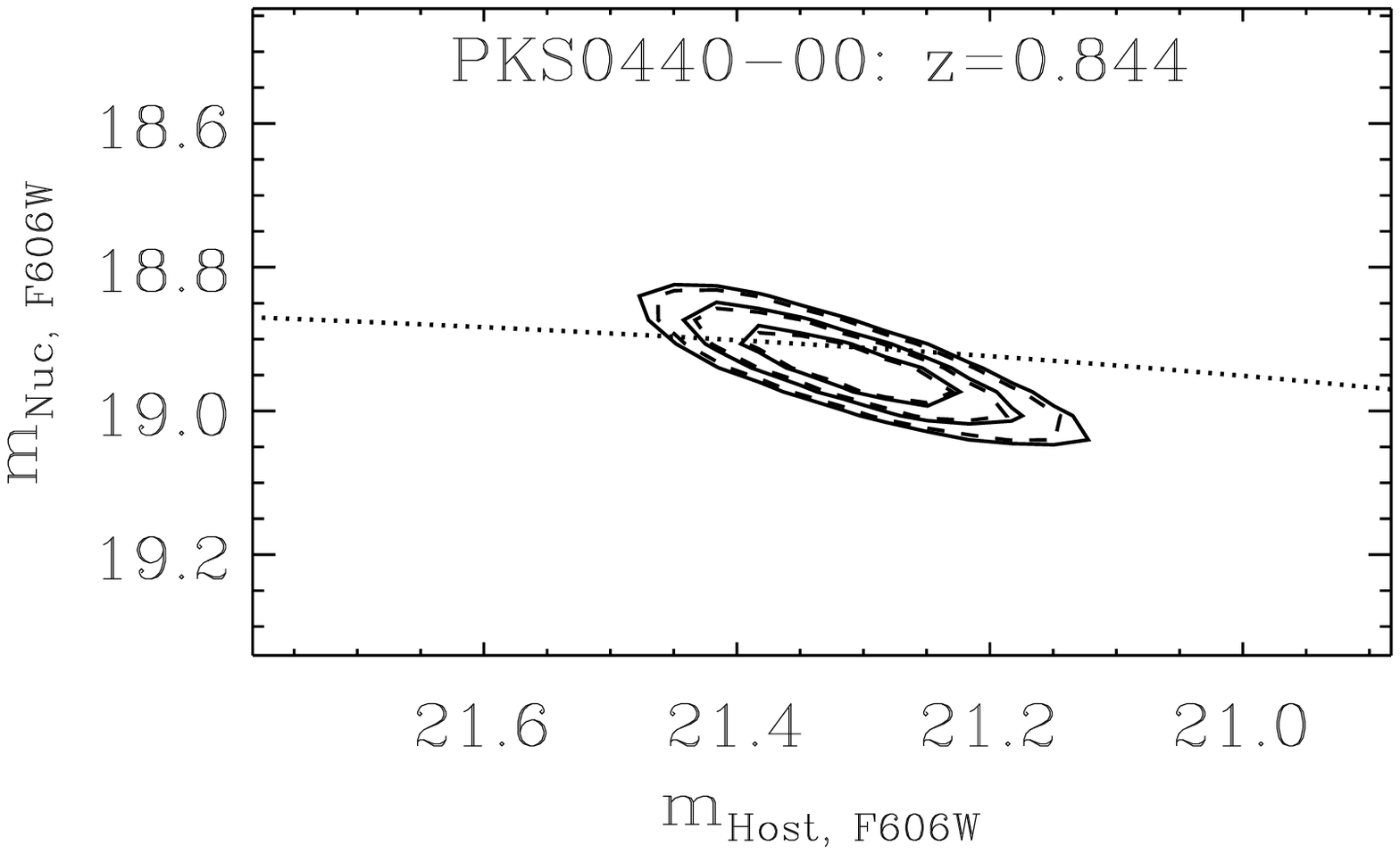}}
{\includegraphics[width=55mm]{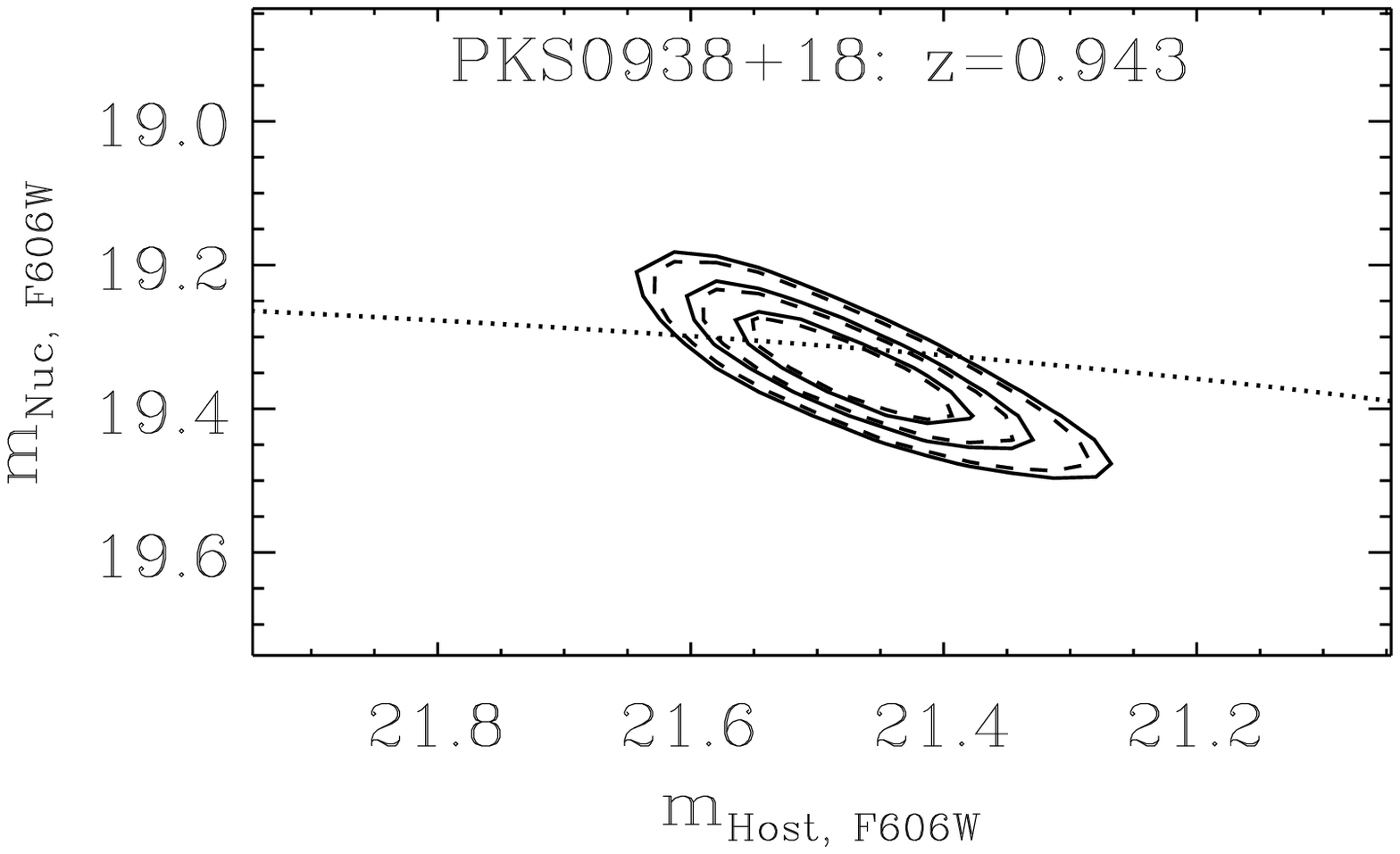}}
{\includegraphics[width=55mm]{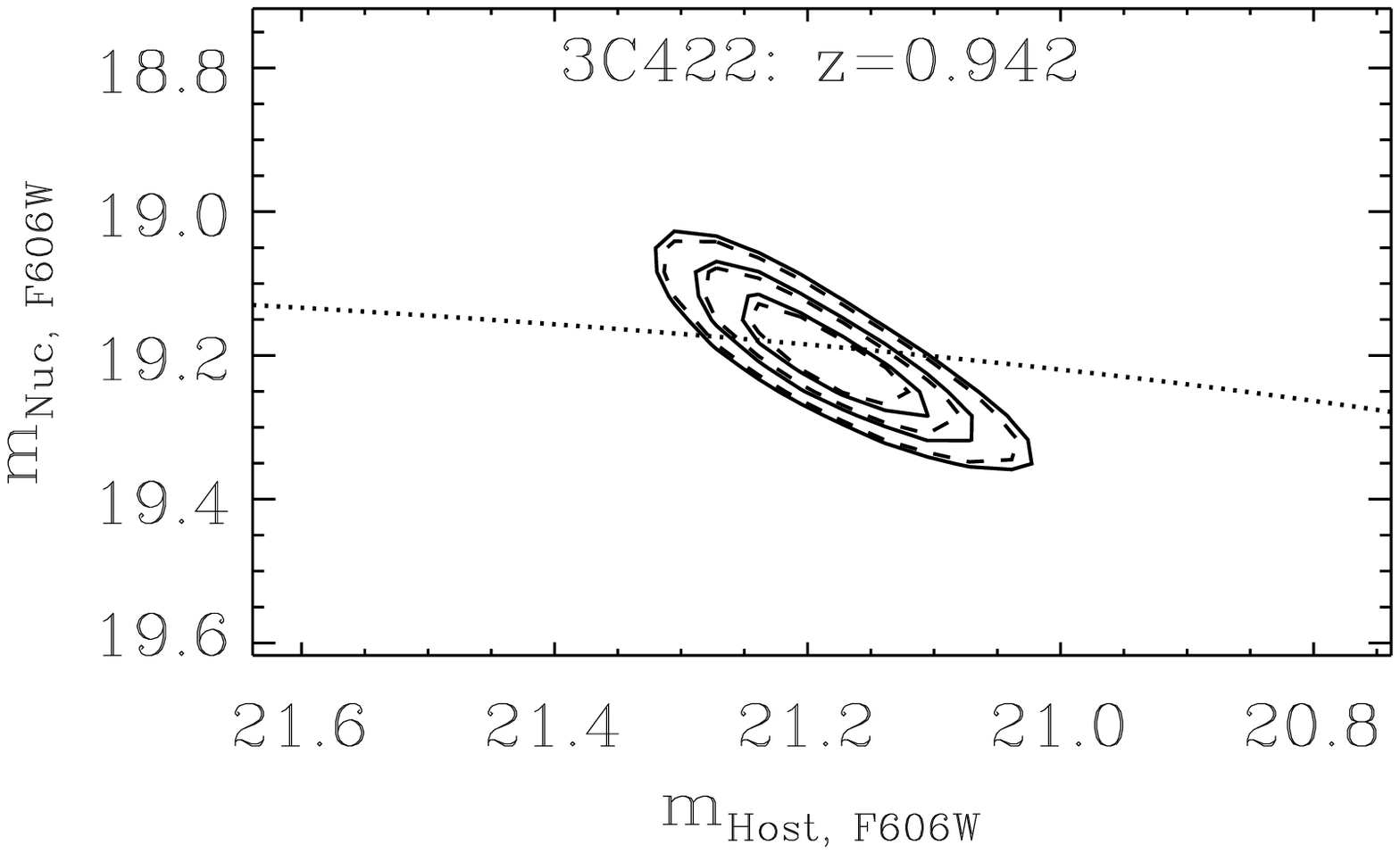}}
\vspace{3mm}

{\includegraphics[width=55mm]{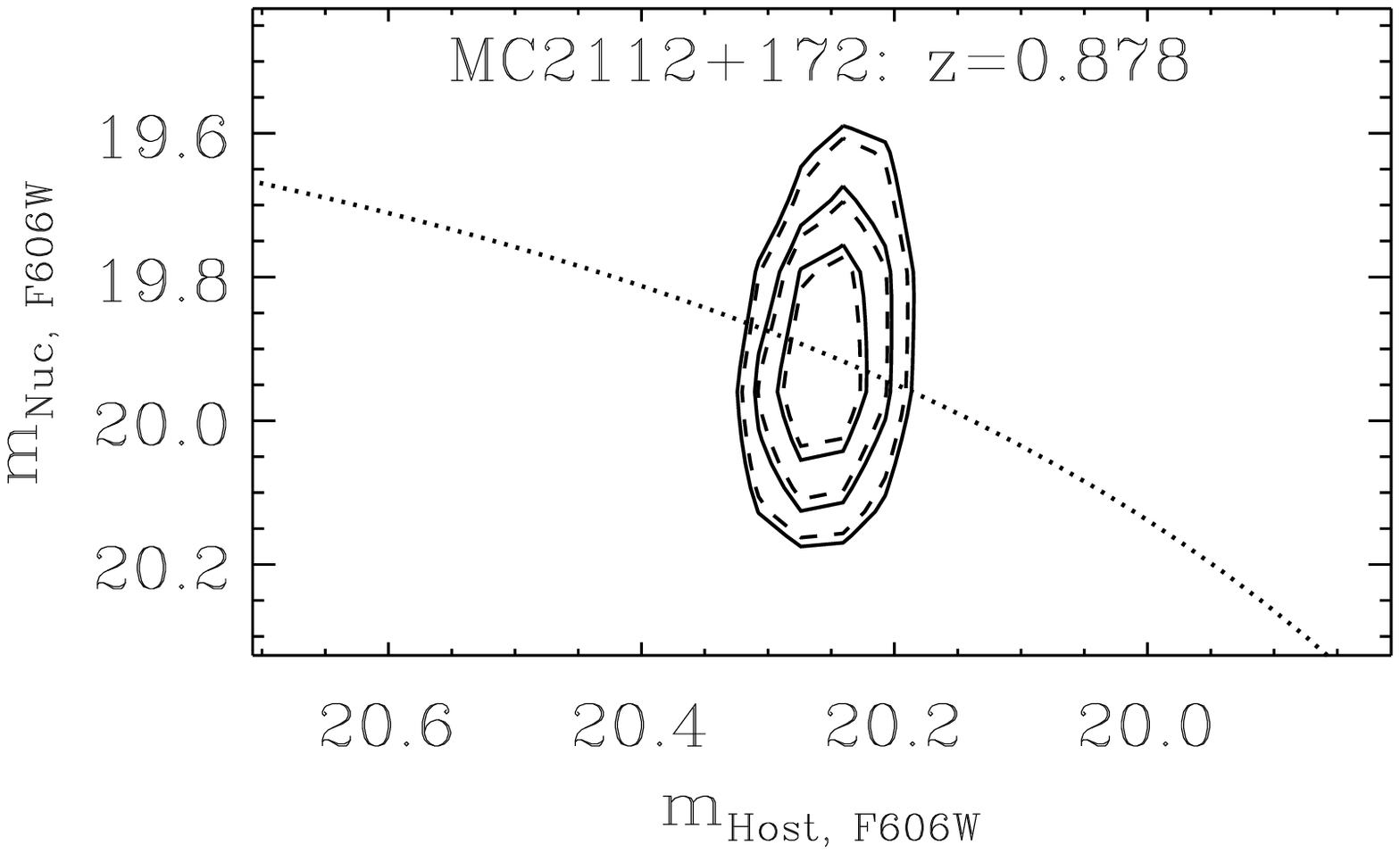}}
{\includegraphics[width=55mm]{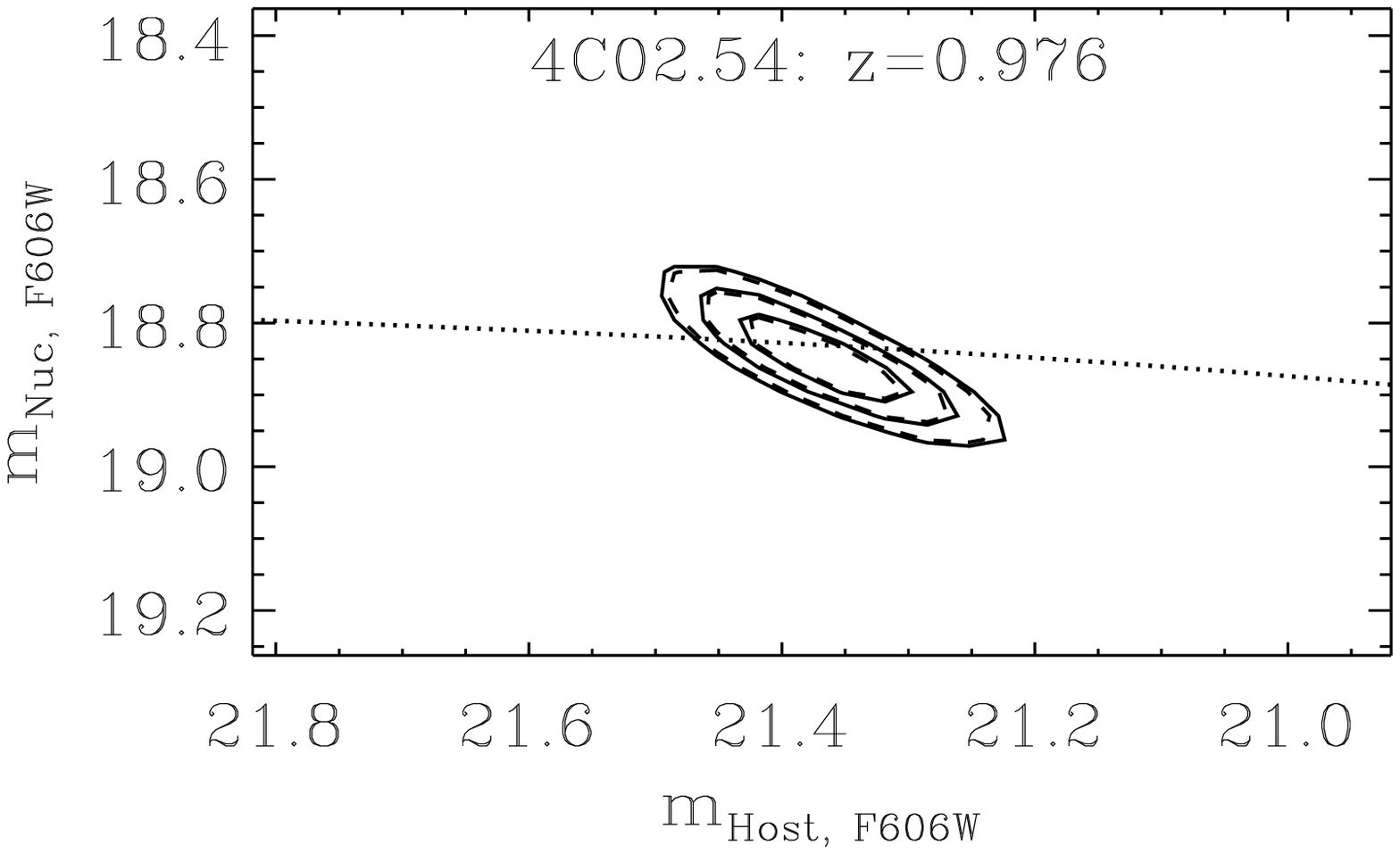}}
{\includegraphics[width=55mm]{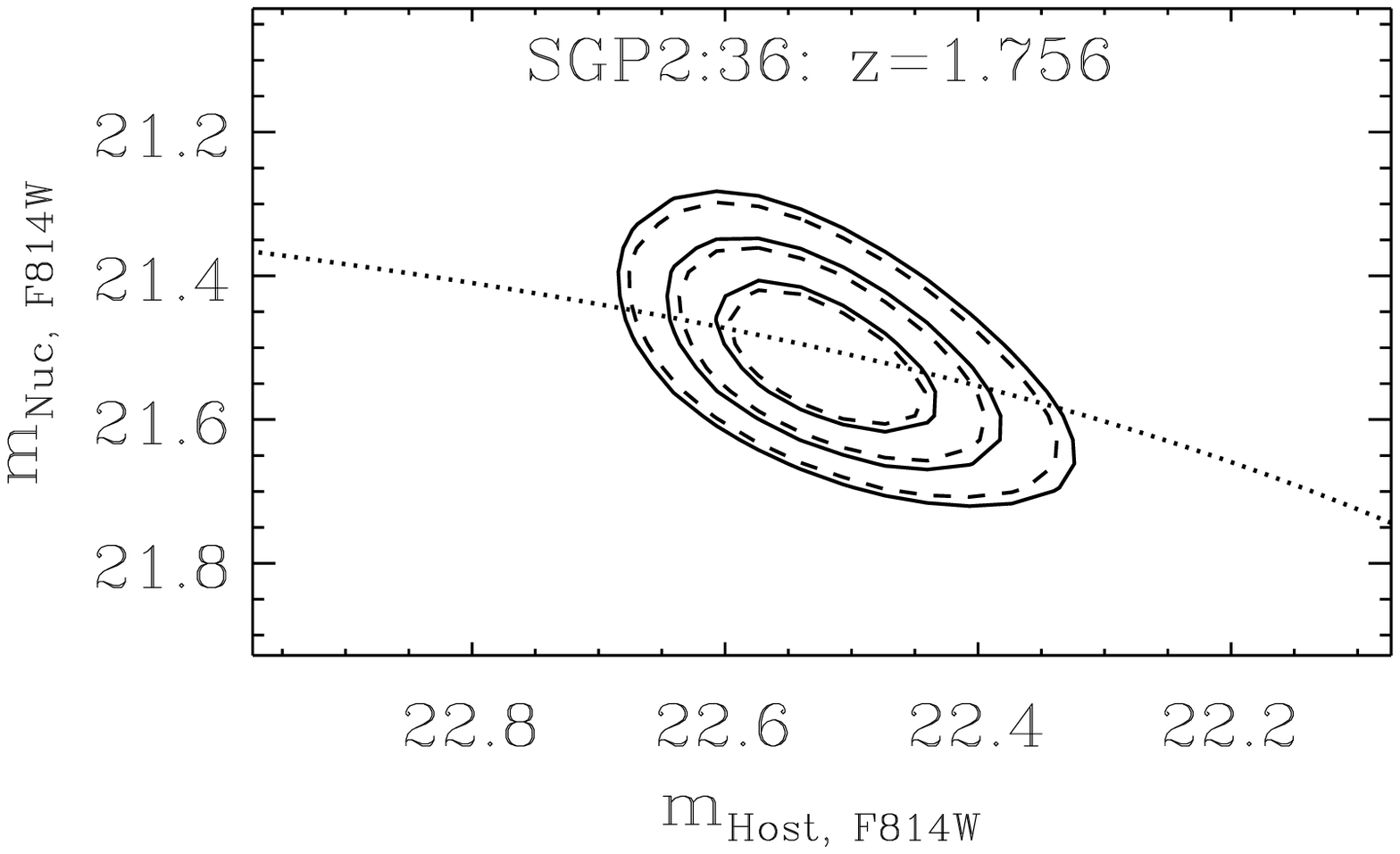}}
\vspace{3mm}

{\includegraphics[width=55mm]{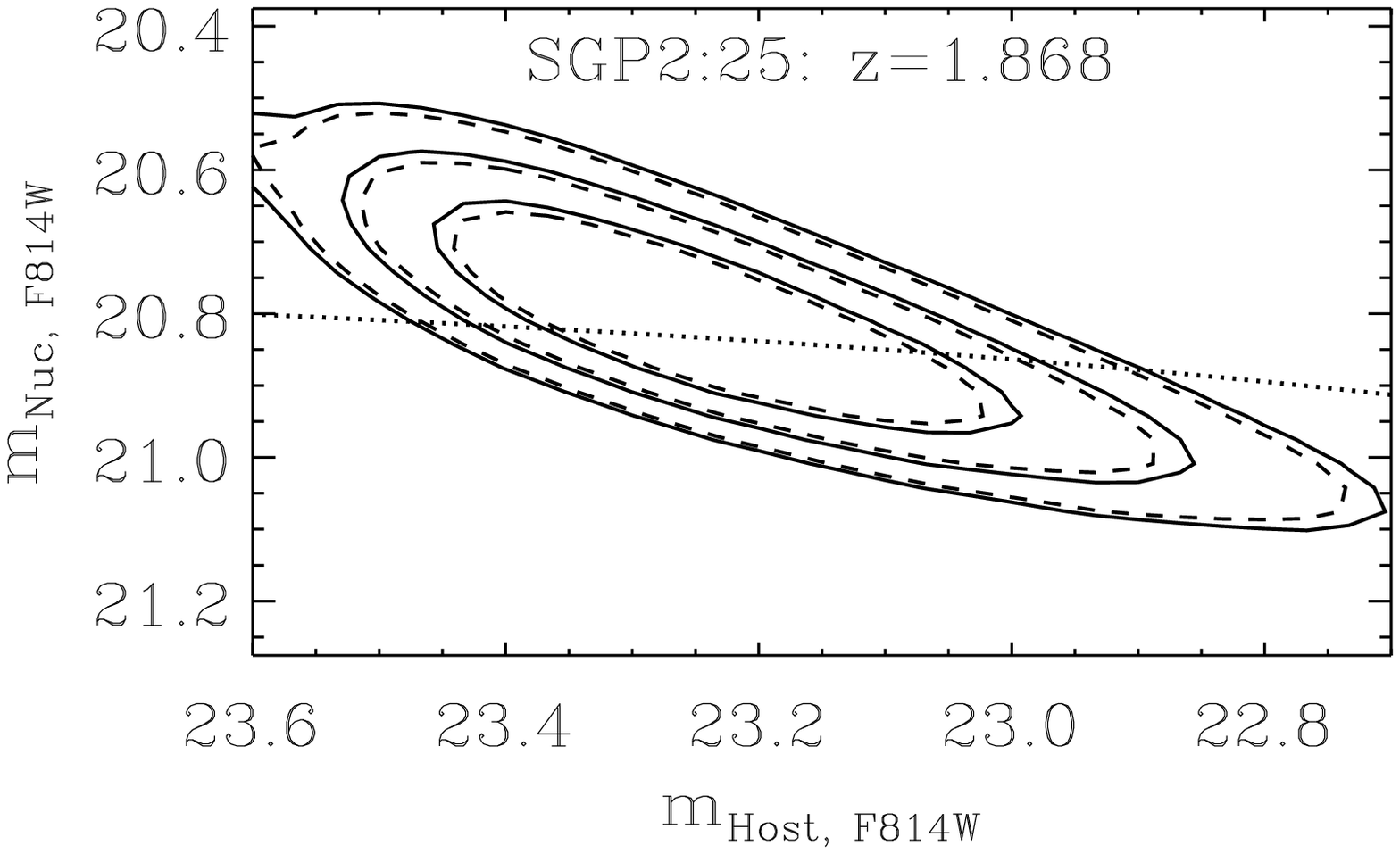}}
{\includegraphics[width=55mm]{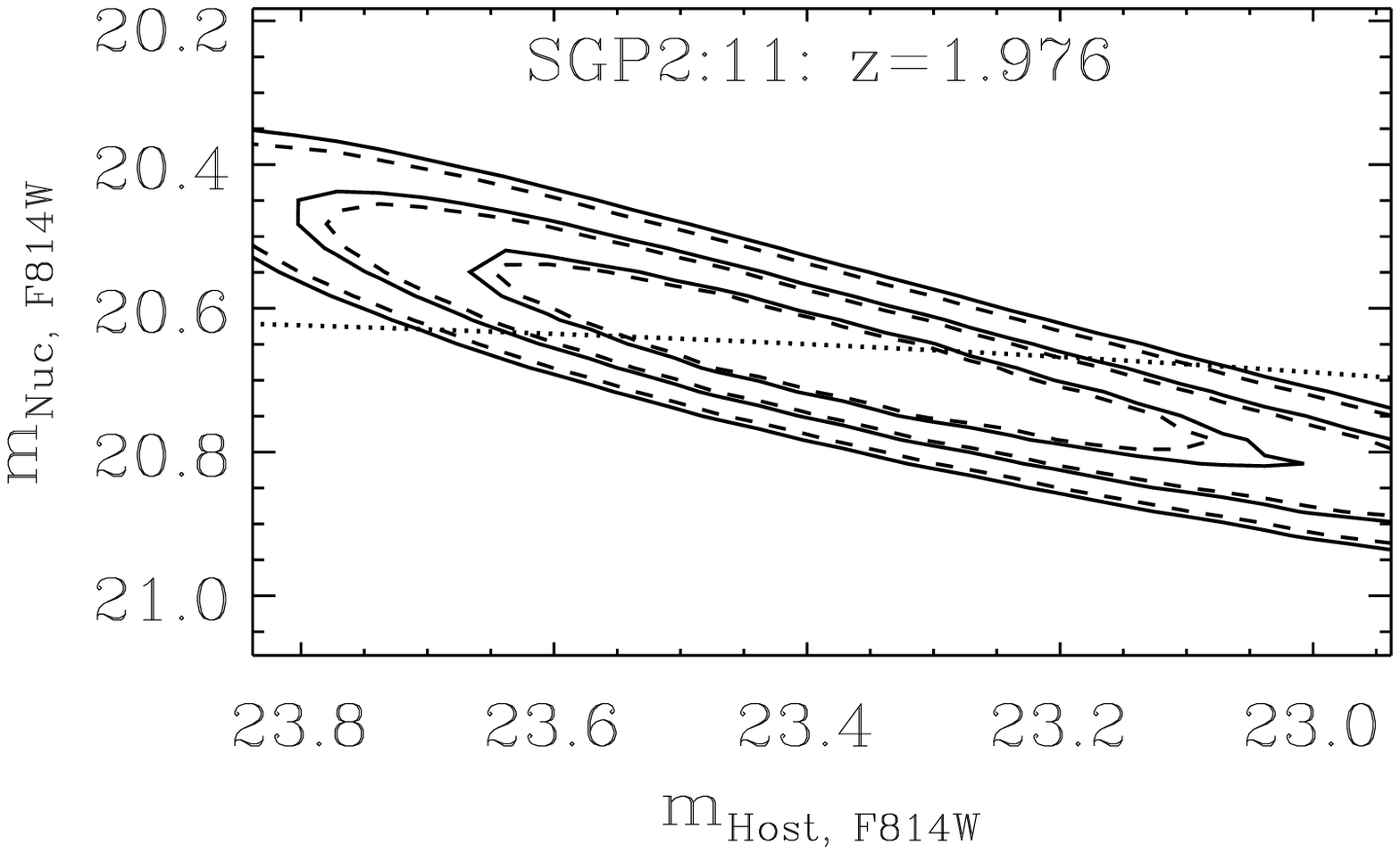}}
{\includegraphics[width=55mm]{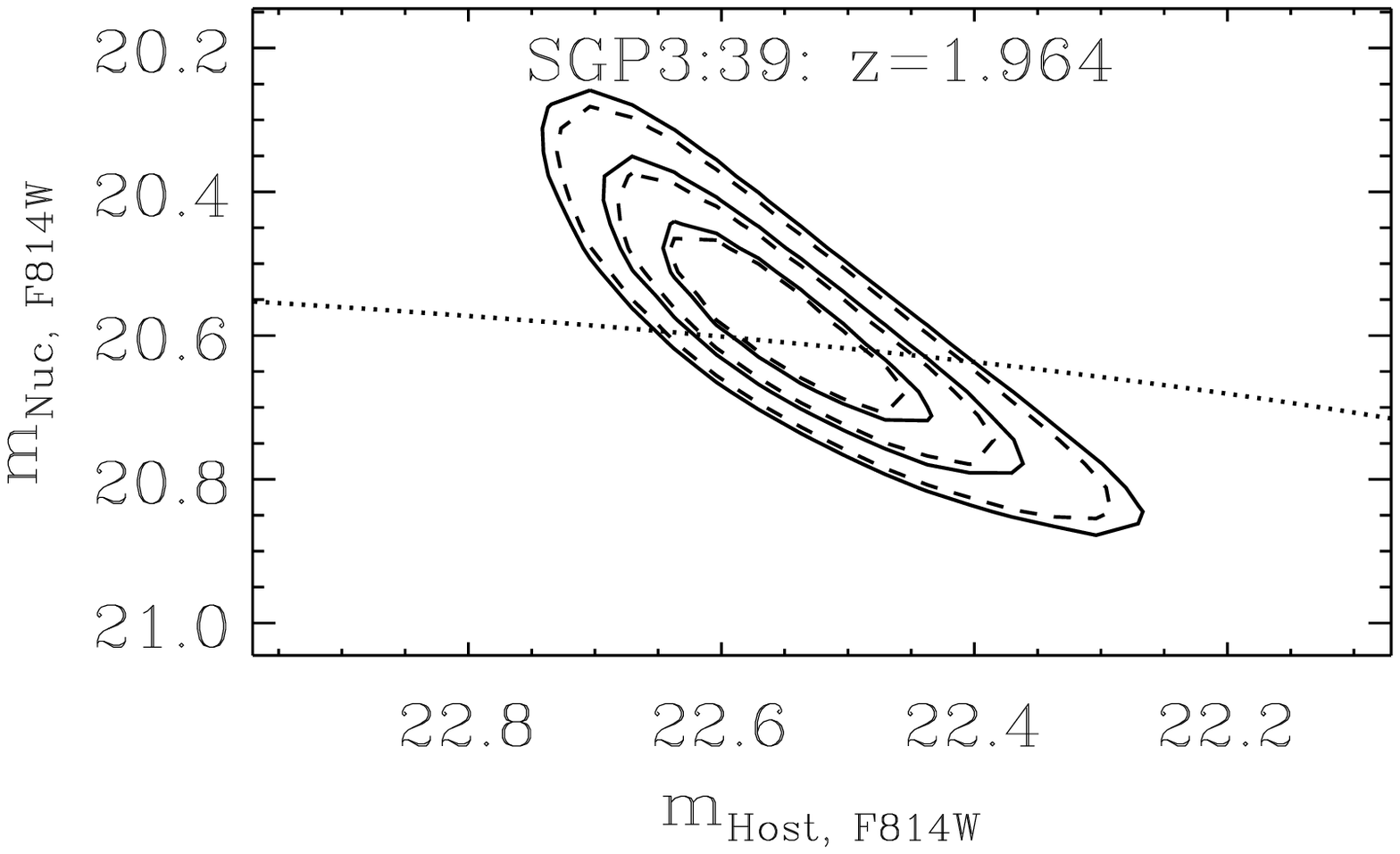}}
\vspace{3mm}

{\includegraphics[width=55mm]{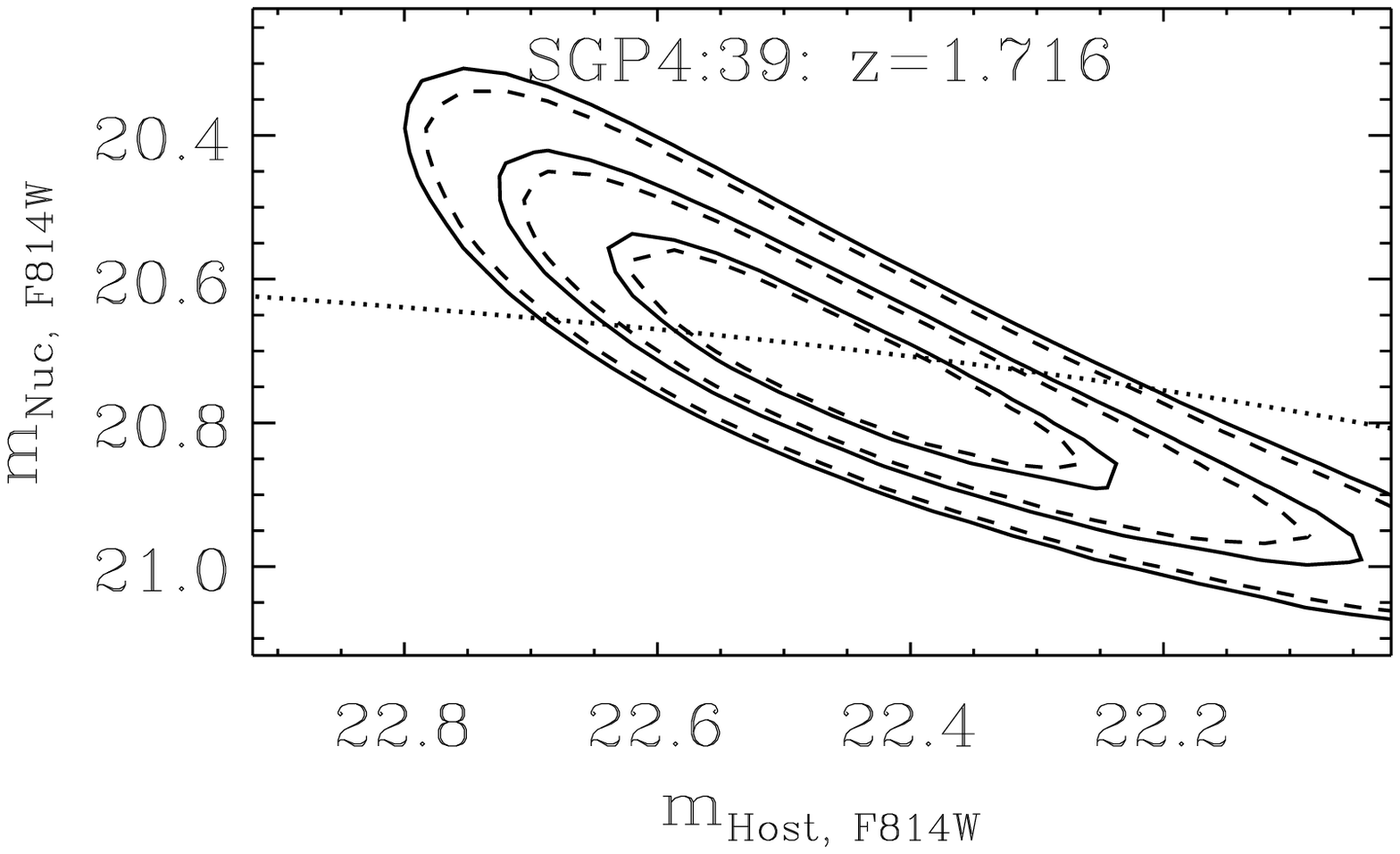}}
{\includegraphics[width=55mm]{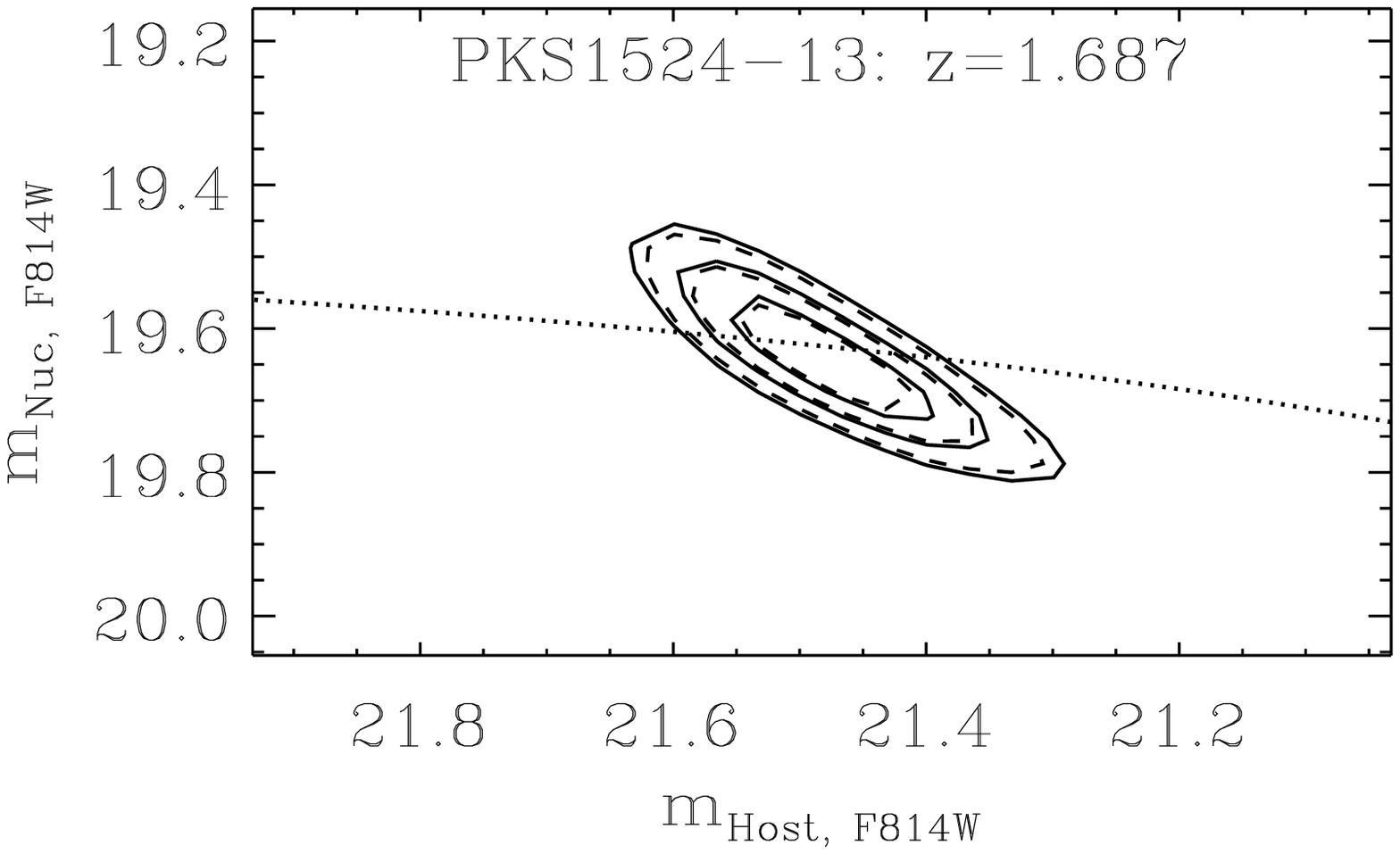}}
{\includegraphics[width=55mm]{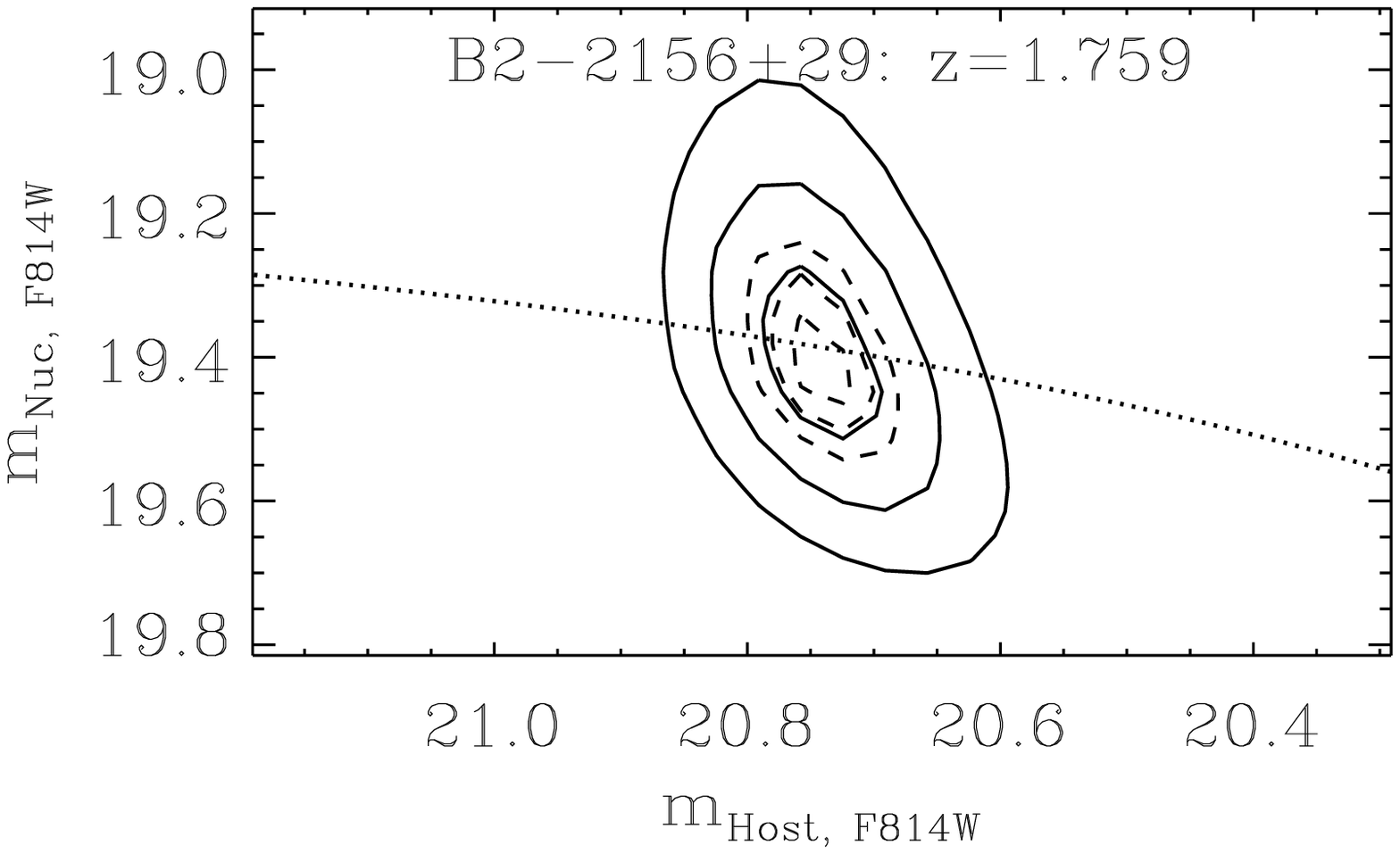}}
\vspace{3mm}

{\includegraphics[width=55mm]{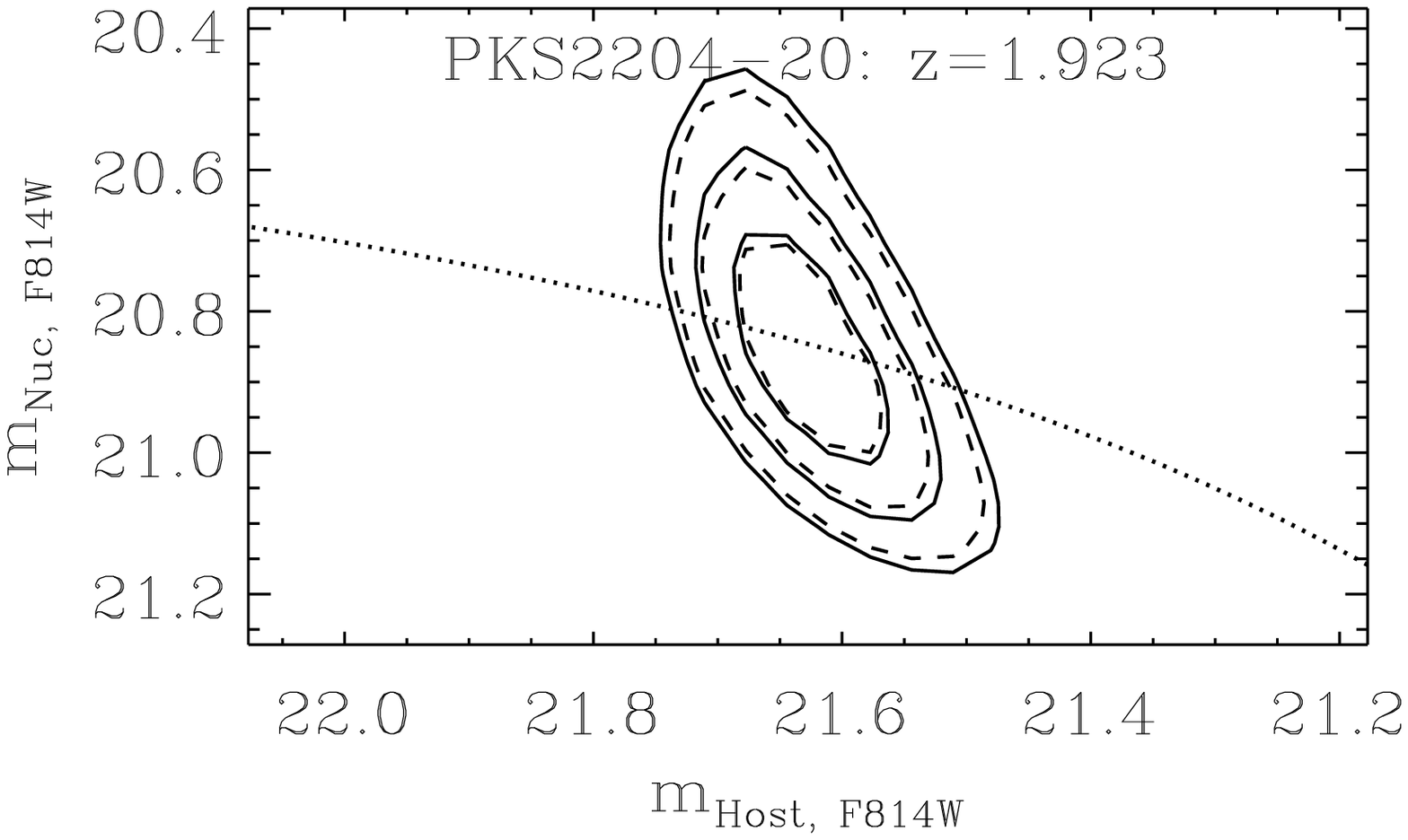}}
{\includegraphics[width=55mm]{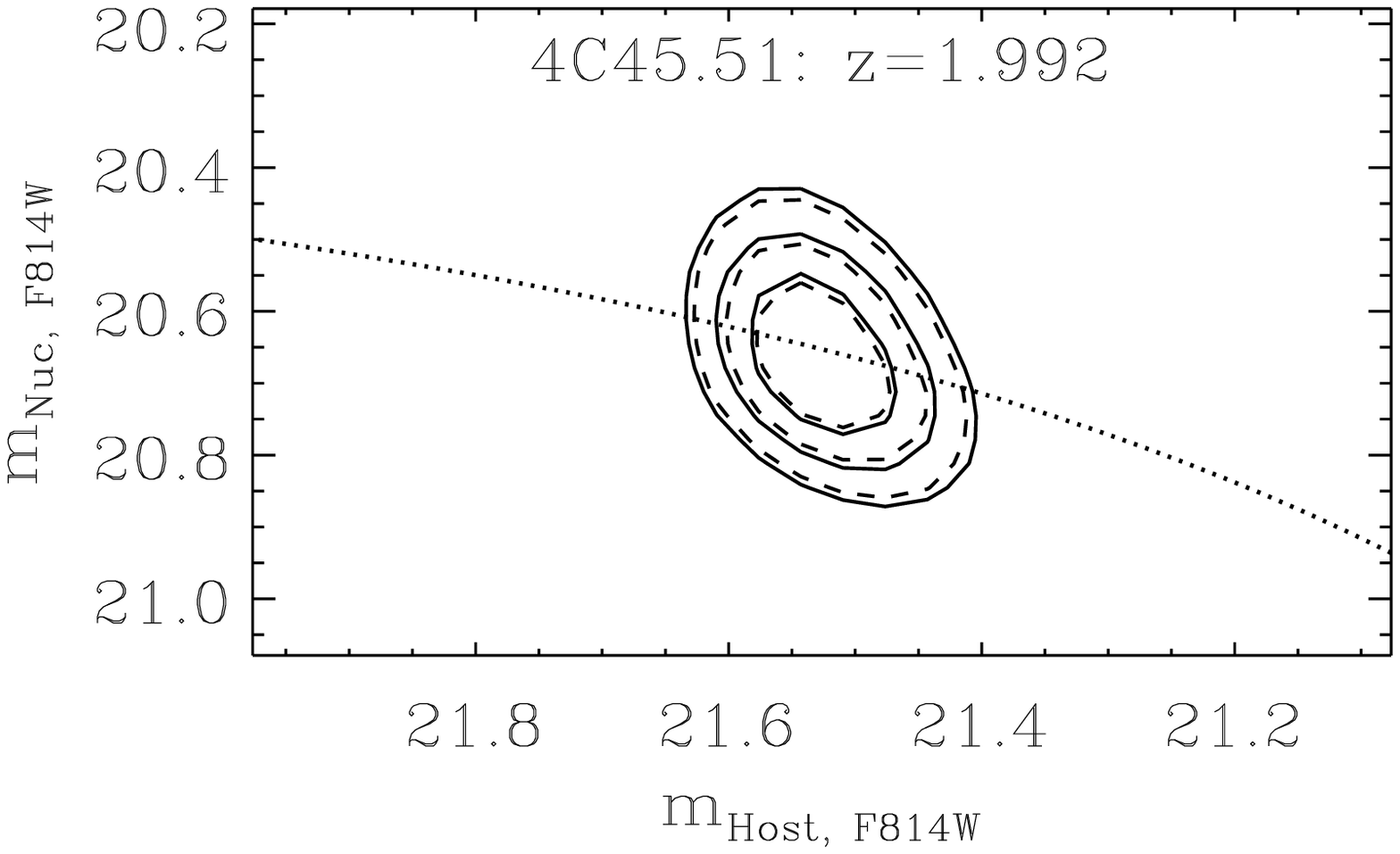}}
\caption{\label{fig-cont} $\Delta\chi^2$ contours (at 1, 2 and 3$\sigma$ for the host galaxy -- solid; and nucleus -- dashed) in the $m_\mathrm{Host}-m_\mathrm{Nuc}$ plane. All magnitudes are AB. The dotted lines indicate constant total flux. The host-galaxy fluxes are constrained to within $\pm0.3$~mag, and the nuclear fluxes generally somewhat better. Those that deviate (MC2112$+$172, B2~2156$+$29, PKS2204$-$20 and 4C45.51) show significant circumnuclear residual flux (Figs.~\ref{fig-h108},~\ref{fig-h1017},~\ref{fig-h1020},~\ref{fig-h1018}), that is confused with the nucleus, thus weakening our constraint on its total flux (see section~\ref{sec-det}).}
\end{figure*}

\subsection{Two-dimensional modelling}
\label{sec-mod}
In order to separate the galaxy and nuclear components of the quasar image we used the 2-D modelling software developed for our previous host-galaxy studies (see M99; F04 for details) and tested and described in detail by~\citet{mclure+00,floydPhDT,floyd+08}. The code uses downhill simplex $\chi^{2}$ minimization to match a synthetic galaxy plus central point source to the HST image, with nuclear luminosity and galaxy effective radius ($R_e$), surface brightness ($\mu_e$), axial ratio ($a/b$) and position angle ($\theta$) as free parameters. The model galaxy's surface brightness profile can be set to either an exponentially decaying disc, or an $r^{1/4}$ de Vaucouleurs law characteristic of an elliptical galaxy. The model is convolved with an appropriate PSF, described in section~\ref{sec-psf} above. 

Accurate characterisation of the uncertainties in the data is critical when fitting across such high dynamic range. The transition from PSF sampling error (see F04 for details of sampling error treatment) to Poissonian error is determined by finding the radius at which the sampling error drops to 1.5 times the mean background Poisson noise. The noise on the central pixel was fixed by the Poisson statistic for that pixel. 

We fit the host galaxy in each image with both an elliptical galaxy ($r^{1/4}$-law profile) and an exponential disc galaxy model. The difference in the $\chi^{2}$ values of the two fits is used to classify the host as disc or bulge-dominated. S\'{e}rsic, and bulge-disc decomposition models are not fitted (c.f. F04) as we are unable to constrain the additional parameters. Each fit is performed out to $\approx$ 3\arcsec\ for the \zo\ objects and $\approx$ 2\arcsec\ at \zt\ (determined from the radial profiles). As discussed further in section~\ref{sec-res}, we find that an Elliptical host galaxy morphology provides the best fit in all cases. 

Note that we adopt the circularly-averaged effective radius, $R_e$ (that is the mean radius of the half-light elliptical isophote) throughout this work. We convert K01's half-light semi-major axis lengths to this standard. This ensures that the size measurement is not positively biased by high eccentricities.

\subsection{Fixed $R_e$ models}
\label{sec-fixlen}
In addition to the conventional modelling, we also fitted each object using a series of fixed-radius models (at $R_e=2,~5$ and 10~kpc). This technique was adopted by K01 for many of their $z\approx 2$ objects, since they were unable to recover reliable measures of the $R_e$ at this redshift. We chose to apply the technique to all the objects in the sample in order to determine the stability of the host-galaxy luminosities recovered by the modelling, and in order to compare host-galaxy models of identical radii at different wavelengths.

\section{Results}
\label{sec-res}
The best fit galaxy and nuclear parameters derived from the 2D modelling procedure are listed in tables~\ref{tab-res1} (for the $z\approx 1$ sample) and~\ref{tab-res2} (for the $z\approx 2$ sample). We have successfully recovered stable ($\pm0.3$~mag) host-galaxy fluxes for all seventeen quasars studied. The stability of the host and nuclear fluxes are shown by the $\chi^2$ contours in Fig.~\ref{fig-cont}. Host-galaxy $V-J$ (\zo) and $I-H$ (\zt) colour data derived from the WFPC2 (this paper) and NICMOS (K01) images are presented in tables~\ref{tab-col1} and~\ref{tab-col2} respectively. The WFPC2 images of the quasars, their best-fit models, modelling residuals and radial profiles are presented along with brief notes on each object in Appendix A (Figs.~\ref{fig-h104}--\ref{fig-h1018}). Finally, the fixed $R_e$ model fits are presented in Appendix B (tables~\ref{tab-fixres1} \&~\ref{tab-fixres2}). 

The fits to the \zo\ objects are formally acceptable at the $4\sigma$ level. 
In general, the \zt\ objects have best fit models that are not formally acceptable, and the resulting parameter estimates are therefore uncertain. However, we find that the host-galaxy luminosity is very stable to changes in the fitted parameters. This can be seen both from the $\chi^2$ contours (Fig.~\ref{fig-cont}), and from the results of the fixed $R_e$ fits (tables~\ref{tab-fixres1} \&~\ref{tab-fixres2}).

\subsection{Host-galaxy morphologies}
\label{sec-morphres}
As mentioned above, in all cases a bulge-dominated host galaxy provides a marginally better fit (up to $3\sigma$) to the data than a disc-dominated host (see tables~\ref{tab-res1} and~\ref{tab-res2}). We note that all of the \zo\ objects were robustly determined to be elliptical in the optical by K01. For this reason, as well as our marginal morphological preference, we show the best-fit elliptical galaxy model in each case. We note, however, that the  smooth $r^{1/4}$-law profile does not form as good a fit to the data as in our previous studies (M99; K01; D03; F04). All the images exhibit significant ``bumps'' in their radial profiles within 2\arcsec\ of the nucleus (see Figs.~\ref{fig-h104}--\ref{fig-h1018}). These bumps are directly traceable to numerous features in the images which may be nearby companions, mergers, or knots of star formation within the host galaxies themselves. We have masked these features out from the fit in order to model just the underlying host-galaxy light distribution. Aside from these caveats, in all cases the best-fit disc and bulge models do give consistent (within 3$\sigma$) values for the total flux from the host galaxy, giving confidence that this parameter has been determined robustly.

\subsection{Fixed $R_e$ fits}
\label{sec-fixlenres}
The results of the fixed $R_e$ modelling are presented in Tables~\ref{tab-fixres1} and~\ref{tab-fixres2}, along with the best-fit models obtained by K01 for reference. Note that K01 presented semi-major axis scale lengths, whereas we use effective radii, $R_e$ (mean radius of the half-light elliptical isophote -- see section~\ref{sec-mod}). We note that several of K01's host-galaxy scale lengths appeared artificially large due to their high eccentricity. We have converted scale lengths to $R_e$ in the tables.

Generally the UV and optical morphologies of the \zo\ host galaxies are indistinguishable (best fit morphology, $R_e$, axial ratio and position angle). However  BVF247, 3C~422 and MC2112 exhibit compact UV host galaxies compared to the optical (and noticeable circumnuclear residuals). At \zt\ the UV host galaxy detections are quite compact compared to the optical. In each case, adjusting $R_e$ over a reasonable range makes little or no difference to the luminosity of the best-fit host galaxy. Furthermore, adopting the fixed-$R_e$ model that best matches each NICMOS (K01) model in place of our own preferred model results in no significant changes to the results. All of the following conclusions are thus robust to the uncertainty in the true size of the host galaxy.

\subsection{Observed fluxes and colours}
The observed optical-IR fluxes and colours for the quasar nuclei and host galaxies, after correction for galactic extinction, are shown in tables~\ref{tab-col1} and~\ref{tab-col2}. NIR fluxes are from the models of K01 (converted to AB), while optical fluxes come from our modelling results above. The mean extinction-corrected apparent host-galaxy colours for the quasar sample are (with associated standard errors): 
\[\langle{V-J}\rangle_{z\approx1} = 1.9\pm0.3\]
\[\langle{I-H}\rangle_{z\approx2} = 1.1\pm0.3\]
For the nuclei they are:
\[\langle{V-J}\rangle_{z\approx1} = -0.1\pm0.2\]
\[\langle{I-H}\rangle_{z\approx2} = 0.4\pm0.2\]

\begin{table}
\begin{center}
\caption{\label{tab-col1} \zo\ host-galaxy and nuclear $V-J$ colour, in AB magnitudes, after correction for galactic extinction.}
\begin{tabular}{lrr}
\hline
\hline
Object&$V-J$&$V-J$)\\
 	& (Nuc) & (Host) \\
\hline
\multicolumn{3}{c}{\bf Radio-Quiet Quasars}\\
BVF225 & $-0.1\pm0.3$ & $0.5\pm0.3$ \\
BVF247 & $-0.4\pm0.3$ & $2.8\pm0.3$ \\
BVF262 & $0.2\pm0.3$ & $1.5\pm0.3$ \\
\multicolumn{3}{c}{\bf Radio-Loud Quasars}\\
PKS0440 & $-0.3\pm0.3$ & $1.8\pm0.3$ \\
PKS0938 & $-1.2\pm0.3$ & $1.3\pm0.3$ \\
3C422 & $0.6\pm0.3$ & $2.2\pm0.3$ \\
MC2112 & $0.0\pm0.3$ & $1.1\pm0.3$ \\
4C02.54 & $0.5\pm0.3$ & $1.3\pm0.3$ \\
\hline
\end{tabular}
\end{center}
\end{table}
\begin{table}
\begin{center}
\caption{\label{tab-col2} \zt\ host-galaxy and nuclear $I-H$ colour, in AB magnitudes, after correction for galactic extinction.}
\begin{tabular}{lrr}
\hline
\hline
Object&$I-H$&$I-H$\\
 	& (Nuc) & (Host) \\
\hline
\multicolumn{3}{c}{\bf Radio-Quiet Quasars}\\
SGP2:36 & $0.2\pm0.3$ & $1.5\pm0.3$ \\
SGP2:25 & $-0.1\pm0.3$ & $1.9\pm0.4$ \\
SGP2:11 & $0.4\pm0.3$ & $1.4\pm0.4$ \\
SGP3:39 & $-0.3\pm0.3$ & $1.4\pm0.3$ \\
SGP4:39 & $0.5\pm0.3$ & $-0.5\pm0.4$ \\
\multicolumn{3}{c}{\bf Radio-Loud Quasars}\\
PKS1524 & $0.1\pm0.3$ & $0.6\pm0.3$ \\
B2~2156 & $0.0\pm0.3$ & $1.4\pm0.3$ \\
PKS2204 & $0.9\pm0.3$ & $-0.4\pm0.3$ \\
4C45.51 & $1.7\pm0.3$ & $2.2\pm0.3$ \\
\hline
\end{tabular}
\end{center}
\end{table}

\subsection{$K$-corrections and intrinsic luminosities}
\label{sec-kc}
The observational bandpasses were chosen to coincide closely with rest-frame $U$ and $V$ for our WFPC2 and NICMOS observations, respectively. We calculate the $U$ and $V$ band luminosities by $K$-correcting the observed fluxes into these emitted bands (see~\citealt{hogg+02}) using the bandpass curves for each filter. For both the host galaxies and the QSO's themselves we assume power-law SED's, $F_\nu\propto\nu^{-\alpha}$, or equivalently $F_\lambda\propto\lambda^{\beta}$, where $\beta=\alpha-2$. For the host galaxies we adopt a power law, $\beta_\mathrm{host}=-0.39$, appropriate for a distant red galaxy~\citep{vandokkum+06}. For the QSO's we have assumed $\beta_\mathrm{QSO}=-1.5$ ($\alpha=0.5$) as defined earlier (section~\ref{sec-ckc}), and consistent with our average measured nuclear colours. although it is clear that there is actually a large range of variation from object to object, with a recent study providing a range $\alpha=0.55\pm0.42$~\citep{kcorrqso}. Since our observed bandpasses coincide closely with the reference emitted bands, the exact slope of the SED for each quasar does not strongly affect our luminosity calculation, it just affects the distribution of flux across the filter bandpass. In fact, varying the index over a very wide range of realistic values ($-1<\beta_\mathrm{host}<1$; $-2<\beta_\mathrm{QSO}<-1$) yields only $\pm0.05$~mag. uncertainty on the host-galaxy luminosity and $\pm0.03$~mag. uncertainty in the quasar nuclear luminosity. 

The absolute $U$ and $V$ band magnitudes for the quasar host galaxies and nuclei are presented in Table~\ref{tab-lum}. As in K01, we find that the hosts of radio-loud quasars are more luminous than those of the radio-quiet quasars, and strongly so at \zt\ (see section~\ref{sec-mass}). However, at \zt\ this is likely to be due to the optical luminosity biases in the RLQ (optically brighter) and RQQ (dimmer) subsamples. As discussed in section~\ref{sec-samp}, the sample was originally designed to match the optical luminosity and redshift distributions of the RL and RQ subsamples, but was based on old and heterogeneous photometric measurements, with large $K$-corrections from observed to rest-frame $V$. The results of K01 (their Fig.~6 and our Fig.~\ref{fig-samp}) show that the subsamples are {\em not} matched in luminosity. This is further discussed in section~\ref{sec-RL}.

The host galaxies' average rest-frame $U-V$ colours (with associated standard error of the mean) are as follows:
\[\langle{U-V}\rangle_{z\approx1} = 1.6\pm0.3\]
\[\langle{U-V}\rangle_{z\approx2} = 0.9\pm0.3\]
In spite of the optical and radio luminosity differences between the RL and RQ samples, the colours of their host galaxies are indistinguishable in the present data.

\begin{table*}
\begin{center}
\caption{\label{tab-lum}Absolute magnitudes and rest-frame $U-V$ colours for the quasar host galaxies and nuclei at $z\approx 1$ and 2.
$U$ magnitudes are F300W and $V$ are Subaru V (all AB, corrected for galactic extinction).}
\begin{tabular}{lrrrrrr}
\hline
\hline
Object	&$M_U$(Nuc)	&$M_U$(Host)	&$M_V$(Nuc)	&$M_V$(Host)	&$U-V$	& $U-V$ \\
		&			&			& 			& 			& (Nuc) 	& (Host) \\
\hline
\multicolumn{7}{c}{\bf Radio-Quiet Quasars}\\
BVF225 	& $-24.7\pm0.1$ & $-21.9\pm0.1$ & $-24.6\pm0.3$ & $-22.4\pm0.3$ & $-0.1\pm0.3$	& $0.48\pm0.3$	\\
BVF247 	& $-22.7\pm0.1$ & $-20.7\pm0.1$ & $-22.3\pm0.3$ & $-23.5\pm0.3$ & $-0.4\pm0.3$	& $2.81\pm0.3$	\\
BVF262 	& $-23.3\pm0.1$ & $-21.3\pm0.1$ & $-23.4\pm0.3$ & $-22.8\pm0.3$ & $0.2\pm0.3$	& $1.49\pm0.3$	\\
\multicolumn{7}{c}{\bf Radio-Loud Quasars}\\
PKS0440 	& $-24.2\pm0.1$ & $-21.7\pm0.1$ & $-23.9\pm0.3$ & $-23.4\pm0.3$ & $-0.3\pm0.3$	& $1.75\pm0.3$	\\
PKS0938 	& $-24.0\pm0.1$ & $-21.8\pm0.1$ & $-22.8\pm0.3$ & $-23.1\pm0.3$ & $-1.2\pm0.3$	& $1.32\pm0.3$	\\
3C422 	& $-24.2\pm0.1$ & $-22.2\pm0.1$ & $-24.8\pm0.3$ & $-24.3\pm0.3$ & $0.6\pm0.3$	& $2.15\pm0.3$	\\
MC2112 	& $-23.6\pm0.1$ & $-23.2\pm0.1$ & $-23.6\pm0.3$ & $-24.3\pm0.3$ & $0.0\pm0.3$	& $1.14\pm0.3$	\\
4C02.54 	& $-24.6\pm0.1$ & $-22.1\pm0.1$ & $-25.1\pm0.3$ & $-23.4\pm0.3$ & $0.5\pm0.3$	& $1.32\pm0.3$	\\
\hline
\multicolumn{7}{c}{\bf Radio-Quiet Quasars}\\
SGP2:36 	& $-23.1\pm0.1$ & $-22.1\pm0.1$ & $-23.2\pm0.3$ & $-23.4\pm0.3$ & $0.1\pm0.3$	& $1.25\pm0.3$	\\
SGP2:25 	& $-23.9\pm0.1$ & $-21.7\pm0.2$ & $-23.7\pm0.3$ & $-23.4\pm0.3$ & $-0.1\pm0.3$	& $1.72\pm0.4$	\\
SGP2:11 	& $-24.2\pm0.1$ & $-21.6\pm0.3$ & $-24.5\pm0.3$ & $-22.8\pm0.3$ & $0.3\pm0.3$	& $1.20\pm0.4$	\\
SGP3:39 	& $-24.2\pm0.1$ & $-22.5\pm0.1$ & $-23.9\pm0.3$ & $-23.6\pm0.3$ & $-0.3\pm0.3$	& $1.19\pm0.3$	\\
SGP4:39 	& $-23.8\pm0.1$ & $-22.1\pm0.2$ & $-24.3\pm0.3$ & $-21.4\pm0.3$ & $0.4\pm0.3$	& $-0.68\pm0.4$	\\
\multicolumn{7}{c}{\bf Radio-Loud Quasars}\\
PKS1524 	& $-25.0\pm0.1$ & $-23.2\pm0.1$ & $-25.0\pm0.3$ & $-23.6\pm0.3$ & $0.0\pm0.3$	& $0.44\pm0.3$	\\
B2~2156 	& $-25.3\pm0.1$ & $-24.0\pm0.1$ & $-25.3\pm0.3$ & $-25.2\pm0.3$ & $-0.1\pm0.3$	& $1.21\pm0.3$	\\
PKS2204 	& $-24.0\pm0.1$ & $-23.3\pm0.1$ & $-24.8\pm0.3$ & $-22.7\pm0.3$ & $0.8\pm0.3$	& $-0.62\pm0.3$	\\
4C45.51 	& $-24.4\pm0.1$ & $-23.7\pm0.1$ & $-26.0\pm0.3$ & $-25.7\pm0.3$ & $1.6\pm0.3$	& $1.98\pm0.3$	\\
\hline
\end{tabular}
\end{center}
\end{table*}

\section{Discussion}
\label{sec-disc}
\subsection{Are we detecting host-galaxy light?}
\label{sec-det}
Our observations were designed to detect the UV light from even a mature stellar population, and it is clear directly from the images themselves (see Figs.~\ref{fig-h104} -- \ref{fig-h1018}) that in general we have an excess of extended UV flux (compared to the PSF). However, in three cases at \zo, and every case at \zt, the rest-frame $U$-band host-galaxy effective radius is significantly smaller than the rest-frame $V$-band equivalent reported by K01 (see tables~\ref{tab-fixres1} \&~\ref{tab-fixres2}). Are we truly detecting the host-galaxy light, or are we being biased by the strength of the AGN, or perhaps by reflected AGN emission?

We find that the host-galaxy and nuclear luminosities are quite stable to changes in the effective radius of the host galaxy. The fixed $R_e$ models (tables~\ref{tab-fixres1} \&~\ref{tab-fixres2}) show a significant increase in $\chi^2$ as we move away from the preferred radius to larger or smaller values, while the host-galaxy luminosity remains stable. However, the stability of the host and nuclear luminosities is demonstrated most clearly by the $\chi^2$ contours in $m_\mathrm{Host}-m_\mathrm{Nuc}$, (Fig.~\ref{fig-cont}). These show that the host-galaxy flux is constrained to within $\pm 0.1$ to $\pm0.3$~mag. The nuclear luminosity is generally somewhat better constrained ($\pm 0.1$ to $\pm0.2$~mag). We also note that the host galaxies generally have similar axial ratios and position angles in each band (tables~\ref{tab-fixres1} \&~\ref{tab-fixres2}). 

In almost all cases, the nuclear $U-V$ colours are close to those expected for a typical QSO SED. Only 4C45.51, PKS2204 and PKS0938 (all radio-loud) strongly deviate from the expected flat nuclear colour. We ran PSF-only fits to all the objects, and found a significantly higher $\chi^2$ for the best PSF-only fits in every case. Finally we note that while our host galaxies are compact, they are still well resolved, and have comparable sizes to the massive ($>10^{11}$~M$_\odot$) galaxies at $1<z<3$ in CANDELS~\citep{bruce+12}. This is discussed further in section~\ref{sec-mass}.

\begin{figure*}
\centering
{\includegraphics[width=85mm]{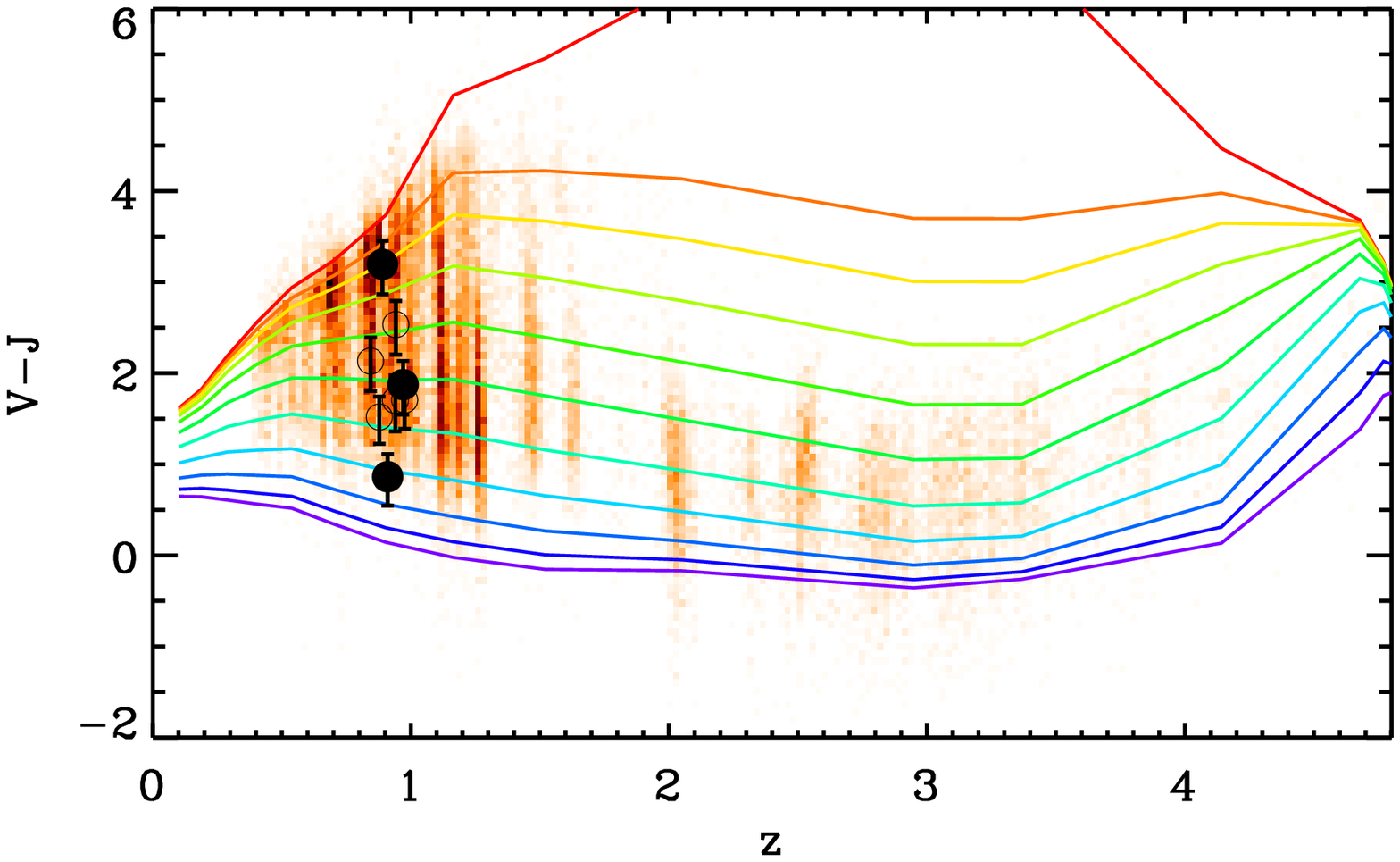}}
\hspace{5mm}
{\includegraphics[width=85mm]{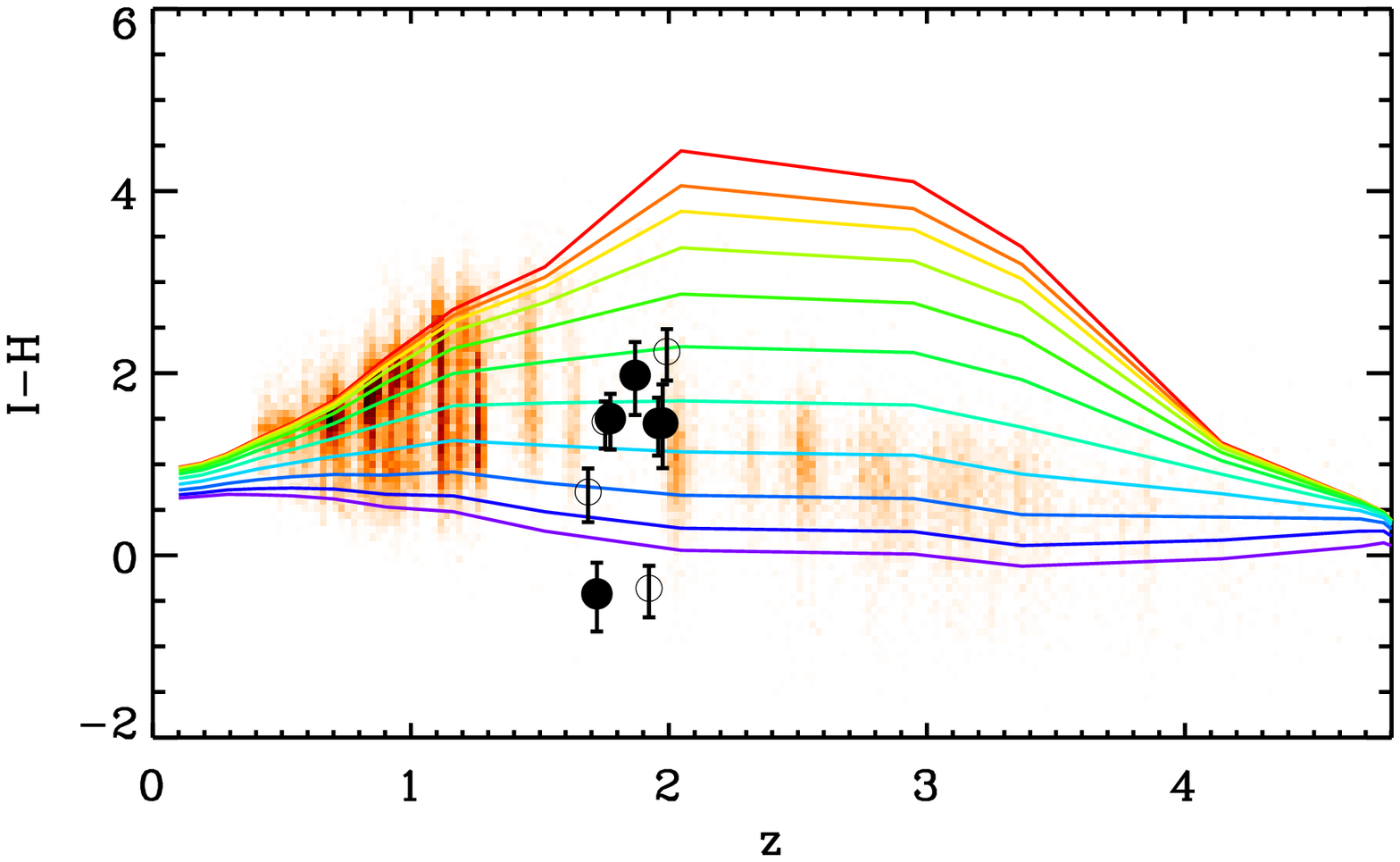}}
\caption{\label{fig-colz} Host-galaxy observed colour with redshift for the RQQs (filled circles) and RLQs (open circles) compared to COSMOS galaxies (density plot) and SSP models (lines). Magnitudes are all extinction-corrected AB in the COSMOS filters. 
{\bf Left:} $V-J$ host colours for the quasars at \zo. 
{\bf Right:} $I-H$ host colours for the quasars at \zt. 
SSP colour evolution curves are shown for a~\citet{BC03} passively evolving population (red line), plus ``frosting'' models with from 0.01\% (orange) up to a maximum (purple) of 5.12\% of the galaxy mass in ongoing star formation (in octaves). All SSP models assume a redshift of formation $z_f=5$, solar metallicity and~\citet{chabrier03} IMF -- see section~\ref{sec-cosmos}. 
All COSMOS galaxies with acceptable best-fit photometric redshifts out to $z=5$ are shown.
Quasar host galaxies are bluer than passive and have similar colours to the full range of COSMOS galaxies at similar redshifts.}
\end{figure*}

\begin{figure*}
\centering
{\includegraphics[width=85mm]{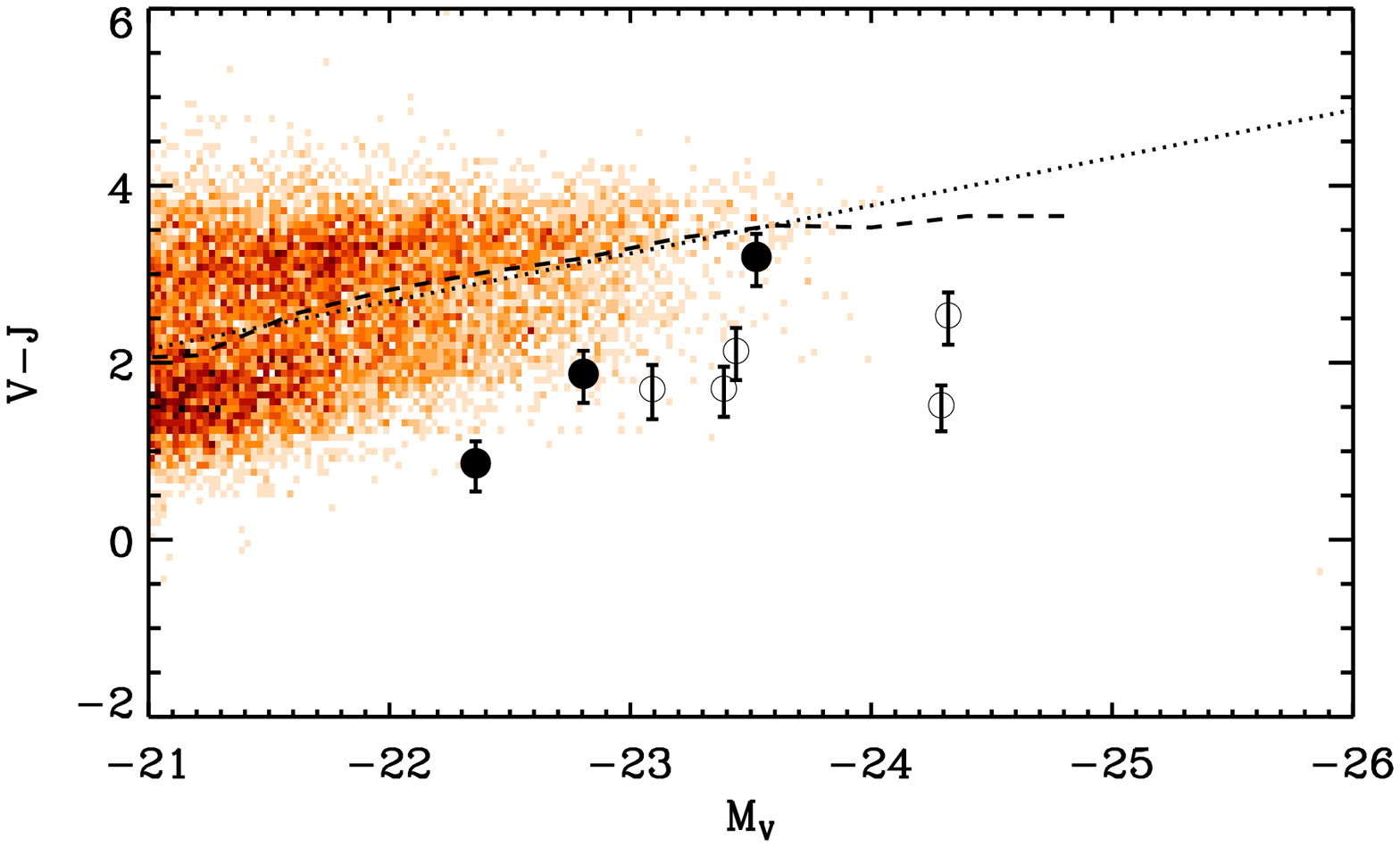}}
\hspace{5mm}
{\includegraphics[width=85mm]{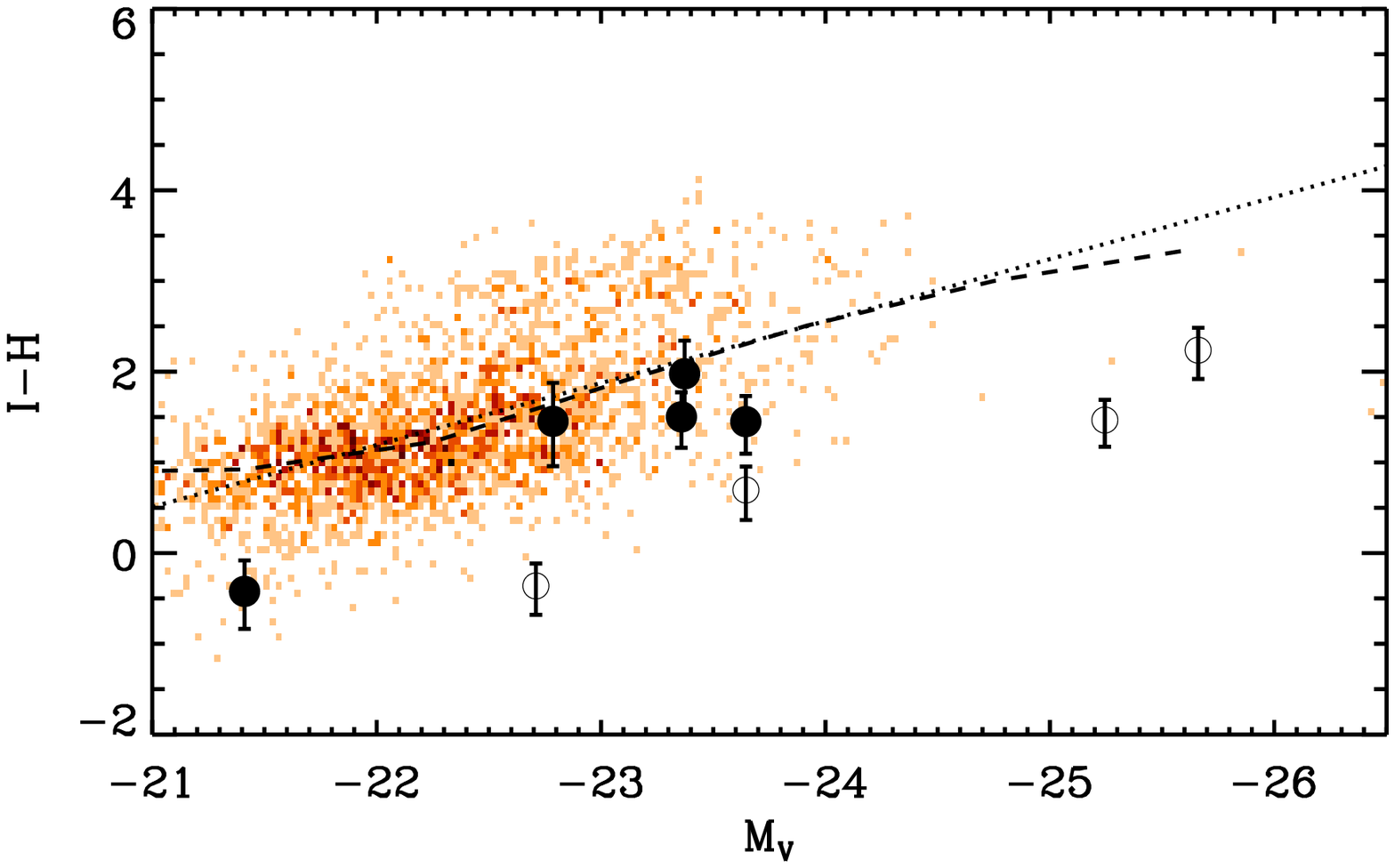}}
\caption{\label{fig-colmag} Host-galaxy observed colour with optical host luminosity (K01) for the RQQs (filled circles) and RLQs (open circles) compared to COSMOS galaxies within $3\sigma$ of the mean quasar redshift (density plot -- see section~\ref{sec-cosmos}). Also show are the best linear fit (dotted line) and median colour at each luminosity (dashed line). Magnitudes are all AB in the COSMOS filters.
{\bf Left:} $V-J$ host colours for the quasars at \zo. 
{\bf Right:} $I-H$ host colours for the quasars at \zt. 
Quasar host galaxies are bluer than expected for galaxies of the same redshift and luminosity -- all fall below the median COSMOS colour (two borderline at \zt).}
\end{figure*}

We are therefore confident that we are constraining the total host-galaxy luminosity in all cases. However, our $R_e$ and $\mu_e$ constraints are sometimes weak, and the $R^{1/4}$-law model is only marginally preferred over an exponential disk at \zt. There is clearly UV-bright substructure in the host galaxies that diminishes the quality of a smooth model profile fit to the data. 

\subsection{Residual features}
We note the significant residual features in MC2112$+$172 (Fig.~\ref{fig-h108}), B2~2156$+$129 (Fig.~\ref{fig-h1019}), and PKS2204$-$20 (Fig.~\ref{fig-h1020}). The former two are objects with compact $U$ band hosts compared to their $V$ band hosts (PKS2204$-$20 had a poorly constrained host-galaxy size in K01). These residuals are not consistent with PSF features, being more extended than the Airy ring and off-centre with respect to the nucleus. Our pixel scale corresponds to $\approx 800$~pc at the distance of our sources, and the feature seen in MC2112$+$172 is up to $\approx 10$~kpc across. This may represent an asymmetric disk component in the host galaxy or a foreground or merging companion. B2~2156$+$129 shows a compact clump of excess flux slightly off-centre (to the northwest), up to $\approx 4$~kpc across. PKS2204$-$20 shows a slight ``bridge'' of material across the nucleus, linking up with the companion to the northwest. 4C45.51 shows a very compact slightly off-centre (to the north) feature that is $\approx 2$~kpc across, although this may be PSF residual. These features explain the unusual contours for these objects' fits (Fig.~\ref{fig-cont}): the relatively poor constraint on the nuclear flux is due to confusion between the nucleus and the contaminant. Although these are the most spectacular examples, all of our images show some excess flux above the smooth model, and the majority (all at \zt\ and three at \zo) show a clear non-PSF, centrally concentrated component.

\subsection{Host-galaxy luminosity and colour}
\label{sec-cosmos}
We now compare our quasar host-galaxy colours and luminosities with the colours and luminosities of typical galaxies at the same redshifts (Figs.~\ref{fig-colz} and~\ref{fig-colmag}). We adopt the data of the COSMOS galaxy survey, since it is sufficiently deep and wide to provide a statistical sample of galaxies at the same redshifts as our quasars. We use the COSMOS galaxy catalogue~\citep{capak+07,  ilbert+09} for redshifts, $V$, $J$ and $I$ band data, and the COSMOS $H$-band catalogue (P. Capak, {\em private communication};~\citealt{bielby+12}) to compare observed colours with our host-galaxy sample. We adopted the best fit photometric redshift, but removed all objects for which the 1-$\sigma$ relative uncertainty in the redshift was greater than 10\%, or for which the quality of the redshift fit was poor ($\chi^2/\nu>2$). We have converted our luminosities and fluxes to the equivalent COSMOS bandpasses for ease of comparison.

\begin{figure*}
\centering
{\includegraphics[width=85mm]{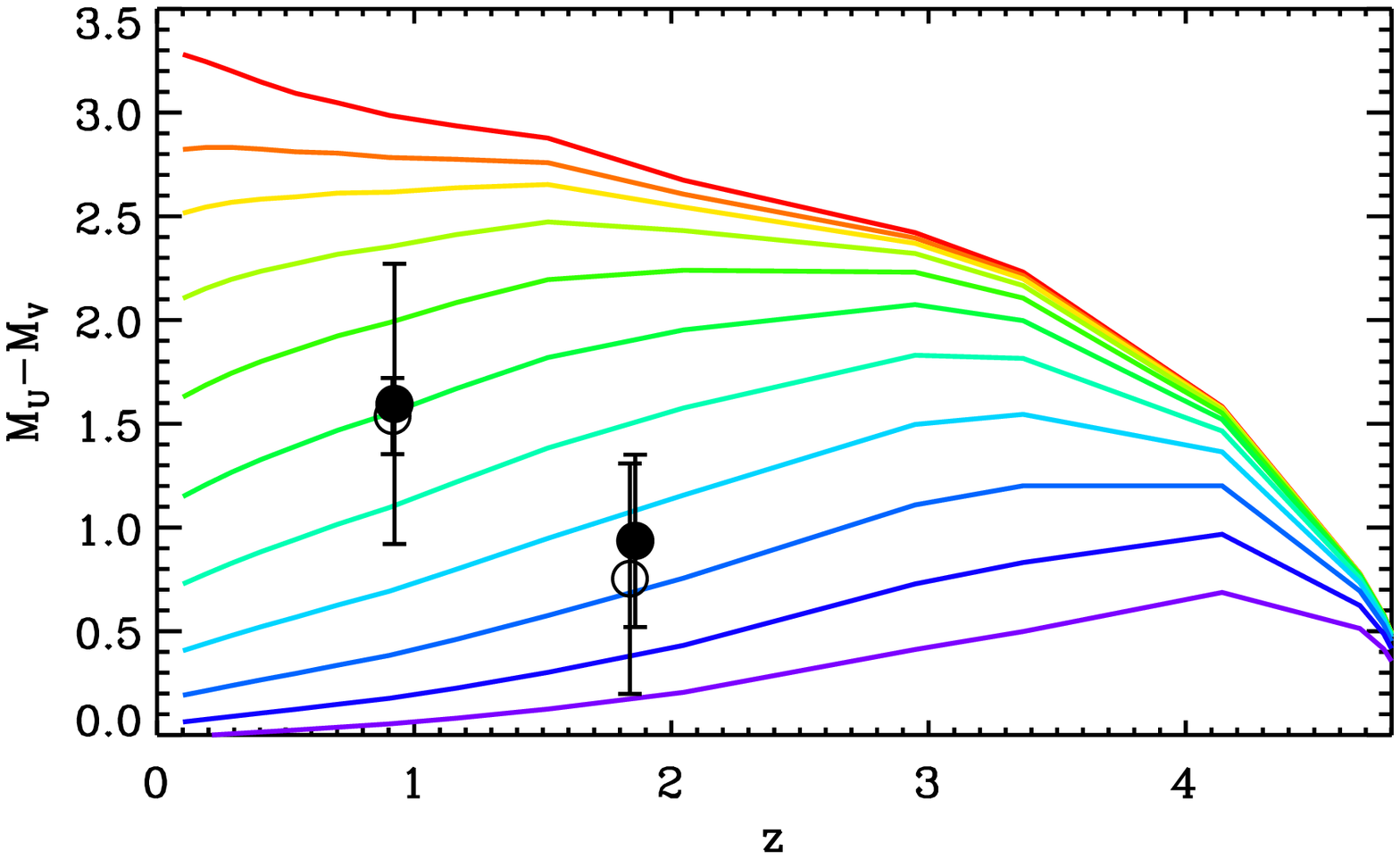}}
\hspace{5mm}
{\includegraphics[width=85mm]{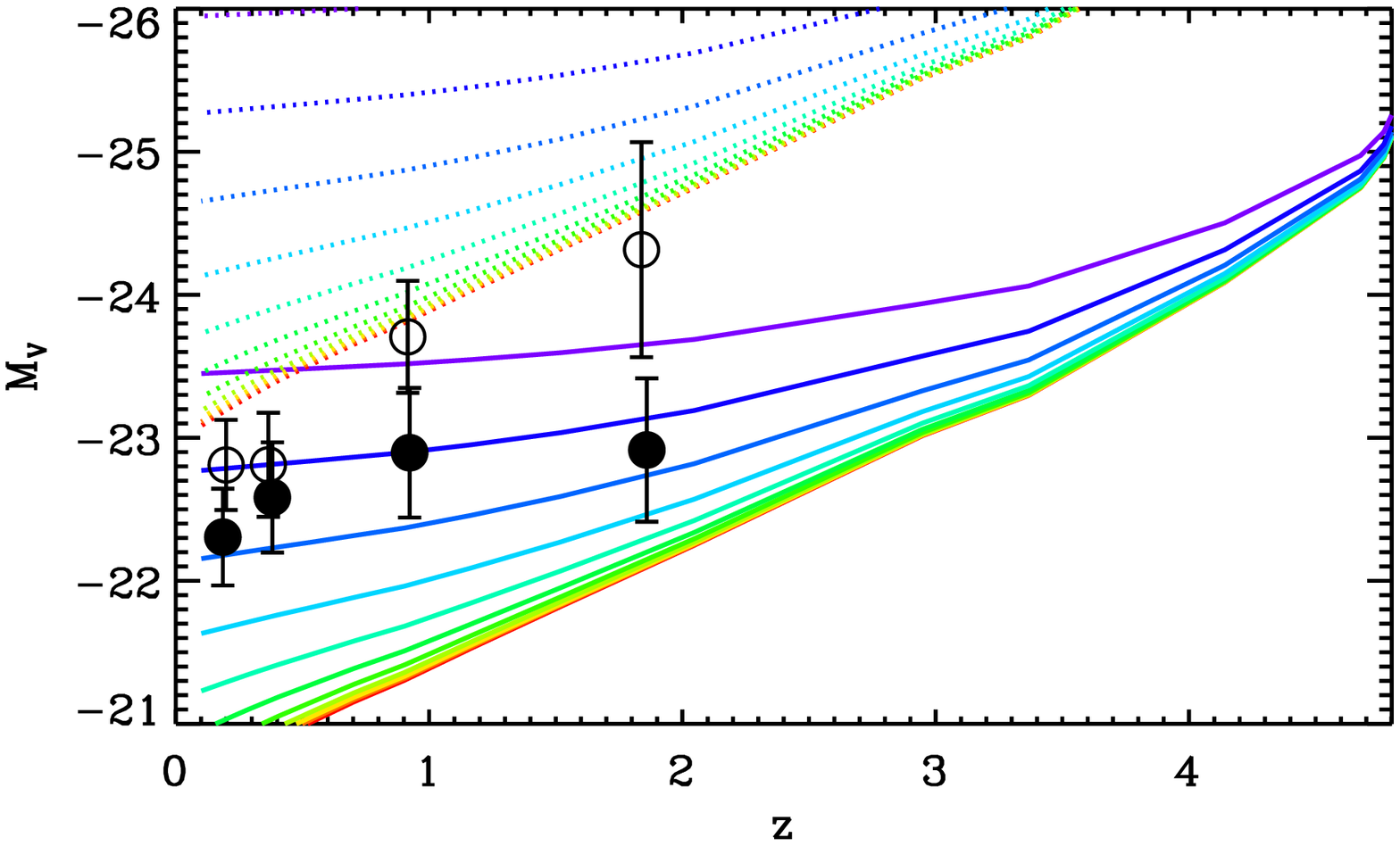}}
\caption{\label{fig-lum} Mean rest-frame $U-V$ colour, and $V$ band luminosity, 
versus redshift, for the quasar host galaxies of RQQ's (filled circles) and RLQ's (open circles) at \zo\ and \zt.
Sample means (by type and by redshift bin) are shown with standard errors. 
All magnitudes are AB, V-band is the COSMOS  V-band and U-band is F300W.
Luminosity and colour evolution curves are shown for a~\citet{BC03} passively evolving SSP (red line), plus ``frosting'' models with 
0.01 (orange), 0.02, 0.04, 0.08, 0.16, 0.32, 0.64, 1.28, 2.56 and 5.12\% (purple) 
of the galaxy mass in ongoing star formation. 
Luminosities are shown for stellar masses of $10^{11}$ (solid lines) and $10^{12}$~M$_\odot$ (dotted lines).
All SSP models assume a redshift of formation $z_f=5$, solar metallicity and~\citet{chabrier03} IMF. 
Quasar host-galaxy colours are consistent with $\sim 0.1\%$ star formation at \zo\ and $\sim1\%$ at \zt.
The colours of the RLQ and RQQ host galaxies are indistinguishable at each epoch, in spite of the large mass differences present in the sample.
}
\end{figure*}

Fig.~\ref{fig-colz} shows the observed galaxy colour against redshift for all such COSMOS galaxies out to $z=5$ (density plot), and for our quasar host galaxies (circles). We also show the colour evolution of several synthetic simple stellar population (SSP) models computed using the stellar spectral synthesis models of~\citet{BC03}. We show an instantaneous burst of star formation at a redshift $z_f=5$, with solar metallicity and the initial mass function (IMF) of~\citet{chabrier03} that includes low mass / sub-stellar objects yielding more accurate mass-to-light ratio determinations. The red line in each case shows the evolution of a passively evolving stellar population. ``Frosting'' models, based on the same underlying old stellar population are shown in colour, with 0.01, 0.02, 0.04, 0.08, 0.16, 0.32, 0.64, 1.28, 2.56 and 5.12\% of the galaxy mass undergoing star formation. The young stellar population is modelled with a single burst of 0.01~Gyr age, and our estimates of the active star forming fractions are therefore lower limits (see section~\ref{sec-mass}).

In Fig.~\ref{fig-colmag}, we plot the observed galaxy colour against luminosity. Here, comparison samples of COSMOS galaxies are selected to be within $3\sigma$ of the mean redshift of our samples; that is $0.780<z<1.058$ at \zo\ and $1.505<z<2.196$ at \zt. 
This cut is in addition to the quality cut described above. The COSMOS data illustrate the well known correlation between galaxy luminosity and colour. A blue cloud and red sequence are clearly apparent at \zo. A hint of the red sequence is still visible at \zt, although at this redshift the most massive galaxies are no longer purely red. We also plot the best linear fit to the COSMOS colour-luminosity data (dotted lines) and the mean colour in 11 equal sized luminosity bins (dashed line).

The quasar host galaxies are extremely bright, sampling the upper end of the galaxy luminosity function ($\sim1-4L^\star$). The majority of such luminous COSMOS galaxies are found to be red at \zo. At \zt\ there is a less clear-cut colour-magnitude division due to low numbers. However, all of our quasar host galaxies are bluer than the median galaxy colour for COSMOS galaxies of the same luminosity and redshift (Fig.~\ref{fig-colmag} -- dashed line). The probability of all 17 quasar host galaxies being bluer than the median, if they are drawn at random from the galaxy distribution at similar redshift and luminosity, is $0.0008$\%. Allowing for the two objects that have error bars that admit redder-than-median colours, the probability (15/17) is $0.12$\%. This result is robust to changes in the host-galaxy scale length, and to matching $R_e$ to the best fit models in K01. We therefore conclude that the host galaxies of quasars are not drawn at random from the massive galaxy population, but are systematically bluer than the average galaxy of the same mass and redshift.

\subsection{Host-galaxy luminosities, masses and star-forming fractions}
\label{sec-mass}
Fig.~\ref{fig-lum} shows the host-galaxy rest-frame $U-V$ colours and $V$ band luminosities. For comparison we show the evolving luminosities and colours of the same set of SSP models as discussed in the previous section. Luminosity evolution is shown for stellar masses of $10^{11}$ (solid lines) and $10^{12}$~M$_\odot$ (dotted lines). From their colours, the quasar host galaxies at \zt\ are $0.5-5$\% star-forming by mass. By \zo\ this has fallen to $0.05-0.5$\%. 

The resulting host-galaxy masses are somewhat lower than would be estimated from the simple passive curves (e.g. K01), since there is significant contribution to the optical luminosity from the young stellar population. This is seen most clearly in Fig.~\ref{fig-lum}. The blue colour of the host galaxies implies a shallower slope to the luminosity evolution, and thus a significantly lower mass than would be inferred assuming a purely passive model. Note that the mass difference is more extreme than one might naively imagine for the few percent ``frosting'' fractions, since the young stellar population contributes strongly to the optical luminosity relative to its low mass. Host-galaxy stellar mass ranges at \zt\ are:
\begin{itemize}
\item $1\times10^{11} - 1.3\times 10^{12}$~M$_\odot$ for the RLQ's.
\item $3\times10^{10} - 3\times10^{11}$~M$_\odot$ for the RQQ's.
\end{itemize}
By \zo\ this has increased to:
\begin{itemize}
\item $4\times10^{11} - 1.3\times10^{12}$~M$_\odot$ for the RLQ's.
\item $2\times10^{11} - 6\times10^{11}$~M$_\odot$ for the RQQ's.
\end{itemize}

\begin{figure}
\centering
{\includegraphics[width=85mm]{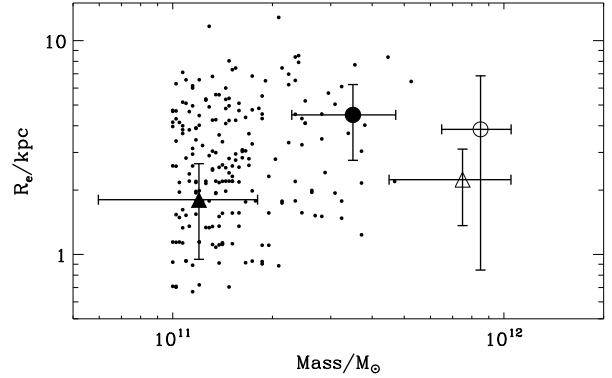}}
\caption{\label{fig-sm} 
Mean galaxy size versus mean stellar mass for our host-galaxy samples (large circles for \zo; triangles for \zt; open for RLQ; closed for RQQ), compared to the $M_\star>10^{11}$~M$_\odot$, $1<z<3$ CANDELS galaxies of~\citet{bruce+12}. 
Uncertainties shown are standard errors on the mean. While compact, our host-galaxy detections are quite consistent with the typical sizes of massive galaxies at the same redshift.}
\end{figure}

There is roughly an order of magnitude scatter in the estimated mass of each subsample thanks to the range of colours seen. This spread in mass dominates over the expected $\sim 0.3-0.6$~dex systematic error in the mass estimates from our SSP modelling~\citep[][]{gallazzi+09,conroy+09}. The masses are shown against effective radius in Fig.~\ref{fig-sm}, together with the massive $1<z<3$ CANDELS galaxies from~\citet{bruce+12}. As we remarked in section~\ref{sec-det} above, our host-galaxy models are quite consistent with this sample of massive galaxies at similar redshifts. 
Our \zo\ host galaxy effective radii are, on average, a factor of 2 larger than those for the lower luminosity AGN of~\cite{sanchez+04} in the same redshift range, but with significant overlap in the populations.

We have modelled the young component with a 0.01Gyr old stellar population, and our colours and luminosities suggest star formation rates $\sim350$~M$_\odot$~yr$^{-1}$ for the RLQ's and $\sim100$~M$_\odot$~yr$^{-1}$ for the RQQ's at \zt. By \zo, these rates have fallen to $\sim150$~M$_\odot$~yr$^{-1}$ for the RLQ's and $\sim50$~M$_\odot$~yr$^{-1}$ for the RQQ's. The ~\citet{kennicutt98} relation for the UV luminosity of star formation gives similar estimates.
Modelling the young component with a range of ages (0.001--0.1~Gyr) gave consistent results for the star-formation rate.

The mean mass of the RLQ's grows $\sim10$\% between \zo\ and \zt. However, the upper limit on the masses of the RLQ host galaxies remains reassuringly unchanged at each redshift, as this is already $10^{12}$~M$_\odot$ at \zt. This gives the illusion that these objects are passively evolving, although it is clear from their colours at each epoch that this is not the case. It is simply that at each epoch, the RLQ's exist in the most massive galaxies, and these have already assembled as much stellar mass as the most massive galaxies known by \zt. 
There is, however, an increase in the host-galaxy masses of the RQQ's observed between \zt\ and \zo. 
An apparent mass difference between the RLQ and RQQ host galaxies remains at each redshift, while the star-forming fractions appear indistinguishable (but see section~\ref{sec-RL}, below). 
At both epochs, there is plenty of time for the host galaxies to evolve onto the red sequence by the present day, but to avoid the production of extremely massive galaxies in the present day Universe, it is clear that the star-formation seen in these objects is not sustainable.

\subsection{Radio-Loudness and Accretion Rate}
\label{sec-RL}
As mentioned earlier (sections~\ref{sec-presamp},~\ref{sec-kc}), the RLQ and RQQ samples were initially selected to have identical observed optical luminosities. However, the results of K01 showed the actual rest-frame optical luminosities to be lower for the RQQ's at \zt, and the host galaxy luminosities to be significantly lower at both redshifts. This is consistent with the well-known anti-correlation between radio-loudness and accretion rate, seen in lower luminosity AGN~\citep[see][]{xu+99,ho02,panessa+07,sikora+07}. Our results demonstrate that this anti-correlation holds even for the most optically luminous RLQÕs and RQQÕs. RLQs (shining at lower Eddington ratios) will have higher black hole masses for a given optical--UV luminosity than their RQQ cousins (which have higher Eddington ratios). Through the well-known scaling relations between black-hole and galaxy bulge mass, we will therefore also observe a higher mass population of host galaxies for the RLQs. 

Given this difference in accretion rate between radio-loud and radio-quiet quasars, the similarity in the colours of their host galaxies (and hence in their specific star formation rates) is perhaps all the more remarkable. While the present data are insufficient to draw strong conclusions, we note that this is consistent with optical quasar activity being associated a particular phase in the evolution of massive galaxies, for example shortly after a starburst, or shortly after truncation of ongoing star formation.

\subsection{Interpretation and future observations}
Quasar host galaxies at $z=1-2$ appear to be exceptional, both in mass (as at lower redshift) and in their colour for their mass (which has been a contentious matter for the past decade at lower redshifts). Our results therefore  fit in well with a growing consensus that quasar and AGN activity is strongly associated with ongoing star formation, although it is too early to establish cause and effect. 

Studies of the host galaxies of lower luminosity AGN and quasars at low redshift ($z\approx0.2$ --~\citealt{jahnke+04a}), intermediate redshift ($0.5<z<1.1$ --~\citealt{sanchez+04}) and high redshift ($1.8<z<2.75$ --~\citealt{jahnke+04b}) present a consistent picture, which this study confirms and extends to higher AGN luminosity. All of the above studies find a mix of morphological types for the host galaxies (as expected at lower AGN luminosities). However, where the host galaxies are found to be elliptical, their colours are consistently found to be bluer-than-average for inactive elliptical galaxies. 

Our results are consistent with those of~\citet{mainieri+11} who showed that the star forming fraction in powerful X-ray selected quasars ($L_\mathrm{bol}>8\times10^{45}$~erg~s$^{-1}$) increases strongly with redshift, with 62\% star forming at $z\sim1$, 71\% at $z\sim 2$, 100\% at $z\sim3$. However, the quasars in this study would appear to be more strongly star forming than those at lower redshift, where such activity has been detected. By stacking radio images of COSMOS X-ray selected AGN at $z<1$,~\citet{pierce+11} showed that the sources detected at 24$\mu$m also have radio emission, and are consistent with star formation rates of up to 40~M$_\odot$~yr$^{-1}$. Evidence of star formation in quasars is also given by the recent detection of large amounts ($\sim10^{10}$~M$_\odot$) of cold molecular gas in the majority of IR-ultraluminous QSO's, and the strong correlation of CO luminosity with FIR luminosity for all QSO's, IRQSO's and ULIRG's~\citep{xia+12}. 

We support the findings of~\citet{shi+09} who propose that type 1 quasars reside in a distinct population of galaxies that exhibits elliptical morphology but harbours a significant fraction of intermediate-age stars and is experiencing intense circumnuclear star formation. \citet{cen11} propose a model for coevolution of galaxies and supermassive black holes whereby a central starburst precedes the quasar phase, which is fuelled by remnants of the starburst. A major prediction is that luminous quasar host galaxies are a special population of early-type galaxies located in the green valley of the galaxy colour-magnitude diagram. Our data are broadly consistent with this model, although our data suggest a rather bluer colour for the host galaxies than would be expected if they lie exclusively in the green valley
We certainly cannot reject it due to possible sources of UV luminosity bias (UV-bright clumps in the host galaxy, and a possible contribution from scattered AGN light). The star-formation levels detected here, as a fraction of galaxy mass, are an order of magnitude greater than those detected by~\citet{nolan+01} in $z\approx0.2$ quasar host galaxies (after correcting for their use of a 0.1~Gyr old young stellar population). It will be interesting to revisit the low redshift samples in the UV to pick out the locations of the young stellar populations and compare them to the tidal features observed in the deep imaging of~\citet{bennert+08}. 
Finally, our estimated star-formation rates are consistent with those found by~\citet{page+12}, although our \zt\ RLQ's would appear to be slightly more strongly star forming than their upper limits ($\sim 350$~M$_\odot$ as opposed to $<300$~M$_\odot$. \citet{page+12} and~\citet{trichas+12} posit the lack of the very highest star-formation rates in quasar host galaxies as evidence for AGN feedback. Our similar range of star formation rates, and the similarity in colour of the RLQ and RQQ host galaxies support this picture. 
It remains possible that the star formation triggers the quasar activity, but in that case we might expect to see even stronger star-formation in the host galaxies.

\section{Conclusions}
\label{sec-conc}
The main result of this study is that UV-luminous host galaxies are present in our entire sample of radio-loud and radio-quiet quasars at \zo\ and \zt.
Our fitted host-galaxy luminosities are stable to within $\pm0.3$~mag. and our conclusions are robust to changes in the preferred UV host-galaxy size.
They occupy elliptical galaxies in the upper regions of the optical luminosity function ($1-4L_\star$), but have bluer than median colours for galaxies of the same redshift and luminosity. They are strongly star forming at \zt\ ($\sim 1\%$ by mass), with weaker ($\sim0.1\%$) star-formation in somewhat more massive galaxies by \zo. The RLQ's and RQQ's are indistinguishable in terms of colour, but the RLQ's are more massive at each epoch, indicative of their lower accretion rates (see section~\ref{sec-RL}).
The most massive RLQ's have stellar masses of $\sim10^{12}$~M$_\odot$, even at \zt.

%

The probability of obtaining galaxies as UV-bright as these quasar hosts from a sample of quiescent galaxies of the same optical luminosity and redshift is low ($<0.12\%$), and we reject the hypothesis that quasar host galaxies are drawn at random from the massive galaxy population. We conclude that while the host galaxies are extremely massive, they remain actively star-forming (or very recently so) at the epoch of the quasar. The relationship between AGN activity and star-formation remains unknown, but it seems clear that optically luminous AGN activity is associated with recent or ongoing star-formation in massive galaxies, whereas radio AGN activity is associated purely with the most massive galaxies.

The host galaxies are more compact than low-redshift analogues, but are consistent with massive quiescent galaxies at same epoch. There is clear evidence for clumpiness in the UV flux. We reiterate that the models are designed to fit the smooth bulk of the starlight. However, we have noted that the radial profiles of the host galaxies presented here are not as smooth as those in the optical. The UV ``bumps'' correspond to compact regions of increased UV luminosity -- perhaps close companion objects or knots of star formation in the host galaxy. These were not apparent in the NICMOS rest-frame V imaging. The UV clumpiness, and compact UV morphology observed in some objects suggest that higher resolution imaging and IFU spectroscopy may reveal compact regions of star formation in the host galaxy, and hint at the possibility of discovering circumnuclear starbursts in some particularly UV-compact objects. Our results therefore support the growing consensus that quasar and AGN activity is strongly associated with recent star formation (possibly its truncation -- see section~\ref{sec-RL}), and suggest that this activity tends to be located in the central regions of the galaxy.

\section*{Acknowledgments}
We thank the referee, Knud Jahnke, for his helpful critique of the paper, and in particular his suggestion to address the dependence of radio-loudness on accretion rate.
DJEF and MJIB acknowledge support from the Australian Research Council via Discovery Project grant DP110102174.
JSD acknowledges the support of the Royal Society via a Wolfson Research Merit award, and also the support of the European Research Council via the award of an Advanced Grant. RJM acknowledges the support of the Royal Society via a University Research Fellowship.
SAB acknowledges supported from the Radcliffe Institute for Advanced Study at Harvard University.
Based on observations with the NASA/ESA {\it Hubble Space Telescope},
(program IDs 9085 and 7447) obtained at the Space Telescope Science
Institute, which is operated by The Association of Universities for
Research in Astronomy, Inc. under NASA contract No. NAS5-26555. 
This research has made use of the NASA/IPAC Extragalactic Database (NED) which is operated by the
Jet Propulsion Laboratory, California Institute of Technology, under contract with NASA. 
We wish to thank Peter Capak, Olivier Ilbert and the COSMOS team for access to their unpublished H-band catalog and Victoria Bruce for her CANDELS galaxy scale length data.
DJEF would like to thank Ned Taylor, Chiara Tonini, Barry Rothberg and Darren Croton for helpful discussions and insights throughout the development of this work.


\bibliographystyle{mn2e}
\bibliography{full_lib,ms,hiz}

\begin{thebibliography}{70}
\expandafter\ifx\csname natexlab\endcsname\relax\def\natexlab#1{#1}\fi

\bibitem[{Bahcall {et~al}\mbox{.}(1997)Bahcall, Kirhakos, Saxe, \&
  Schneider}]{bahcall+97}
Bahcall J.~N., Kirhakos S., Saxe D.~H., Schneider D.~P., 1997, \apj, 479, 642

\bibitem[{Bennert {et~al}\mbox{.}(2008)Bennert, Canalizo, Jungwiert, Stockton,
  Schweizer, Peng, \& Lacy}]{bennert+08}
Bennert N., Canalizo G., Jungwiert B., Stockton A., Schweizer F., Peng C.~Y.,
  Lacy M., 2008, \apj, 677, 846

\bibitem[{{Bielby} {et~al}\mbox{.}(2012){Bielby}, {Hudelot}, {McCracken},
  {Ilbert}, {Daddi}, {Le F{\`e}vre}, {Gonzalez-Perez}, {Kneib}, {Marmo},
  {Mellier}, {Salvato}, {Sanders}, \& {Willott}}]{bielby+12}
{Bielby} R. {et~al.}, 2012, \aap, 545, A23

\bibitem[{{Boroson} \& {Oke}(1982)}]{borosonoke82}
{Boroson} T.~A., {Oke} J.~B., 1982, \nat, 296, 397

\bibitem[{Boyle {et~al}\mbox{.}(1990)Boyle, Fong, Shanks, \&
  Peterson}]{Boyle:UVQSO}
Boyle B., Fong R., Shanks T., Peterson B., 1990, \mnras, 243, 1

\bibitem[{Brotherton {et~al}\mbox{.}(1999)Brotherton, van Breugel, Stanford,
  Smith, Boyle, Miller, Shanks, Croom, \& Filippenko}]{brotherton+99}
Brotherton M.~S. {et~al.}, 1999, \apj, 520, L87

\bibitem[{Bruce {et~al}\mbox{.}(2012)Bruce, Dunlop, Cirasuolo, McLure, Targett,
  Bell, Croton, Dekel, Faber, Ferguson, Grogin, Kocevski, Koekemoer, Koo, Lai,
  Lotz, McGrath, Newman, \& Van Der~Wel}]{bruce+12}
Bruce V.~A. {et~al.}, 2012, arXiv:1206.4322

\bibitem[{Bruzual \& Charlot(2003)}]{BC03}
Bruzual G., Charlot S., 2003, \mnras, 344, 1000

\bibitem[{Canalizo {et~al}\mbox{.}(2007)Canalizo, Bennert, Jungwiert, Stockton,
  Schweizer, Lacy, \& Peng}]{canalizo+07}
Canalizo G., Bennert N., Jungwiert B., Stockton A., Schweizer F., Lacy M., Peng
  C., 2007, \apj, 669, 801

\bibitem[{Capak {et~al}\mbox{.}(2007)Capak, Aussel, Ajiki, Mccracken, Mobasher,
  Scoville, Shopbell, Taniguchi, Thompson, Tribiano, Sasaki, Blain, Brusa, \&
  et~al.}]{capak+07}
Capak P. {et~al.}, 2007, \apjs, 172, 99

\bibitem[{{Cen}(2012)}]{cen11}
{Cen} R., 2012, \apj, 755, 28

\bibitem[{Chabrier(2003)}]{chabrier03}
Chabrier G., 2003, Publications of the Astronomical Society of the Pacific,
  115, 763

\bibitem[{Cid~Fernandes {et~al}\mbox{.}(2004)Cid~Fernandes, Gu, Melnick,
  Terlevich, Terlevich, Kunth, Rodrigues~Lacerda, \& Joguet}]{cidfernandes+04}
Cid~Fernandes R., Gu Q., Melnick J., Terlevich E., Terlevich R., Kunth D.,
  Rodrigues~Lacerda R., Joguet B., 2004, \mnras, 355, 273

\bibitem[{{Conroy} {et~al}\mbox{.}(2009){Conroy}, {Gunn}, \&
  {White}}]{conroy+09}
{Conroy} C., {Gunn} J.~E., {White} M., 2009, \apj, 699, 486

\bibitem[{Croom {et~al}\mbox{.}(2001)Croom, Smith, Boyle, Shanks, Loaring,
  Miller, \& Lewis}]{Croom:2dfQSO}
Croom S., Smith R., Boyle B., Shanks T., Loaring N., Miller L., Lewis I., 2001,
  \mnras, 322, L29

\bibitem[{Croom {et~al}\mbox{.}(2004)Croom, Smith, Boyle, Shanks, Miller,
  Outram, \& Loaring}]{croom+04}
Croom S.~M., Smith R.~J., Boyle B.~J., Shanks T., Miller L., Outram P.~J.,
  Loaring N.~S., 2004, \mnras, 349, 1397

\bibitem[{Davies {et~al}\mbox{.}(2007)Davies, S{\'a}nchez, Genzel, Tacconi,
  Hicks, Friedrich, \& Sternberg}]{davies+07}
Davies R.~I., S{\'a}nchez F.~M., Genzel R., Tacconi L.~J., Hicks E. K.~S.,
  Friedrich S., Sternberg A., 2007, \apj, 671, 1388

\bibitem[{Disney {et~al}\mbox{.}(1995)Disney, Boyce, Blades, Boksenberg, Cane,
  Deharveng, Macchetto, Mackay, Sparks, \& Phillipps}]{disney+95}
Disney M.~J. {et~al.}, 1995, \nat, 376, 150

\bibitem[{Dunlop {et~al}\mbox{.}(1989)Dunlop, Peacock, Savage, Lilly, Heasley,
  \& Simon}]{Dunlop:PKS_QSO}
Dunlop J., Peacock J., Savage A., Lilly S., Heasley J., Simon A., 1989, \mnras,
  238, 1171

\bibitem[{Dunlop {et~al}\mbox{.}(2003)Dunlop, McLure, Kukula, Baum, O'Dea, \&
  Hughes}]{dunlop+03}
Dunlop J.~S., McLure R.~J., Kukula M.~J., Baum S.~A., O'Dea C.~P., Hughes
  D.~H., 2003, \mnras, 340, 1095

\bibitem[{Floyd(2005)}]{floydPhDT}
Floyd D., 2005, PhD thesis, University of Edinburgh

\bibitem[{Floyd {et~al}\mbox{.}(2008)Floyd, Axon, Baum, Capetti, Chiaberge,
  Macchetto, Madrid, Miley, O'dea, Perlman, Quillen, Sparks, \&
  Tremblay}]{floyd+08}
Floyd D. J.~E. {et~al.}, 2008, \apjs, 177, 148

\bibitem[{Floyd {et~al}\mbox{.}(2004)Floyd, Kukula, Dunlop, Mclure, Miller,
  Percival, Baum, \& O'dea}]{floyd+04}
Floyd D. J.~E., Kukula M.~J., Dunlop J.~S., Mclure R.~J., Miller L., Percival
  W.~J., Baum S.~A., O'dea C.~P., 2004, \mnras, 355, 196

\bibitem[{{Gallazzi} \& {Bell}(2009)}]{gallazzi+09}
{Gallazzi} A., {Bell} E.~F., 2009, \apjs, 185, 253

\bibitem[{Gonz{\'a}lez~Delgado {et~al}\mbox{.}(2001)Gonz{\'a}lez~Delgado,
  Heckman, \& Leitherer}]{gonzalezdelgado+01}
Gonz{\'a}lez~Delgado R.~M., Heckman T., Leitherer C., 2001, \apj, 546, 845

\bibitem[{Gu {et~al}\mbox{.}(2001)Gu, Huang, de~Diego, Dultzin-Hacyan, Lei, \&
  Ben{\'\i}tez}]{gu+01}
Gu Q.~S., Huang J.~H., de~Diego J.~A., Dultzin-Hacyan D., Lei S.~J.,
  Ben{\'\i}tez E., 2001, \aap, 374, 932

\bibitem[{Guyon {et~al}\mbox{.}(2006)Guyon, Sanders, \& Stockton}]{guyon+06}
Guyon O., Sanders D.~B., Stockton A., 2006, \apjs, 166, 89

\bibitem[{Hamilton {et~al}\mbox{.}(2002)Hamilton, Casertano, \&
  Turnshek}]{hamilton+02}
Hamilton T.~S., Casertano S., Turnshek D.~A., 2002, \apj, 576, 61

\bibitem[{Heckman {et~al}\mbox{.}(1997)Heckman, Gonzalez-Delgado, Leitherer,
  Meurer, Krolik, Wilson, Koratkar, \& Kinney}]{heckman+97}
Heckman T.~M., Gonzalez-Delgado R., Leitherer C., Meurer G.~R., Krolik J.,
  Wilson A.~S., Koratkar A., Kinney A., 1997, \apj, 482, 114

\bibitem[{{Hewitt} \& {Burbidge}(1989)}]{HB_QSO_89}
{Hewitt} A., {Burbidge} G., 1989, in A new optical catalog of QSO (1989), p.~0

\bibitem[{{Ho}(2002)}]{ho02}
{Ho} L.~C., 2002, \apj, 564, 120

\bibitem[{Hogg {et~al}\mbox{.}(2002)Hogg, Baldry, Blanton, \&
  Eisenstein}]{hogg+02}
Hogg D.~W., Baldry I.~K., Blanton M.~R., Eisenstein D.~J., 2002,
  arXiv:astro-ph/0210394

\bibitem[{Hooper {et~al}\mbox{.}(1997)Hooper, Impey, \& Foltz}]{hooper+97}
Hooper E.~J., Impey C.~D., Foltz C.~B., 1997, \apjl, 480, 95

\bibitem[{Ilbert {et~al}\mbox{.}(2009)Ilbert, Capak, Salvato, Aussel,
  Mccracken, Sanders, Scoville, Kartaltepe, Arnouts, Le~Floc'h, Mobasher,
  Taniguchi, \& et~al.}]{ilbert+09}
Ilbert O. {et~al.}, 2009, \apj, 690, 1236

\bibitem[{{Jahnke} {et~al}\mbox{.}(2009){Jahnke}, {Bongiorno}, {Brusa},
  {Capak}, {Cappelluti}, {Cisternas}, {Civano}, {Colbert}, {Comastri}, {Elvis},
  {Hasinger}, {Ilbert}, {Impey}, {Inskip}, {Koekemoer}, {Lilly}, {Maier},
  {Merloni}, {Riechers}, {Salvato}, {Schinnerer}, {Scoville}, {Silverman},
  {Taniguchi}, {Trump}, \& {Yan}}]{jahnke+09}
{Jahnke} K. {et~al.}, 2009, \apjl, 706, L215

\bibitem[{Jahnke {et~al}\mbox{.}(2004{\natexlab{a}})Jahnke, Kuhlbrodt, \&
  Wisotzki}]{jahnke+04a}
Jahnke K., Kuhlbrodt B., Wisotzki L., 2004{\natexlab{a}}, \mnras, 352, 399

\bibitem[{Jahnke {et~al}\mbox{.}(2004{\natexlab{b}})Jahnke, S{\'a}nchez,
  Wisotzki, Barden, Beckwith, Bell, Borch, Caldwell, H{\"a}ussler, Heymans,
  Jogee, McIntosh, Meisenheimer, Peng, Rix, Somerville, \& Wolf}]{jahnke+04b}
Jahnke K. {et~al.}, 2004{\natexlab{b}}, \apj, 614, 568

\bibitem[{Jahnke {et~al}\mbox{.}(2007)Jahnke, Wisotzki, Courbin, \&
  Letawe}]{jahnke+07}
Jahnke K., Wisotzki L., Courbin F., Letawe G., 2007, \mnras, 378, 23

\bibitem[{Kauffmann {et~al}\mbox{.}(2003)Kauffmann, Heckman, Tremonti,
  Brinchmann, Charlot, White, Ridgway, Brinkmann, Fukugita, Hall, Ivezi{\'c},
  Richards, \& Schneider}]{kauffmann+03}
Kauffmann G. {et~al.}, 2003, \mnras, 346, 1055

\bibitem[{Kennefick \& Bursick(2008)}]{kcorrqso}
Kennefick J., Bursick S., 2008, \aj, 136, 1799

\bibitem[{Kennicutt~Jr.(1998)}]{kennicutt98}
Kennicutt~Jr. R.~C., 1998, \araa, 36, 189

\bibitem[{Kotilainen \& Ward(1994)}]{kotilainenward94}
Kotilainen J.~K., Ward M.~J., 1994, \mnras, 266, 953

\bibitem[{Kukula {et~al}\mbox{.}(2001)Kukula, Dunlop, Mclure, Miller, Percival,
  Baum, \& O'dea}]{kukula+01}
Kukula M.~J., Dunlop J.~S., Mclure R.~J., Miller L., Percival W.~J., Baum
  S.~A., O'dea C.~P., 2001, \mnras, 326, 1533

\bibitem[{{Mainieri} {et~al}\mbox{.}(2011){Mainieri}, {Bongiorno}, {Merloni},
  {Aller}, {Carollo}, {Iwasawa}, {Koekemoer}, {Mignoli}, {Silverman},
  {Bolzonella}, {Brusa}, {Comastri}, {Gilli}, {Halliday}, {Ilbert}, {Lusso},
  {Salvato}, {Vignali}, {Zamorani}, {Contini}, {Kneib}, {Le F{\`e}vre},
  {Lilly}, {Renzini}, {Scodeggio}, {Balestra}, {Bardelli}, {Caputi}, {Coppa},
  {Cucciati}, {de la Torre}, {de Ravel}, {Franzetti}, {Garilli}, {Iovino},
  {Kampczyk}, {Knobel}, {Kova{\v c}}, {Lamareille}, {Le Borgne}, {Le Brun},
  {Maier}, {Nair}, {Pello}, {Peng}, {Perez Montero}, {Pozzetti},
  {Ricciardelli}, {Tanaka}, {Tasca}, {Tresse}, {Vergani}, {Zucca}, {Aussel},
  {Capak}, {Cappelluti}, {Elvis}, {Fiore}, {Hasinger}, {Impey}, {Le Floc'h},
  {Scoville}, {Taniguchi}, \& {Trump}}]{mainieri+11}
{Mainieri} V. {et~al.}, 2011, \aap, 535, A80

\bibitem[{Marshall {et~al}\mbox{.}(1984)Marshall, Huchra, Tananbaum, Avni,
  Braccesi, Zitelli, \& Zamorani}]{marshall+84_QSOsamp}
Marshall H.~L., Huchra J.~P., Tananbaum H., Avni Y., Braccesi A., Zitelli V.,
  Zamorani G., 1984, Astrophysical Journal, 283, 50

\bibitem[{McLeod {et~al}\mbox{.}(1999)McLeod, Rieke, \&
  Storrie-Lombardi}]{mcleod+99}
McLeod K.~K., Rieke G.~H., Storrie-Lombardi L.~J., 1999, \apjl, 511, L67

\bibitem[{McLure {et~al}\mbox{.}(2000)McLure, Dunlop, \& Kukula}]{mclure+00}
McLure R.~J., Dunlop J.~S., Kukula M.~J., 2000, \mnras, 318, 693

\bibitem[{McLure {et~al}\mbox{.}(1999)McLure, Kukula, Dunlop, Baum, O'Dea, \&
  Hughes}]{mclure+99}
McLure R.~J., Kukula M.~J., Dunlop J.~S., Baum S.~A., O'Dea C.~P., Hughes
  D.~H., 1999, \mnras, 308, 377

\bibitem[{Nolan {et~al}\mbox{.}(2001)Nolan, Dunlop, Kukula, Hughes, Boroson, \&
  Jimenez}]{nolan+01}
Nolan L.~A., Dunlop J.~S., Kukula M.~J., Hughes D.~H., Boroson T., Jimenez R.,
  2001, \mnras, 323, 308

\bibitem[{Page {et~al}\mbox{.}(2012)Page, Symeonidis, Vieira, Altieri, Amblard,
  Arumugam, Aussel, Babbedge, Blain, Bock, Boselli, Buat,
  Castro-Rodr{\'\i}guez, Cava, Chanial, Clements, Conley, Conversi, Cooray,
  Dowell, Dubois, Dunlop, Dwek, Dye, Eales, Elbaz, Farrah, Fox, Franceschini,
  Gear, Glenn, Griffin, Halpern, Hatziminaoglou, Ibar, Isaak, Ivison, Lagache,
  Levenson, Lu, Madden, Maffei, Mainetti, Marchetti, Nguyen, O'Halloran,
  Oliver, Omont, Panuzzo, Papageorgiou, Pearson, Perez-Fournon, Pohlen,
  Rawlings, Rigopoulou, Riguccini, Rizzo, Rodighiero, Roseboom, Rowan-Robinson,
  Portal, Schulz, Scott, Seymour, Shupe, Smith, Stevens, Trichas, Tugwell,
  Vaccari, Valtchanov, Viero, Vigroux, Wang, Ward, Wright, Xu, \&
  Zemcov}]{page+12}
Page M.~J. {et~al.}, 2012, \nat, 485, 213

\bibitem[{{Panessa} {et~al}\mbox{.}(2007){Panessa}, {Barcons}, {Bassani},
  {Cappi}, {Carrera}, {Ho}, \& {Pellegrini}}]{panessa+07}
{Panessa} F., {Barcons} X., {Bassani} L., {Cappi} M., {Carrera} F.~J., {Ho}
  L.~C., {Pellegrini} S., 2007, \aap, 467, 519

\bibitem[{{Pierce} {et~al}\mbox{.}(2011){Pierce}, {Ballantyne}, \&
  {Ivison}}]{pierce+11}
{Pierce} C.~M., {Ballantyne} D.~R., {Ivison} R.~J., 2011, \apj, 742, 45

\bibitem[{Ridgway {et~al}\mbox{.}(2001)Ridgway, Heckman, Calzetti, \&
  Lehnert}]{ridgway+01}
Ridgway S.~E., Heckman T.~M., Calzetti D., Lehnert M., 2001, \apj, 550, 122

\bibitem[{Riffel {et~al}\mbox{.}(2009)Riffel, Storchi-Bergmann, Dors, \&
  Winge}]{riffel+09}
Riffel R.~A., Storchi-Bergmann T., Dors O.~L., Winge C., 2009, \mnras, 393, 783

\bibitem[{R{\"o}nnback {et~al}\mbox{.}(1996)R{\"o}nnback, Van~Groningen,
  Wanders, \& {\"O}rndahl}]{ronnback+96}
R{\"o}nnback J., Van~Groningen E., Wanders I., {\"O}rndahl E., 1996, \mnras,
  283, 282

\bibitem[{S{\'a}nchez {et~al}\mbox{.}(2004)S{\'a}nchez, Jahnke, Wisotzki,
  McIntosh, Bell, Barden, Beckwith, Borch, Caldwell, H{\"a}ussler, Jogee,
  Meisenheimer, Peng, Rix, Somerville, \& Wolf}]{sanchez+04}
S{\'a}nchez S.~F. {et~al.}, 2004, \apj, 614, 586

\bibitem[{Schlegel {et~al}\mbox{.}(1998)Schlegel, Finkbeiner, \&
  Davis}]{schlegel+98}
Schlegel D.~J., Finkbeiner D.~P., Davis M., 1998, \apj, 500, 525

\bibitem[{Schneider {et~al}\mbox{.}(2010)Schneider, Richards, Hall, Strauss,
  Anderson, Boroson, Ross, Shen, Brandt, Fan, Inada, Jester, Knapp, Krawczyk,
  Thakar, Vanden~Berk, Voges, Yanny, York, Bahcall, Bizyaev, Blanton,
  Brewington, \& et~al.}]{SDSS_QSO_DR7}
Schneider D. {et~al.}, 2010, VizieR Online Data Catalog, 7260, 0

\bibitem[{Shi {et~al}\mbox{.}(2009)Shi, Rieke, Ogle, Jiang, \&
  Diamond-Stanic}]{shi+09}
Shi Y., Rieke G.~H., Ogle P., Jiang L., Diamond-Stanic A.~M., 2009, \apj, 703,
  1107

\bibitem[{Sikora {et~al}\mbox{.}(2007)Sikora, Stawarz, \& Lasota}]{sikora+07}
Sikora M., Stawarz L., Lasota J.-P., 2007, \apj, 658, 815

\bibitem[{Stickel \& Kuhr(1993)}]{Stickel:radio}
Stickel M., Kuhr H., 1993, Astronomy and Astrophysics Supplement Series, 101,
  521

\bibitem[{Taylor {et~al}\mbox{.}(1996)Taylor, Dunlop, Hughes, \&
  Robson}]{taylor+96}
Taylor G.~L., Dunlop J.~S., Hughes D.~H., Robson E.~I., 1996, \mnras, 283, 930

\bibitem[{{Thomas} {et~al}\mbox{.}(1995){Thomas}, {Webster}, \&
  {Drinkwater}}]{thomas+95}
{Thomas} P.~A., {Webster} R.~L., {Drinkwater} M.~J., 1995, \mnras, 273, 1069

\bibitem[{{Trichas} {et~al}\mbox{.}(2012){Trichas}, {Green}, {Silverman},
  {Aldcroft}, {Barkhouse}, {Cameron}, {Constantin}, {Ellison}, {Foltz},
  {Haggard}, {Jannuzi}, {Kim}, {Marshall}, {Mossman}, {P{\'e}rez},
  {Romero-Colmenero}, {Ruiz}, {Smith}, {Smith}, {Torres}, {Wik}, {Wilkes}, \&
  {Wolfgang}}]{trichas+12}
{Trichas} M. {et~al.}, 2012, \apjs, 200, 17

\bibitem[{van Dokkum {et~al}\mbox{.}(2006)van Dokkum, Quadri, Marchesini,
  Rudnick, Franx, Gawiser, Herrera, Wuyts, Lira, Labb{\'e}, Maza, Illingworth,
  F{\"o}rster~Schreiber, Kriek, Rix, Taylor, Toft, Webb, \& Yi}]{vandokkum+06}
van Dokkum P.~G. {et~al.}, 2006, \apjl, 638, L59

\bibitem[{Vanden~Berk {et~al}\mbox{.}(2001)Vanden~Berk, Richards, Bauer,
  Strauss, Schneider, Heckman, York, Hall, Fan, Knapp, Anderson, Annis,
  Bahcall, \& et~al.}]{VandenBerk:2001p3354}
Vanden~Berk D.~E. {et~al.}, 2001, \aj, 122, 549

\bibitem[{Vanden~Berk {et~al}\mbox{.}(2006)Vanden~Berk, Shen, Yip, Schneider,
  Connolly, Burton, Jester, Hall, Szalay, \& Brinkmann}]{vandenberk+06}
Vanden~Berk D.~E. {et~al.}, 2006, \apj, 131, 84

\bibitem[{Veron-Cetty \& Veron(1993)}]{VCV93}
Veron-Cetty M., Veron P., 1993, Astronomy and Astrophysics Supplement Series,
  100, 521

\bibitem[{{Xia} {et~al}\mbox{.}(2012){Xia}, {Gao}, {Hao}, {Tan}, {Mao},
  {Omont}, {Flaquer}, {Leon}, \& {Cox}}]{xia+12}
{Xia} X.~Y. {et~al.}, 2012, \apj, 750, 92

\bibitem[{Xu {et~al}\mbox{.}(1999)Xu, Livio, \& Baum}]{xu+99}
Xu C., Livio M., Baum S., 1999, \aj, 118, 1169

\end{thebibliography}

\appendix

\section{Images, models and notes on each object}
In this section we present the following for each object (left-to-right and moving downwards): 
A 10\arcsec\ square ``postage stamp'' image centred on the quasar; The best fit convolved model; The model-subtracted residuals; The radial profile.
In the radial profile, the best-fit model is shown as a solid line, with the PSF component indicated by a dotted line. Data points are azimuthal averages with 1$\sigma$ error bars. Note the significant deviations from smooth profiles in the majority of cases. These bumps are traceable to numerous features in the image which may be nearby companions or mergers, or knots of star formation within the host galaxies themselves.

\begin{figure}
{\includegraphics[width=40mm]{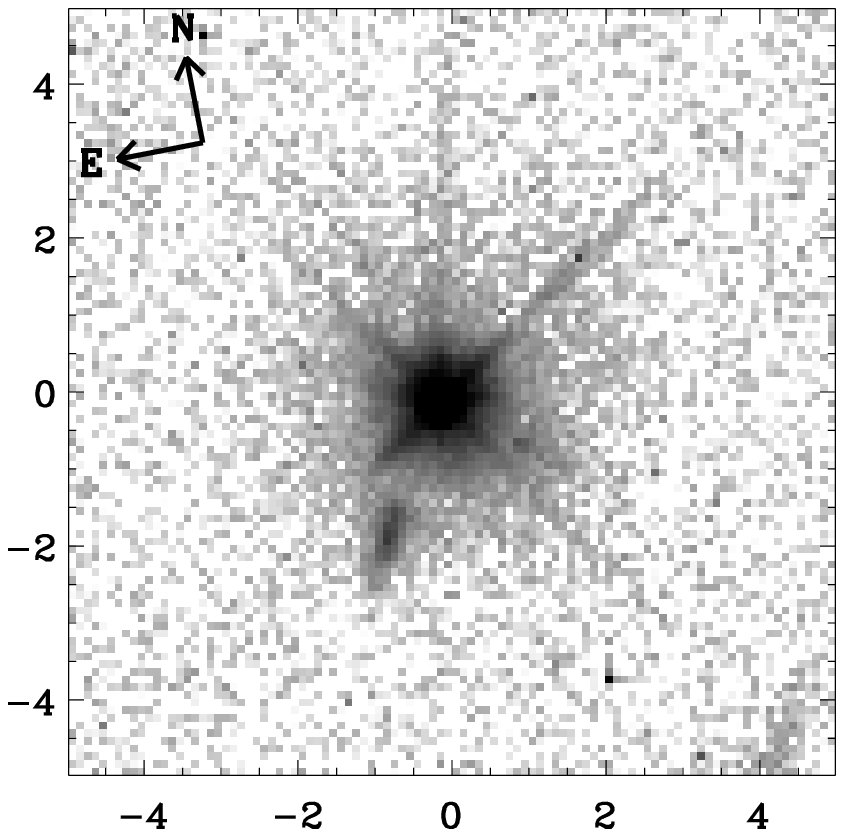}}
{\includegraphics[width=40mm]{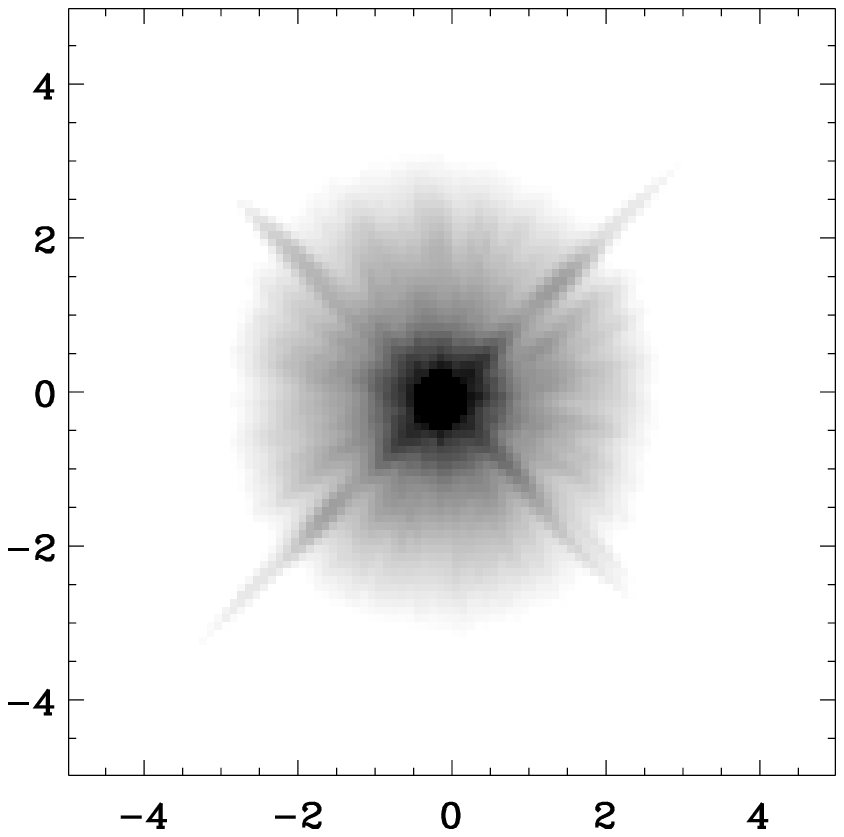}}

{\includegraphics[width=40mm]{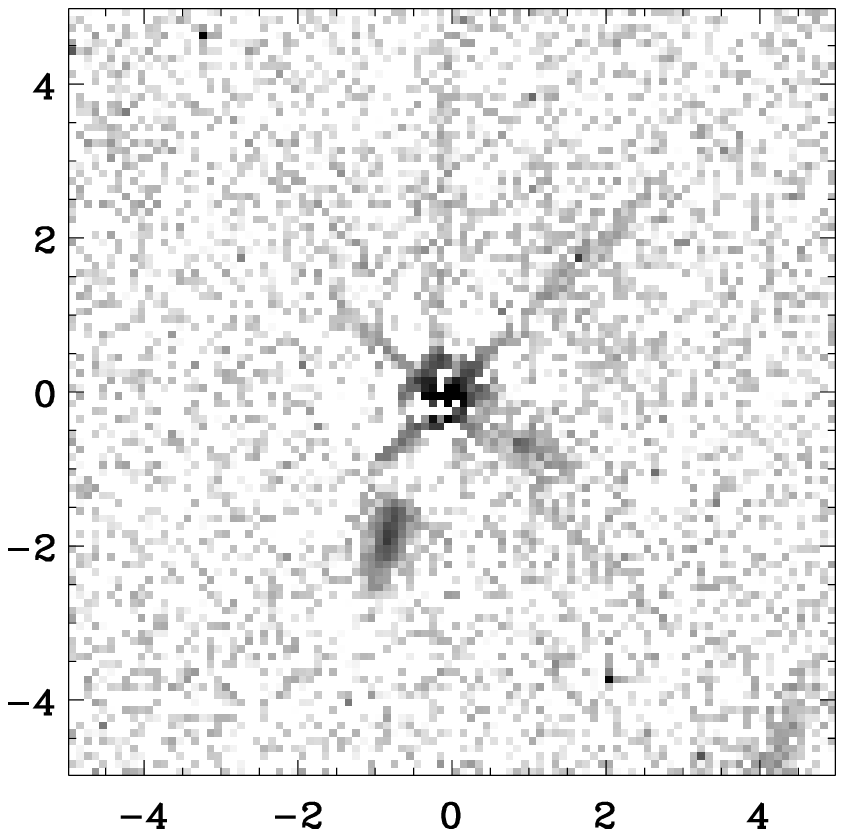}}
{\includegraphics[angle=270,origin=c,width=40mm]{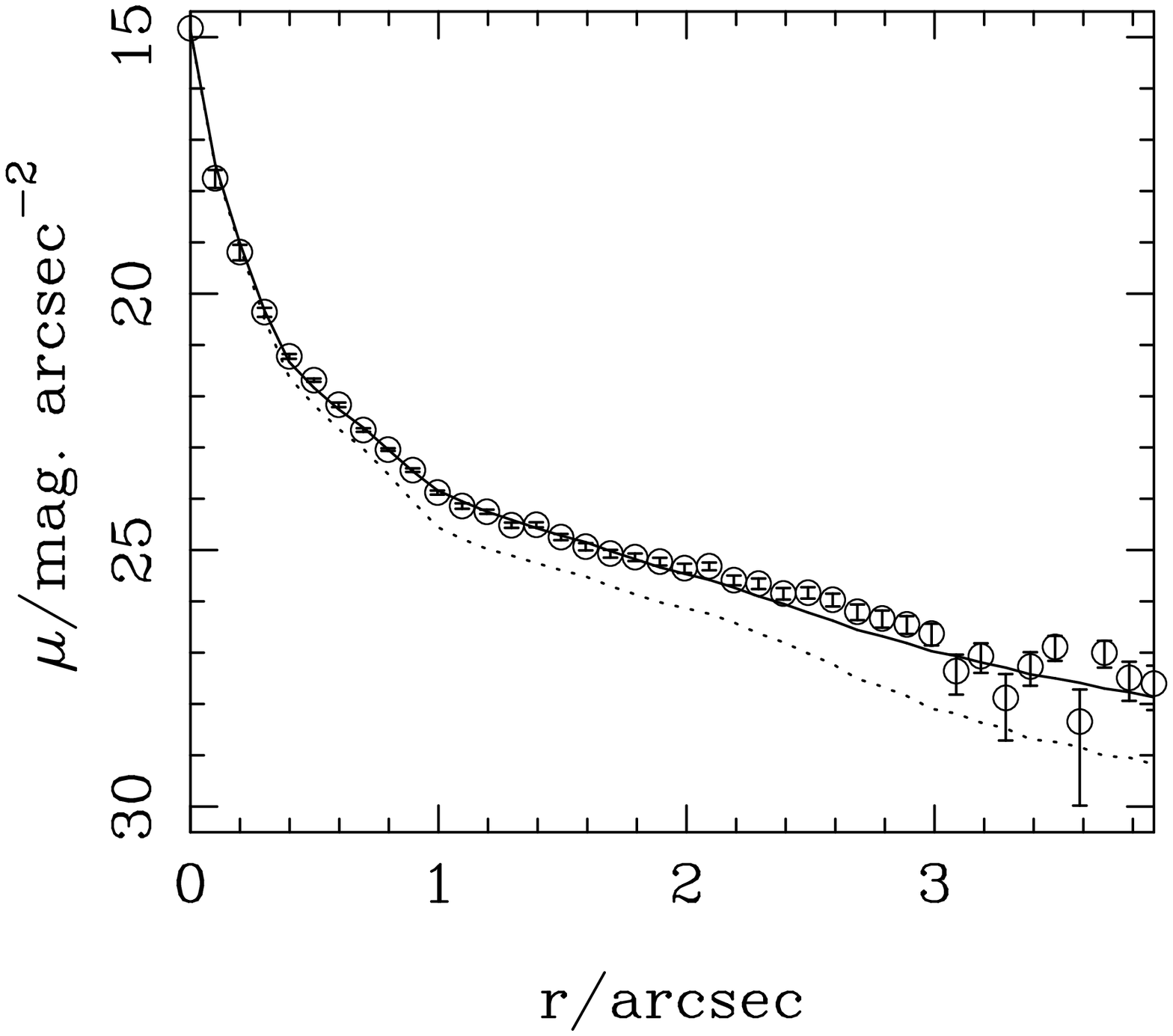}}
\caption{\label{fig-h104} The RQQ BVF225 at $z=0.910$. A prominent nucleus and an interacting companion are the most noticeable features of this object. The field contains a number of other small galaxies greater than 5~arcsec from the quasar. It is strongly nuclear dominated in the rest-frame optical ($[L_\mathrm{nuc}/ L_\mathrm{host}]_V=7.6$) and in the rest-frame $U$ ($[L_\mathrm{nuc}/L_\mathrm{host}]_U= 12.6$). Significant PSF-related circumnuclear flux is visible in the residuals.}
\end{figure}

\begin{figure}
{\includegraphics[width=40mm]{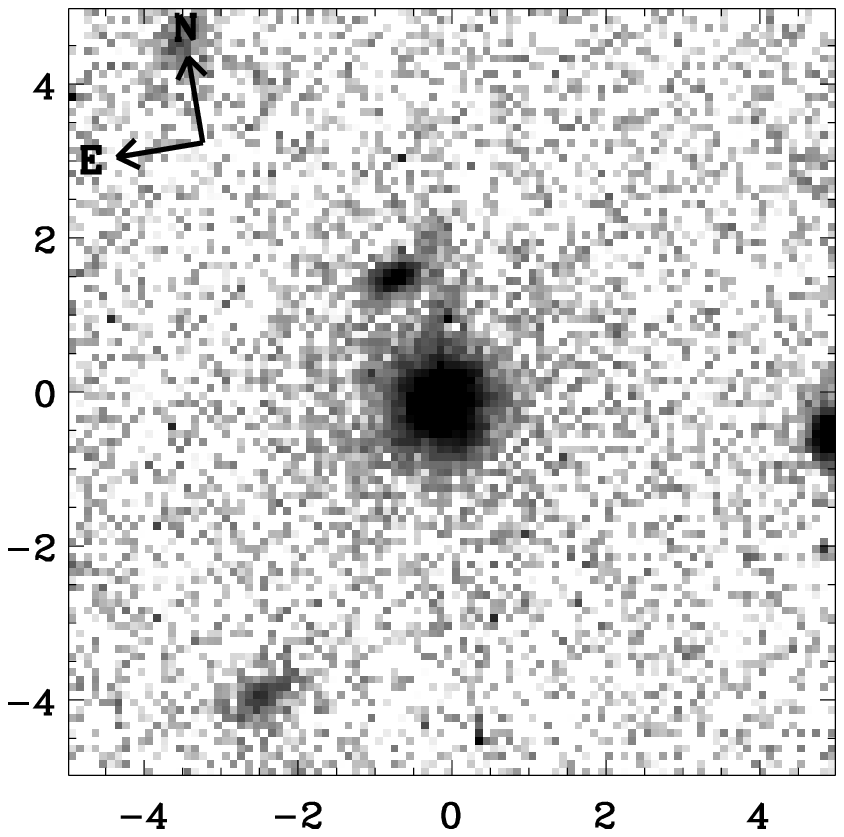}}
{\includegraphics[width=40mm]{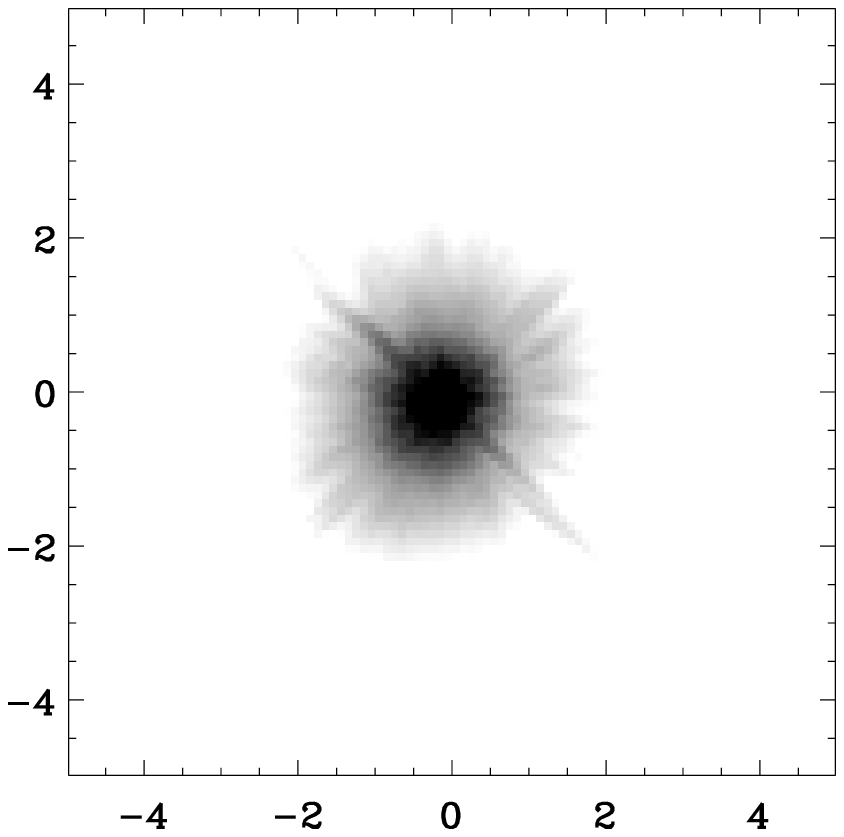}}

{\includegraphics[width=40mm]{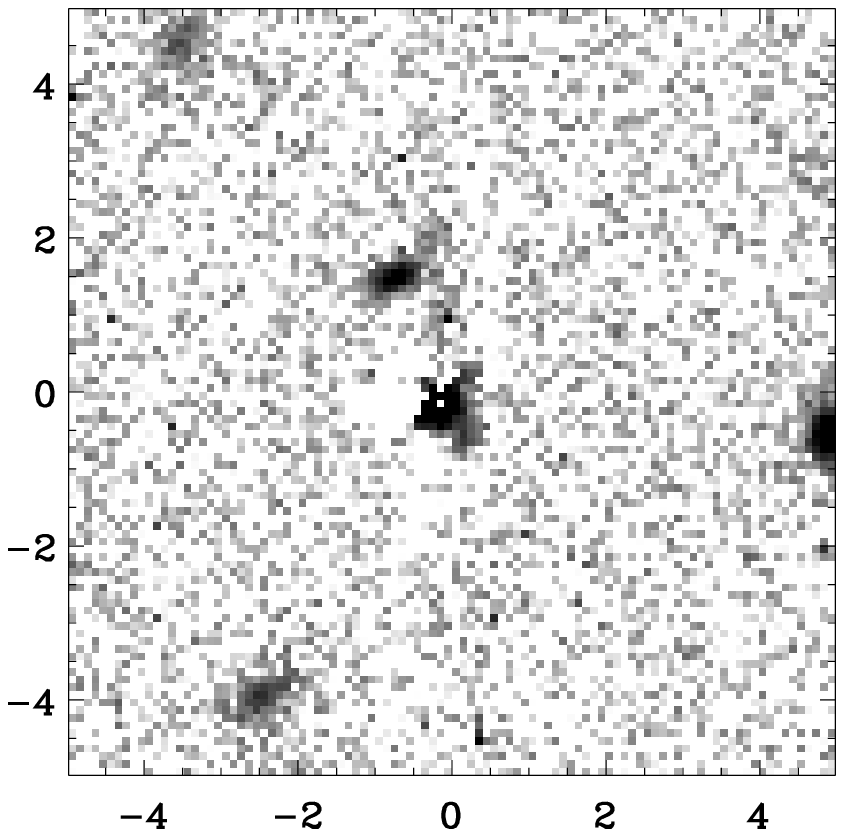}}
{\includegraphics[angle=270,origin=c,width=40mm]{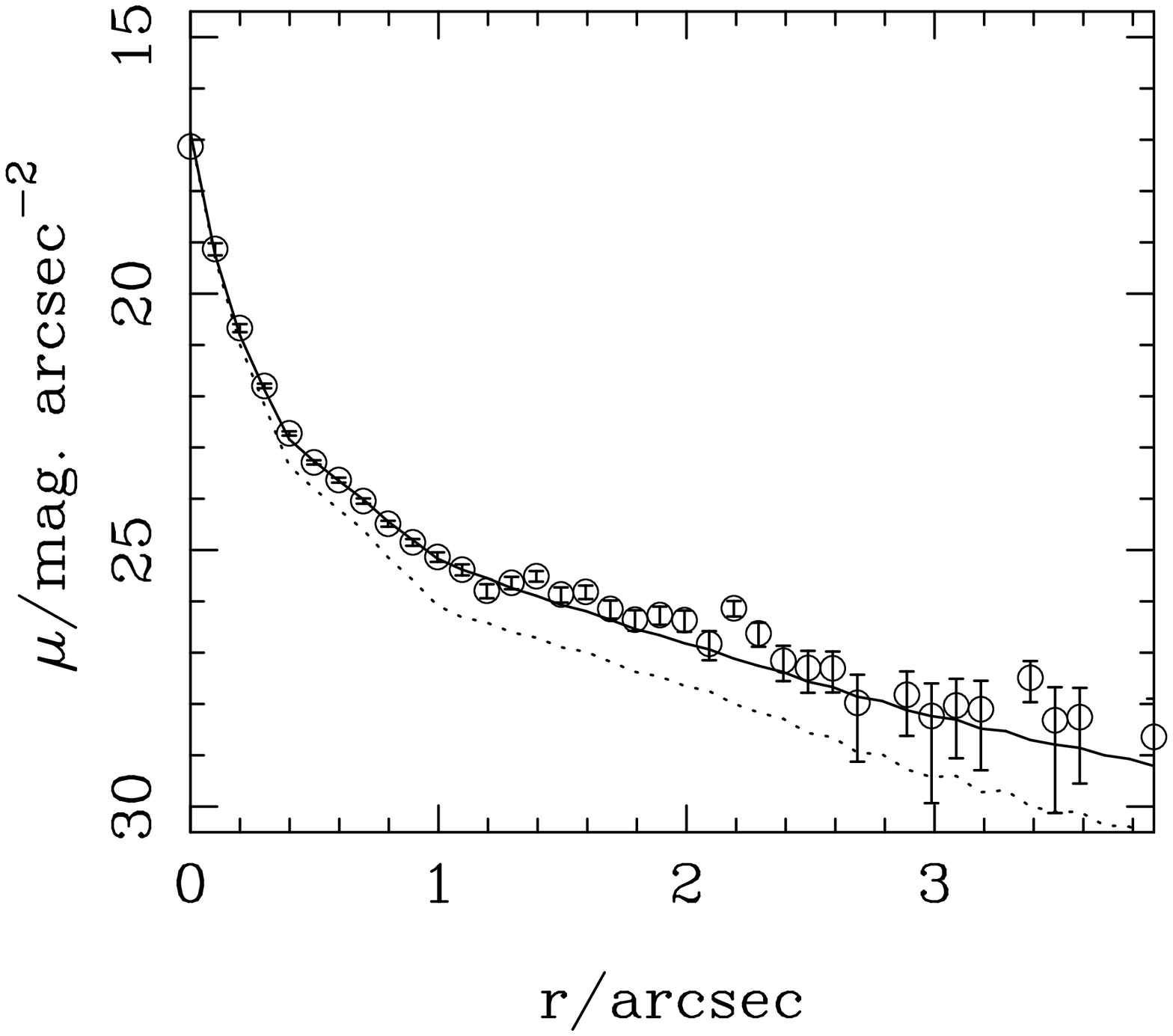}}
\caption{\label{fig-h103} The RQQ BVF247 at $z=0.890$. A number of candidate companions are visible, most noticeably a small, possibly interacting object some 2\arcsec\ to the North. Whilst the host is quite prominent in rest-frame $V$ , it is heavily nuclear dominated in the UV: $(L_\mathrm{nuc}/L_\mathrm{host})_V=0.3$;~$(L_\mathrm{nuc}/L_\mathrm{host})_U=6.1$. There is some excess circumnuclear flux. }
\end{figure}

\begin{figure}
{\includegraphics[width=40mm]{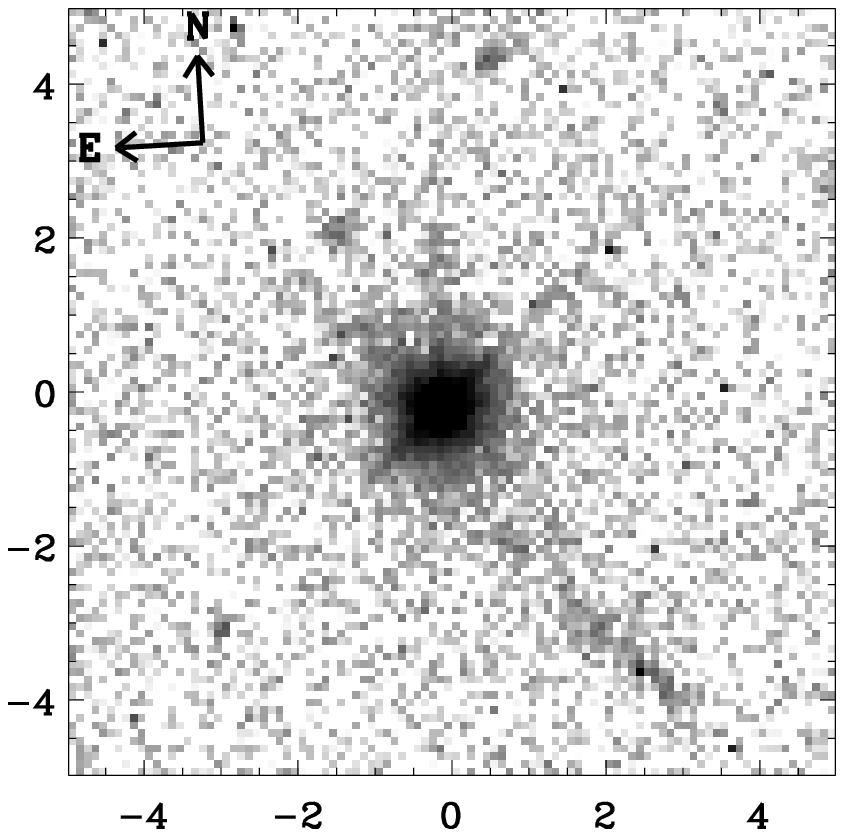}}
{\includegraphics[width=40mm]{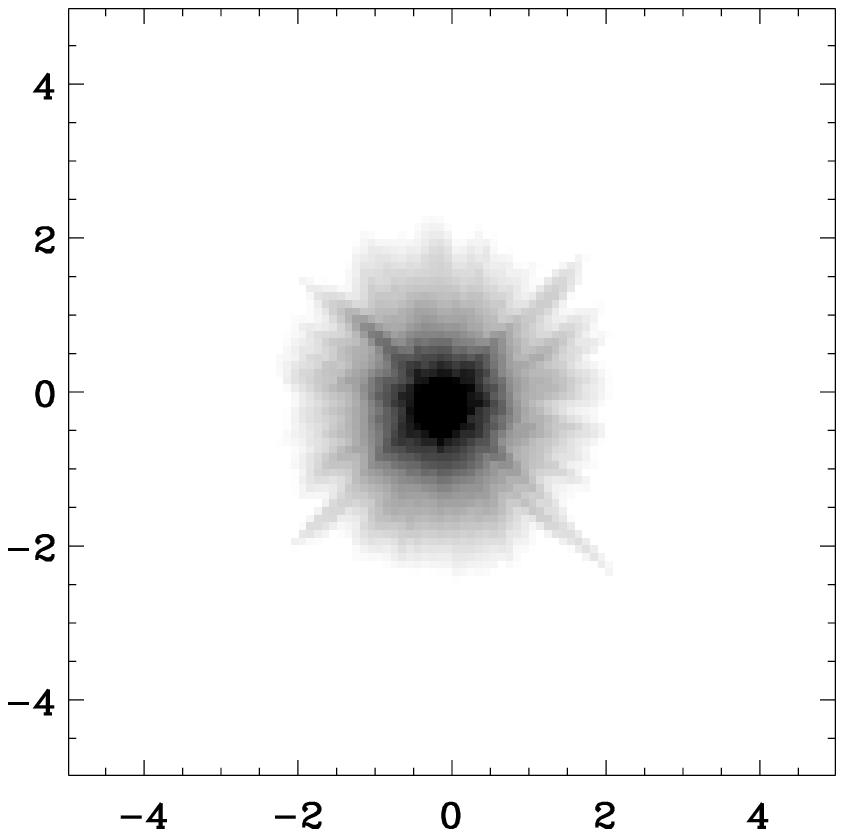}}

{\includegraphics[width=40mm]{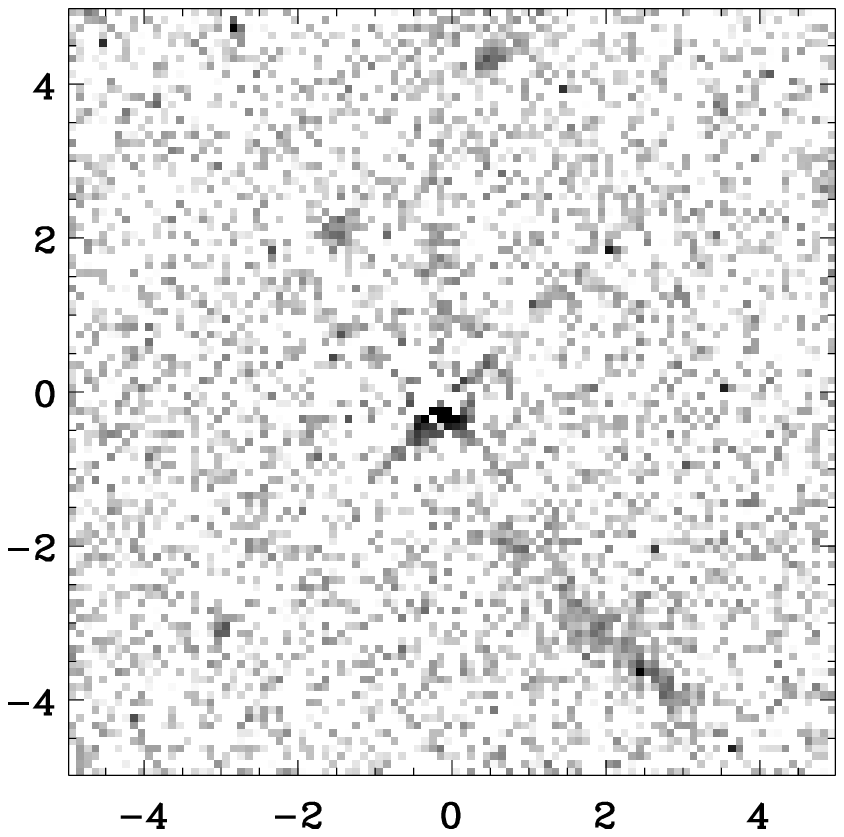}}
{\includegraphics[angle=270,origin=c,width=40mm]{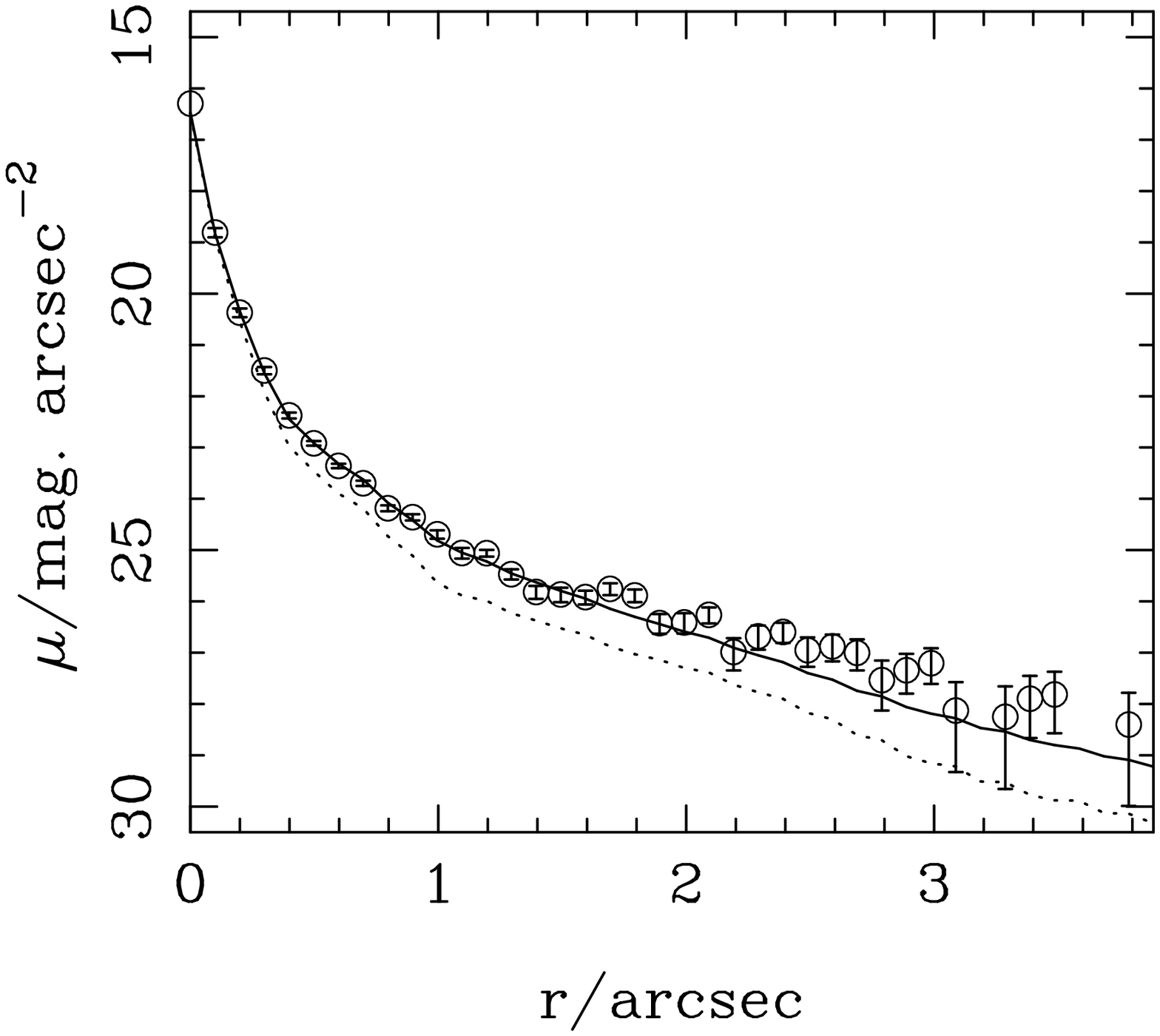}}
\caption{\label{fig-h106} The RQQ BVF262 at $z=0.970$. 
Nuclear domination is more extreme in $U$ as expected: $[L_\mathrm{nuc}/L_\mathrm{host}]_V=1.8$;~$[L_\mathrm{nuc}/ L_\mathrm{host}]_U= 5.9$.}
\end{figure}

\begin{figure}
{\includegraphics[width=40mm]{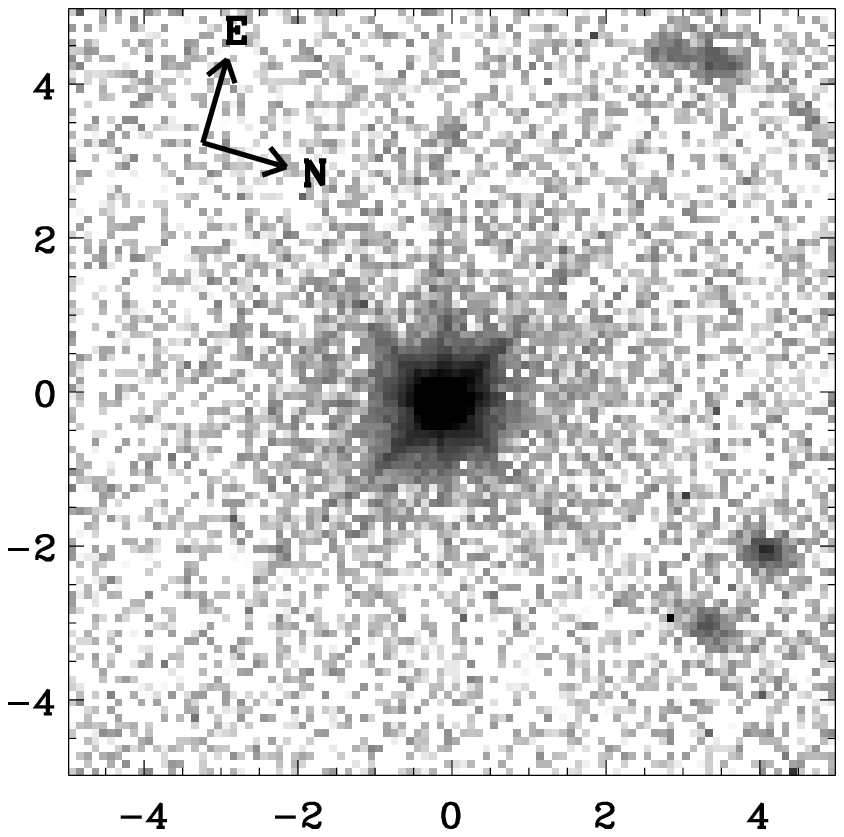}}
{\includegraphics[width=40mm]{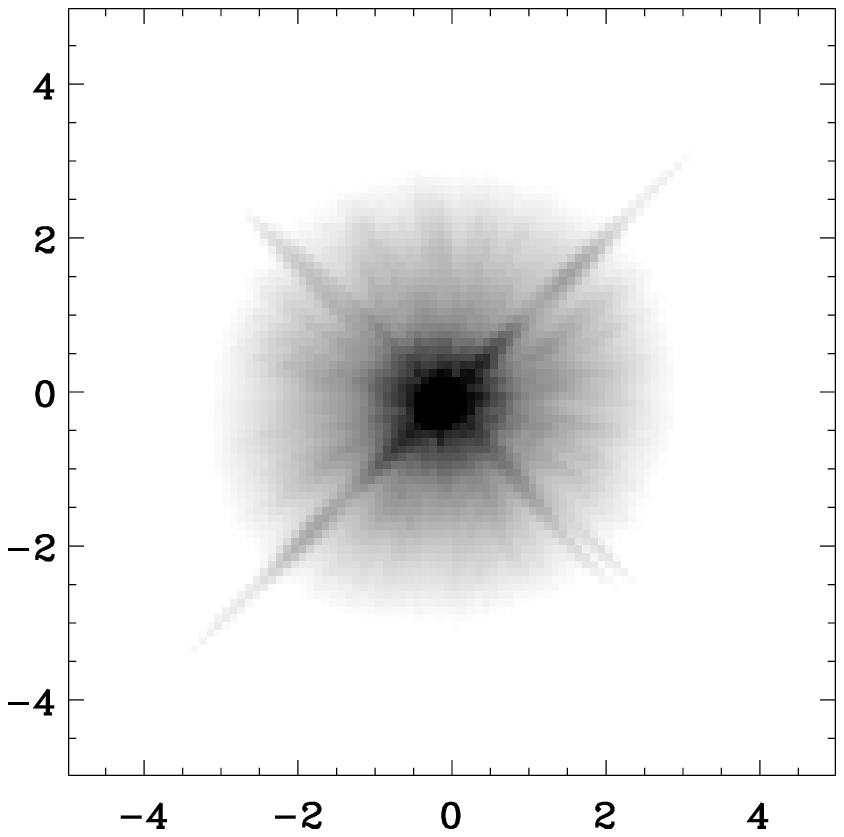}}

{\includegraphics[width=40mm]{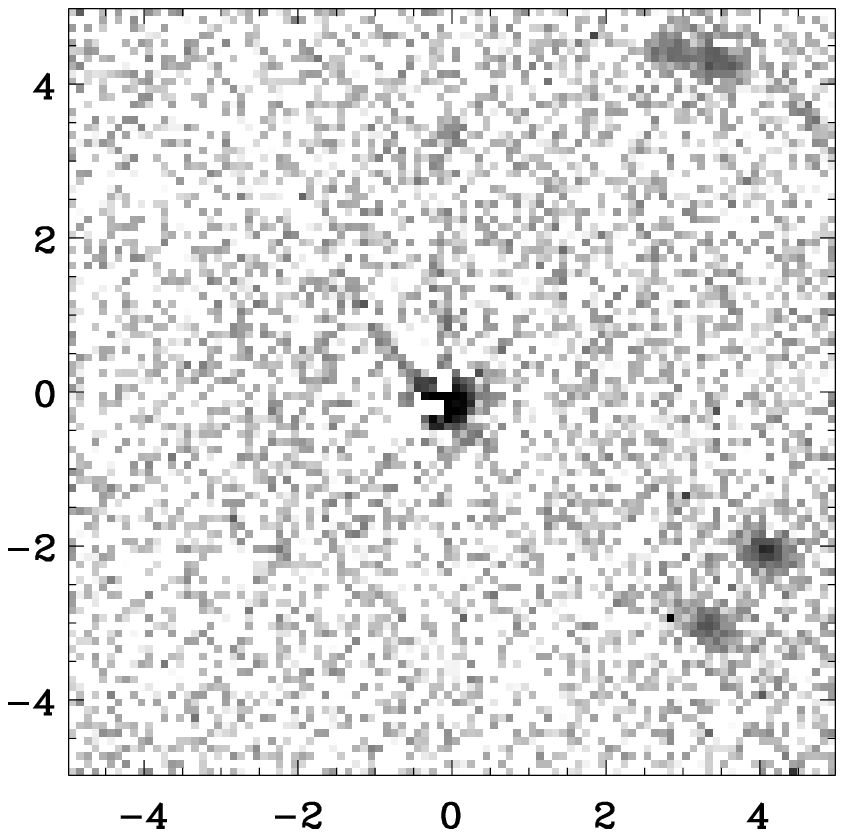}}
{\includegraphics[angle=270,origin=c,width=40mm]{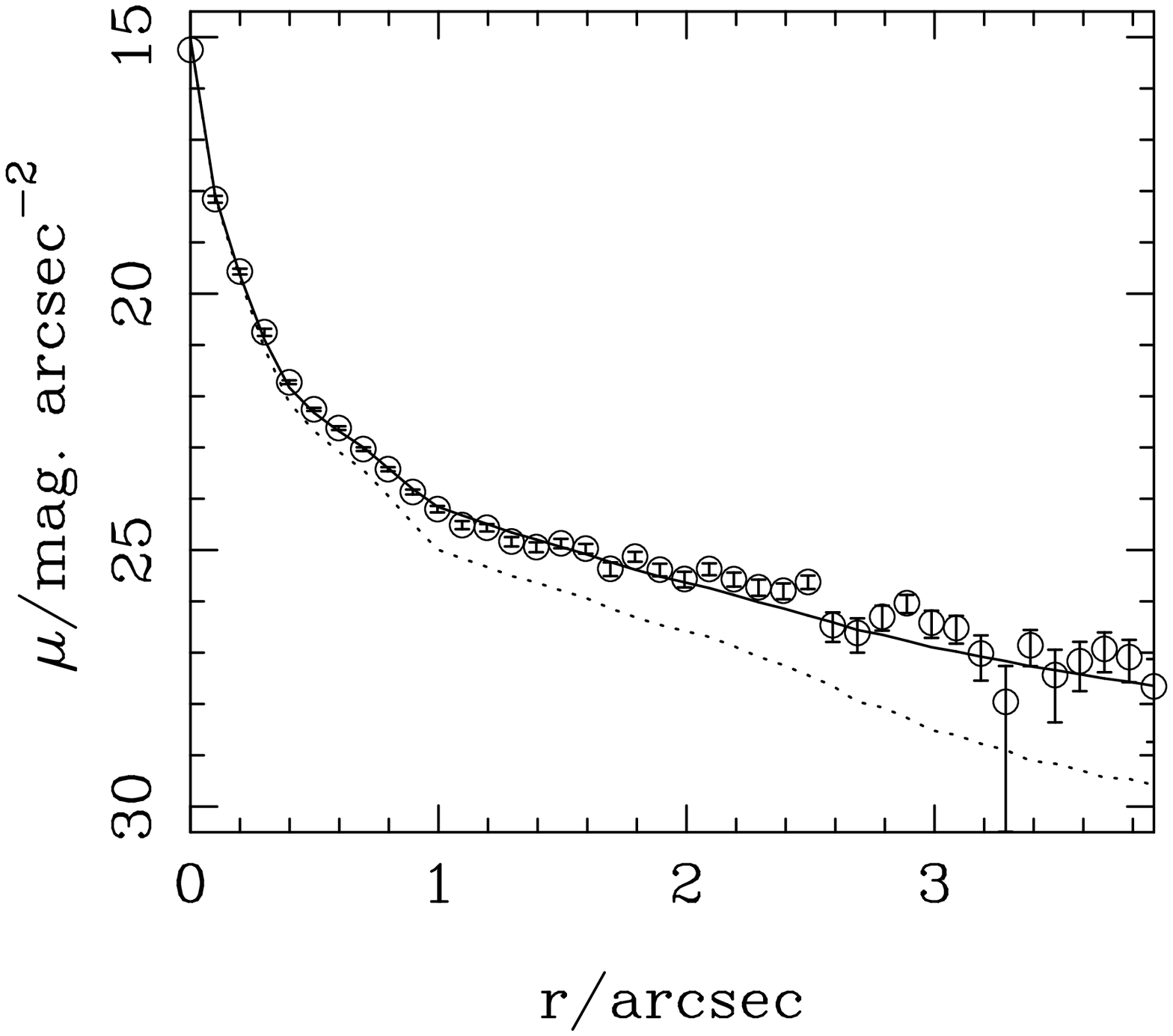}}
\caption{\label{fig-h107} The RLQ PKS0440--00 at $z=0.844$. A high-powered radio quasar, with a Gigahertz-peaked spectrum.
We note a number of very compact candidate companions in the field, but there are no obvious mergers or interactions. 
The object is more nuclear-dominated at $U$ as expected: $[L_\mathrm{nuc}/ L_\mathrm{host}]_V=1.4$;~$[L_\mathrm{nuc}/ L_\mathrm{host}]_U=9.2$}
\end{figure}

\begin{figure}
{\includegraphics[width=40mm]{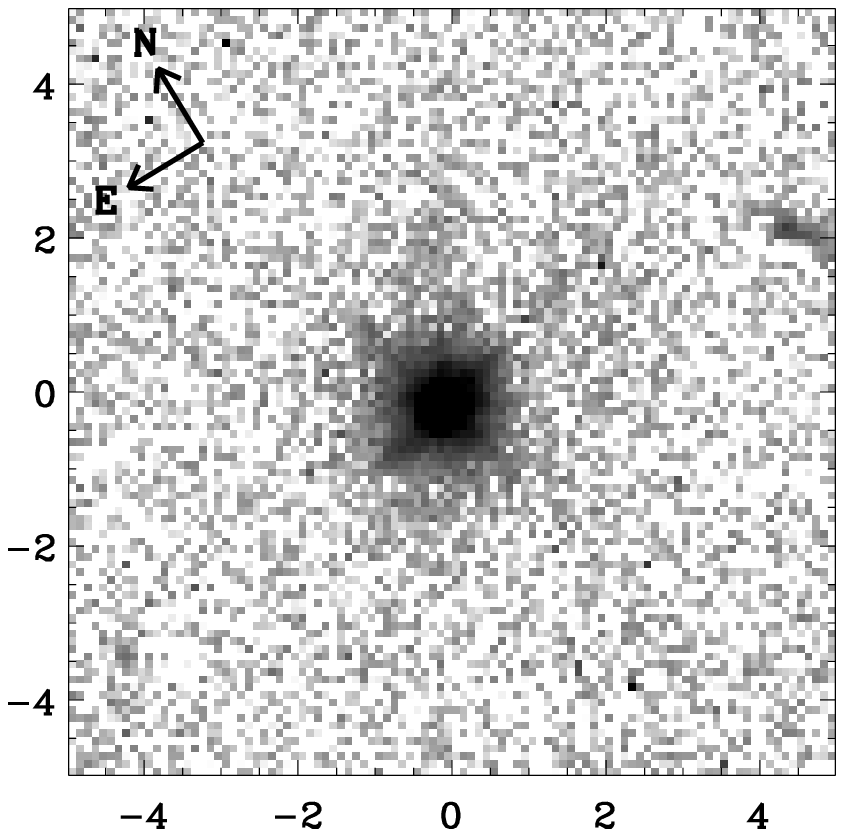}}
{\includegraphics[width=40mm]{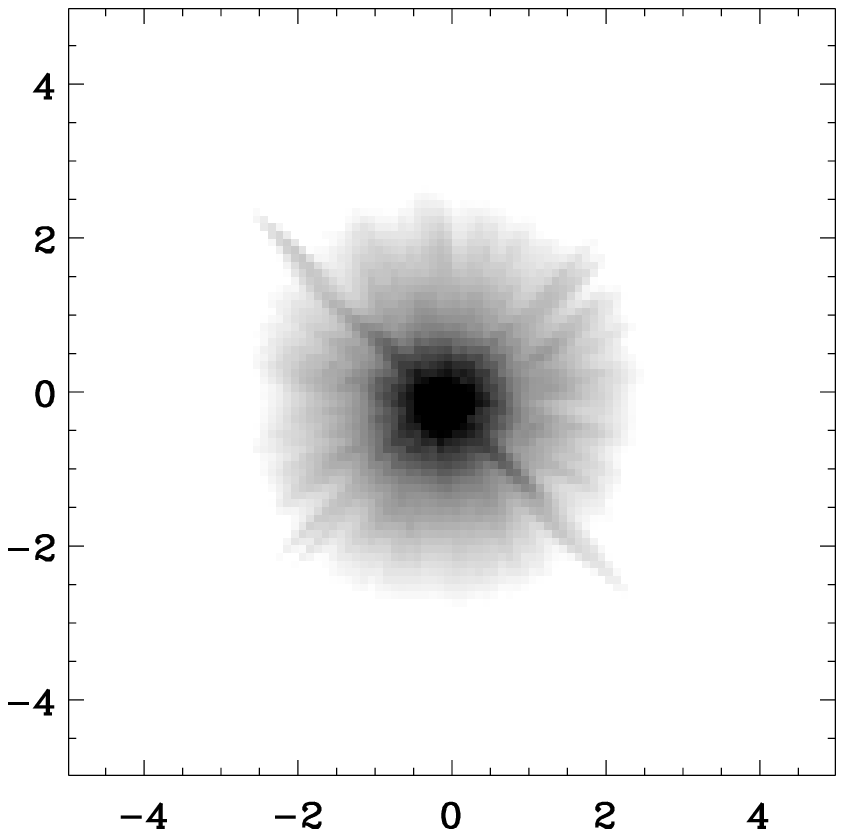}}

{\includegraphics[width=40mm]{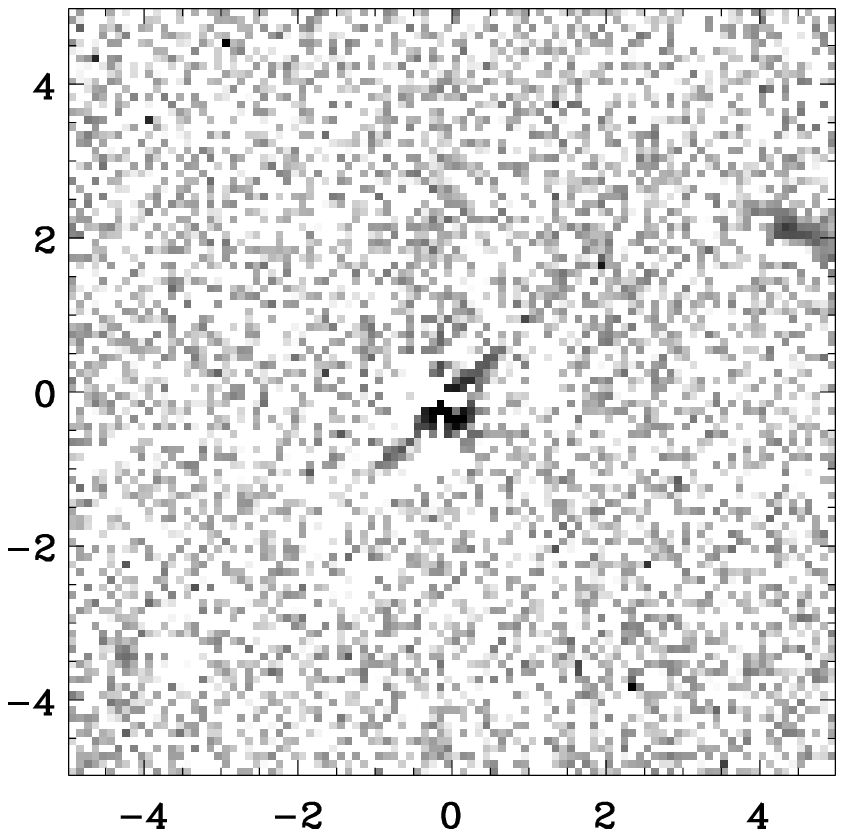}}
{\includegraphics[angle=270,origin=c,width=40mm]{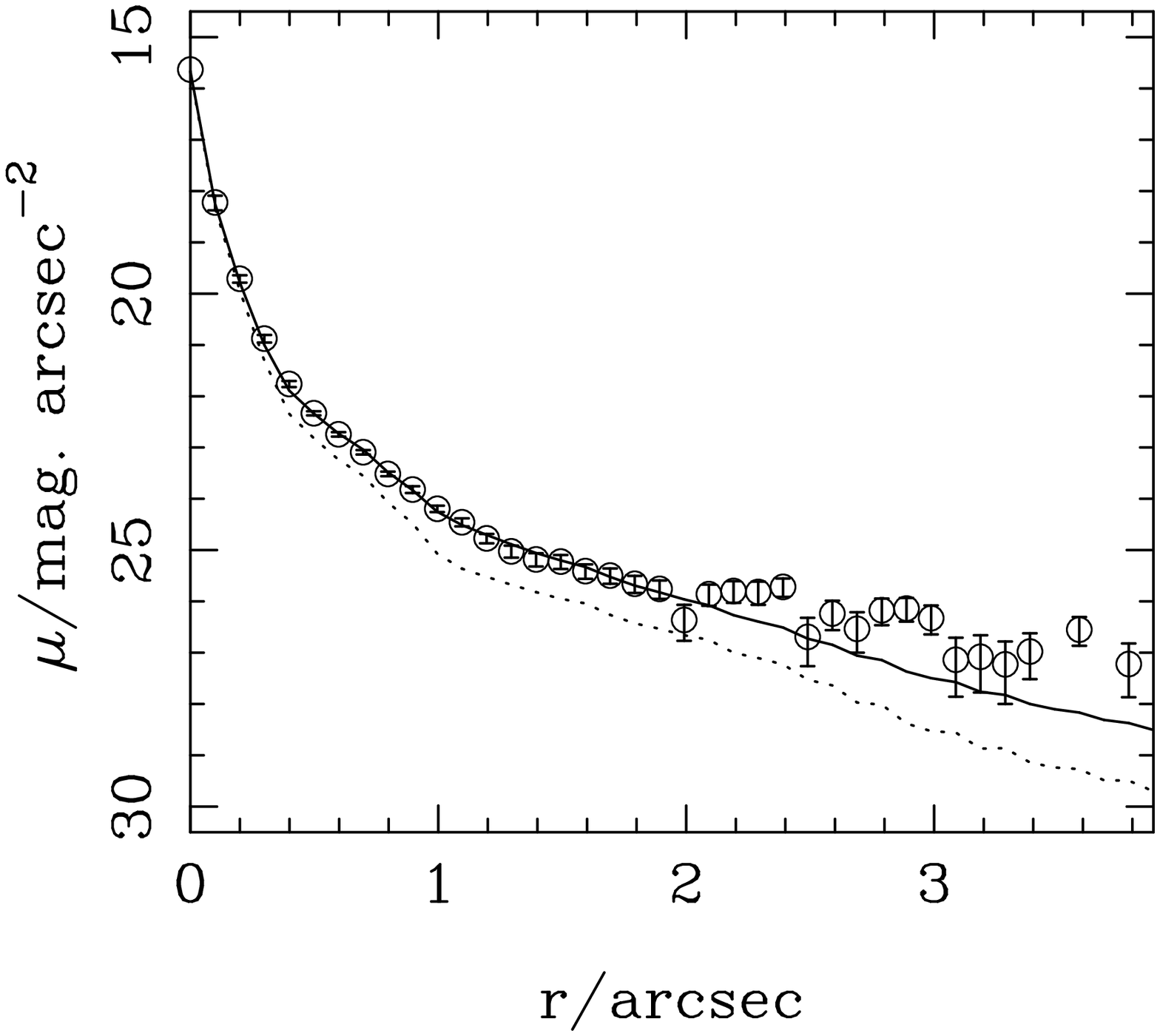}}
\caption{\label{fig-h1010} The RLQ PKS0938+18 at $z=0.940$.
We see the expected increase in nuclear domination by a factor of $\sim3$ from $V$ to $U$: 
$[L_\mathrm{nuc}/ L_\mathrm{host}]_V= 2.3$;~$[L_\mathrm{nuc}/ L_\mathrm{host}]_U= 7.4$.
There is a loose collection of small candidate companions out to a radius of around 20\arcsec. In
addition, there is some very low surface brightness ``fuzz'', 3\arcsec\ NW of the nucleus, which is left 
over in the modelling residuals and may indicate a merger.}
\end{figure}

\begin{figure}
{\includegraphics[width=40mm]{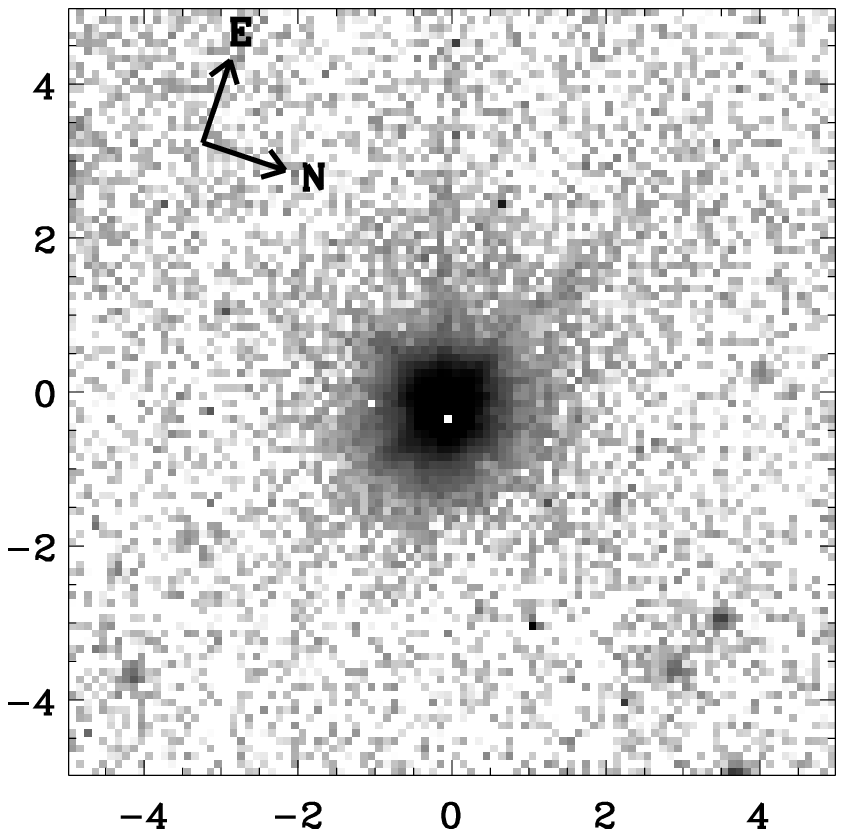}}
{\includegraphics[width=40mm]{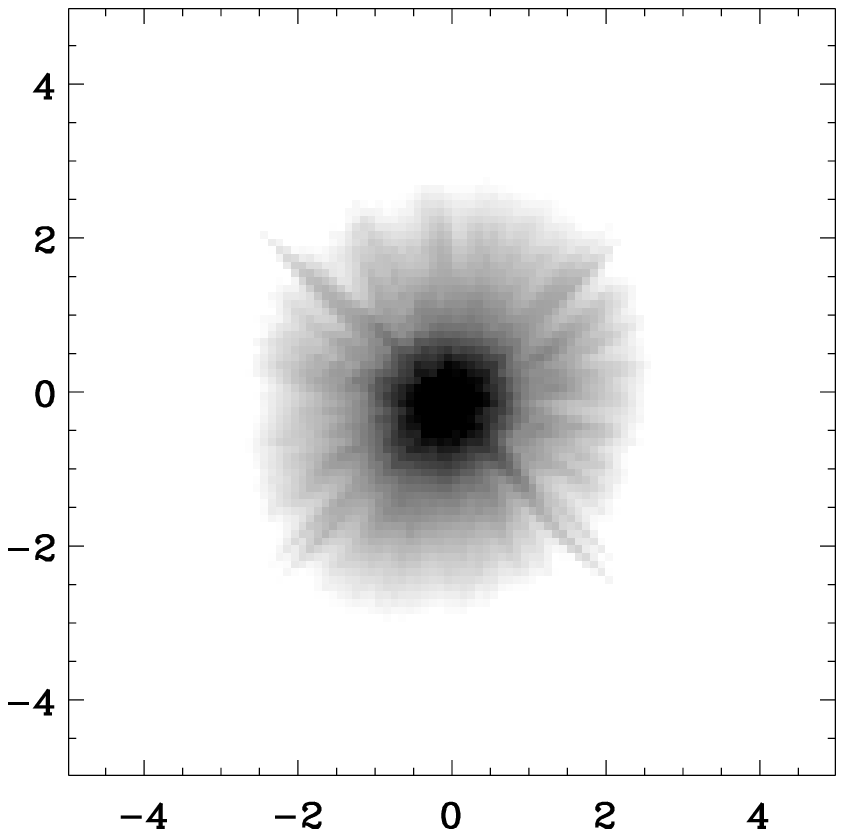}}

{\includegraphics[width=40mm]{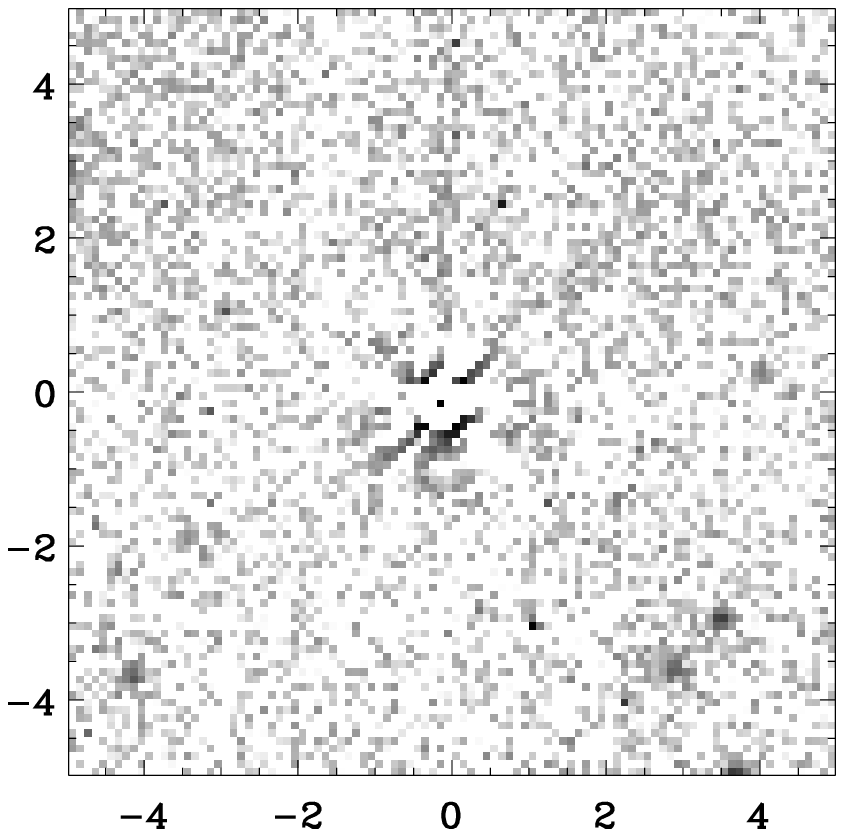}}
{\includegraphics[angle=270,origin=c,width=40mm]{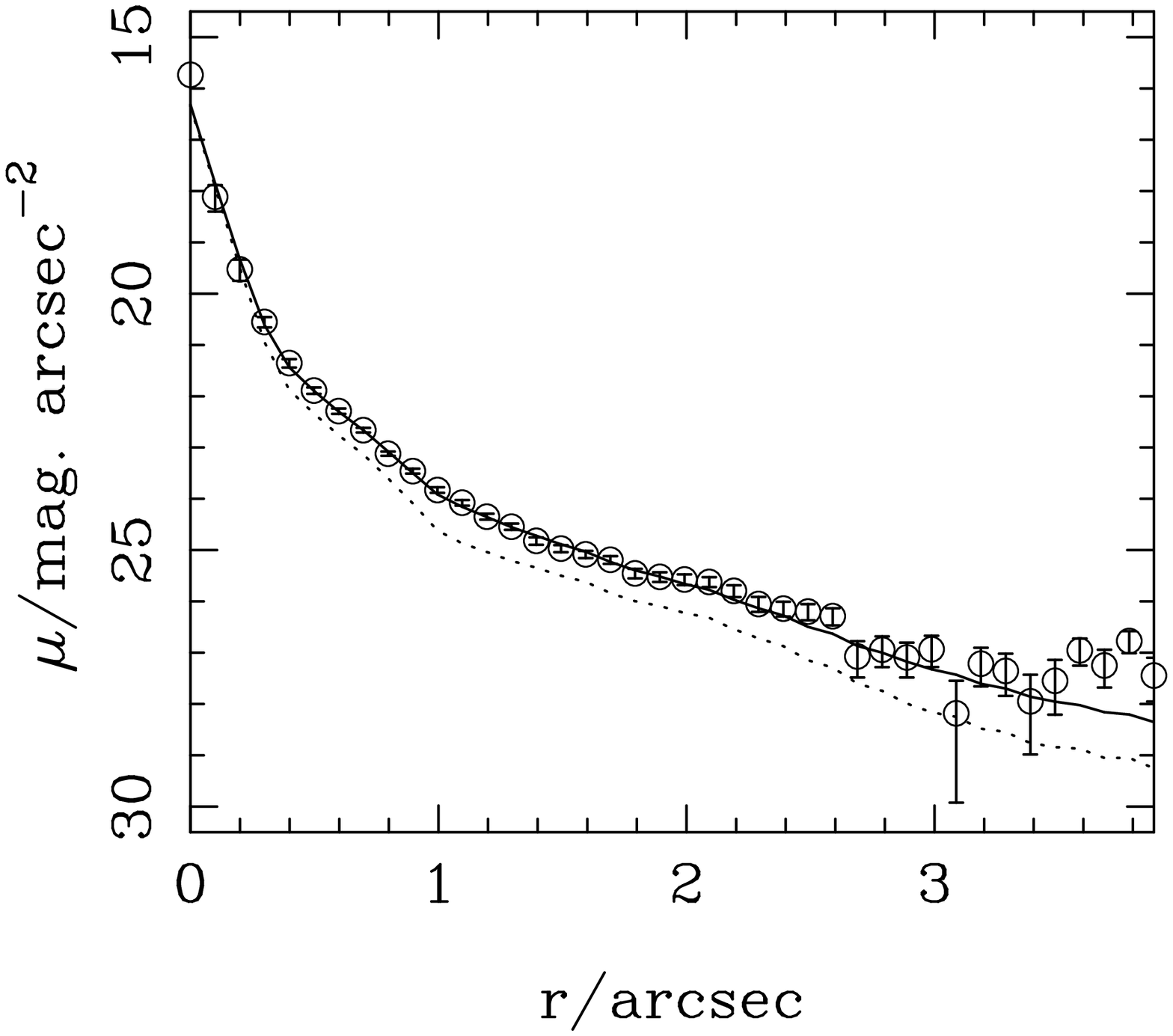}}
\caption{\label{fig-h109} The RLQ 3C422 at $z=0.942$. A typical compact steep-spectrum object.
There is one obvious, comparably large companion 6\arcsec\ to the south, as well as a number 
of smaller objects dotted around the field. 
The quasar is nuclear-dominated at $U$: $[L_\mathrm{nuc}/ L_\mathrm{host}]_V=1.4$;~$[L_\mathrm{nuc}/ L_\mathrm{host}]_U=6.2$.
Some PSF-related circumnuclear flux is visible in the residuals}
\end{figure}

\begin{figure}
{\includegraphics[width=40mm]{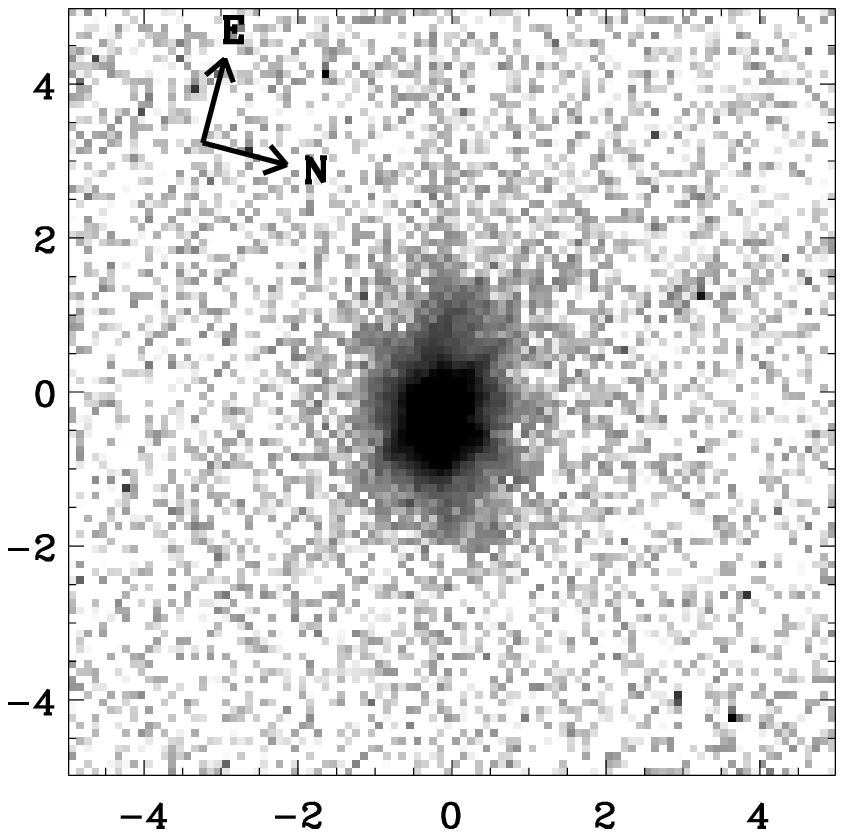}}
{\includegraphics[width=40mm]{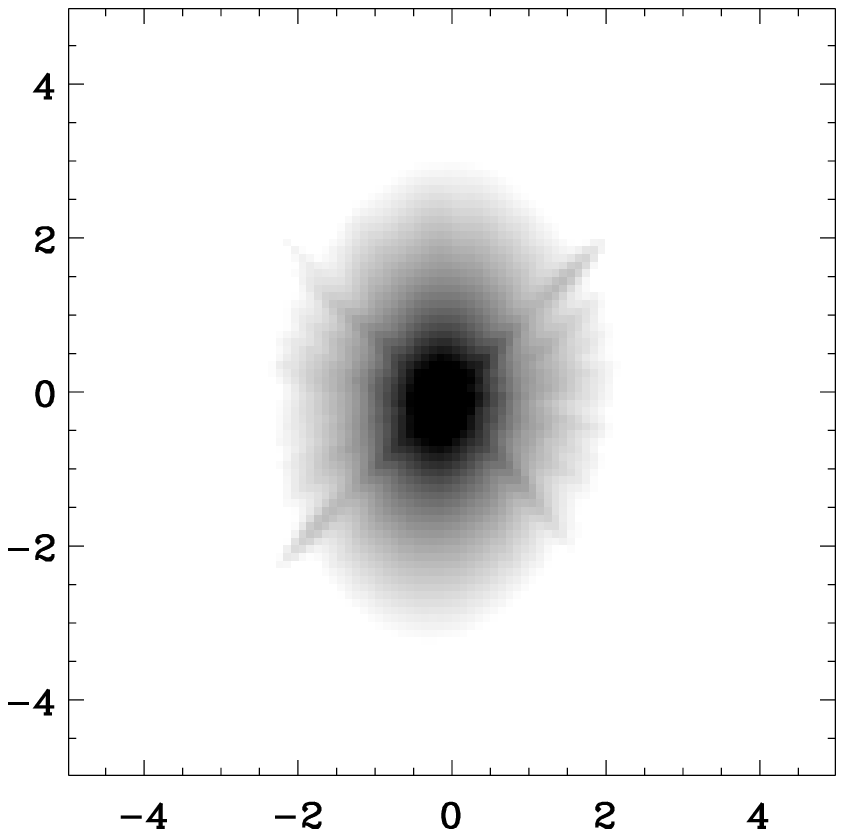}}

{\includegraphics[width=40mm]{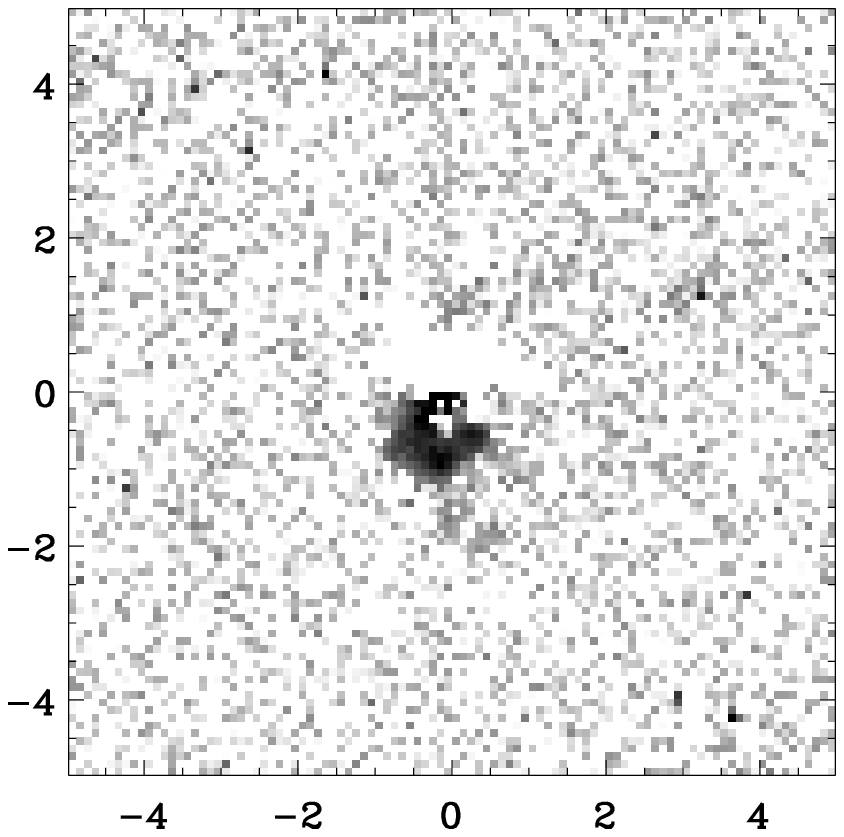}}
{\includegraphics[angle=270,origin=c,width=40mm]{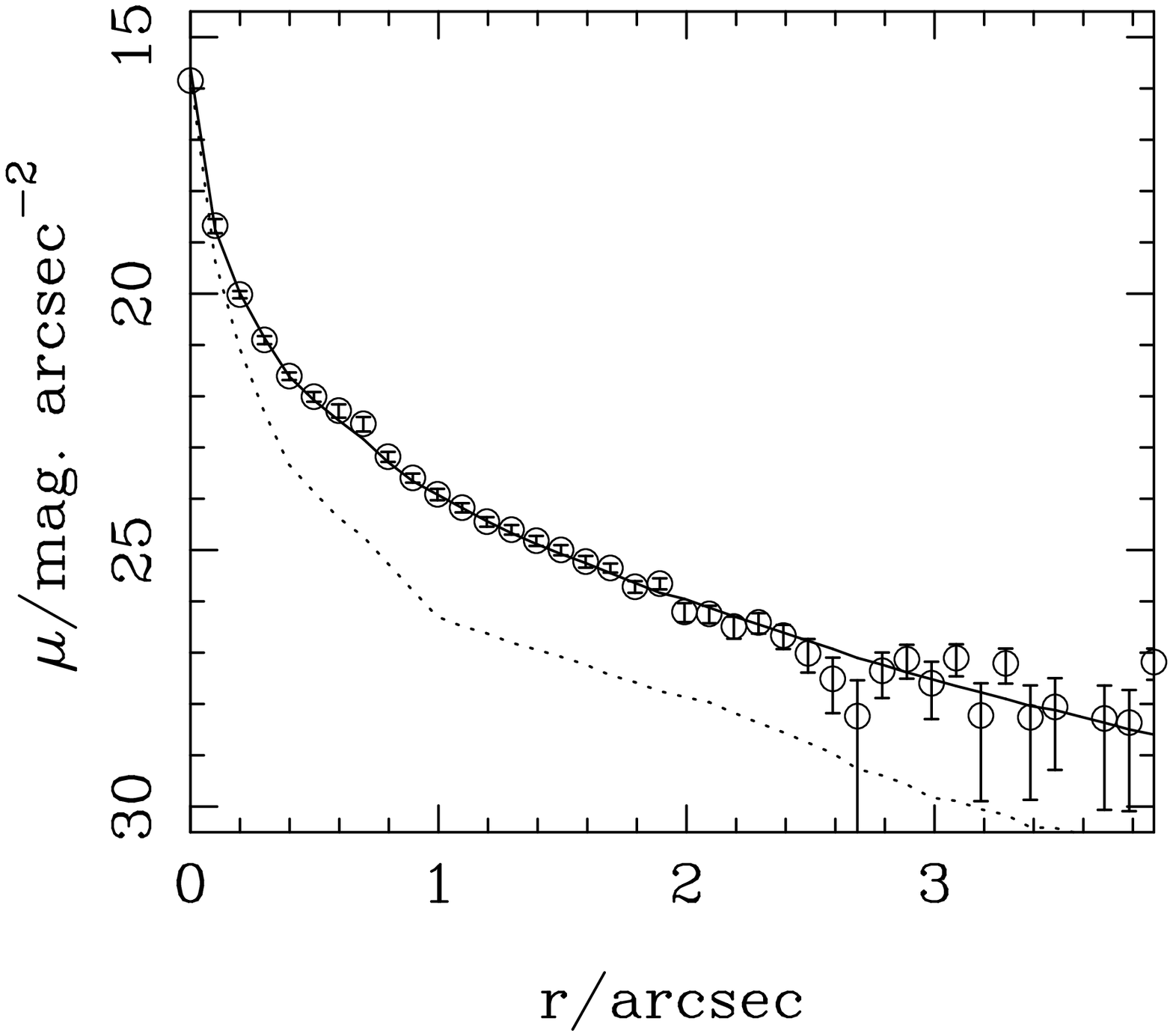}}
\caption{\label{fig-h108} The RLQ MC2112+172 at $z=0.878$. An elongated object in our WFPC2 image, suggesting a significant ongoing merger event, possibly with a small object visible to the NNE. There are a number of small candidate companions in the field. The residuals show significant (apparently non-PSF) excess flux west of the nucleus. The host galaxy is prominent, with a slightly stronger nuclear component in $U$: $[L_\mathrm{nuc}/ L_\mathrm{host}]_V=0.5$;~$[L_\mathrm{nuc}/ L_\mathrm{host}]_U=1.4$.}
\end{figure}

\begin{figure}
{\includegraphics[width=40mm]{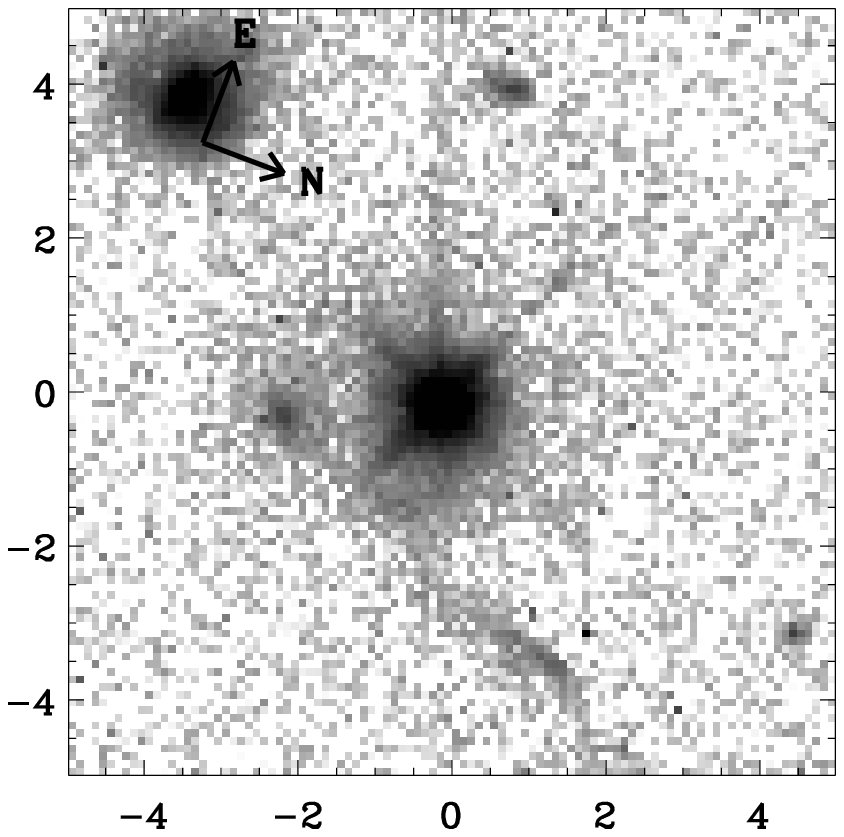}}
{\includegraphics[width=40mm]{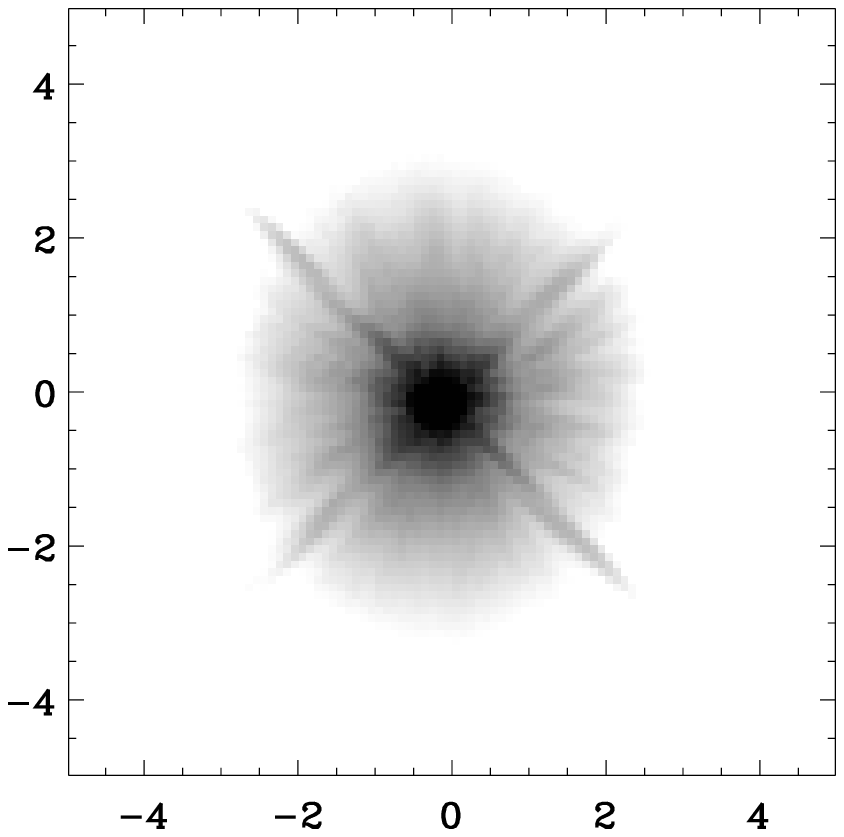}}

{\includegraphics[width=40mm]{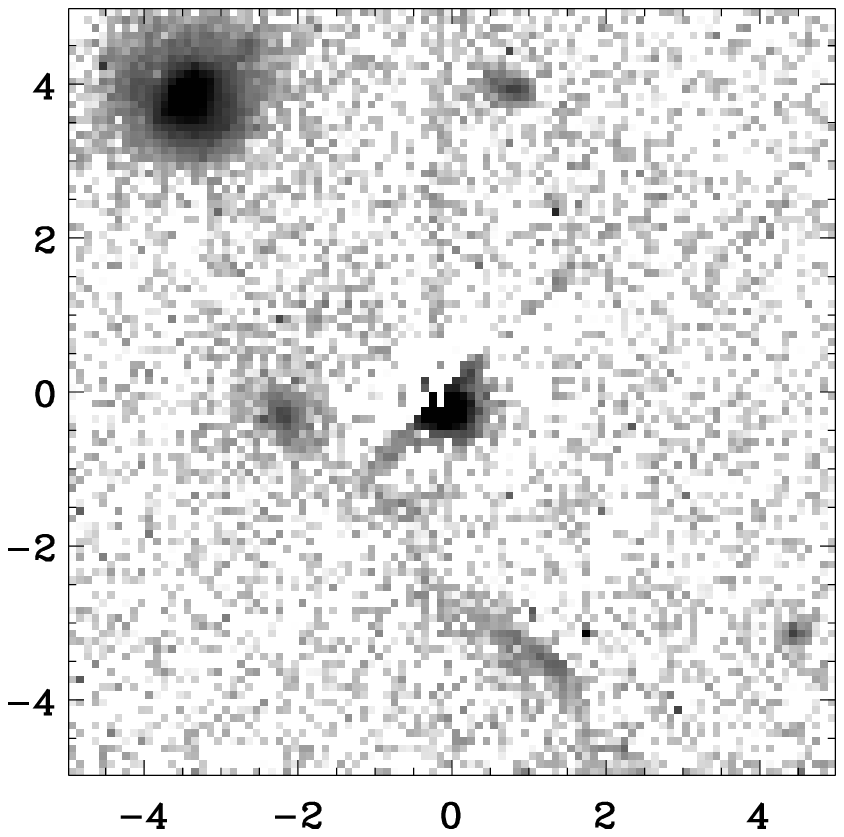}}
{\includegraphics[angle=270,origin=c,width=40mm]{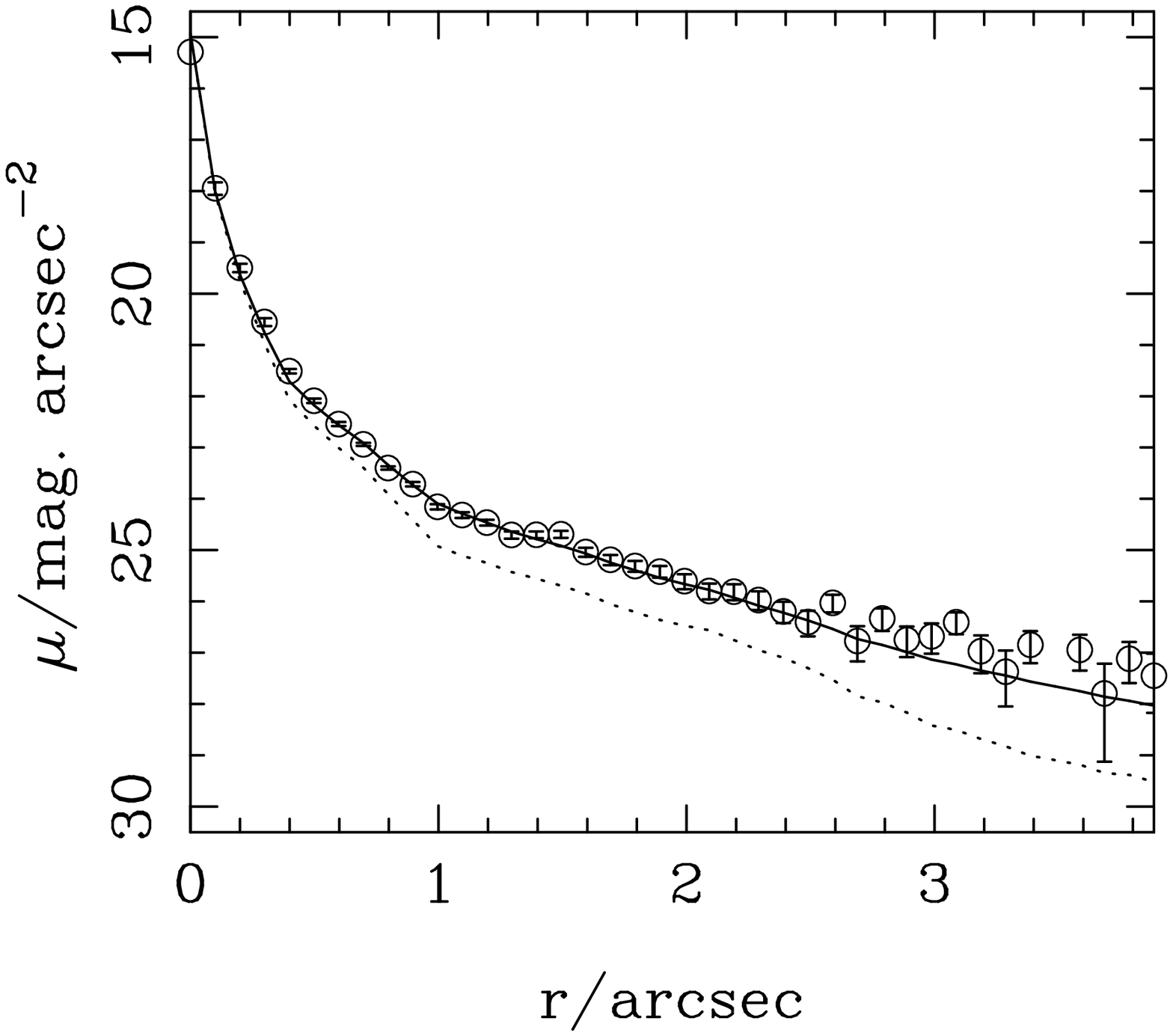}}
\caption{\label{fig-h1011} The RLQ 4C02.54 at $z=0.976$. In our image there is clear evidence for a merging companion 3\arcsec\ to the northwest, in addition to a number of other galaxies in close proximity to the quasar. If the large object 5\arcsec\ to the southeast is found to be at the same redshift as the quasar, then it too is a large elliptical galaxy. We find a higher level of nuclear dominance in rest-frame $U$: $[L_\mathrm{nuc}/ L_\mathrm{host}]_V=4.8$;~$[L_\mathrm{nuc}/ L_\mathrm{host}]_U=10.2$. There are significant residuals to the immediate north of the nucleus.}
\end{figure}


\begin{figure}
{\includegraphics[width=40mm]{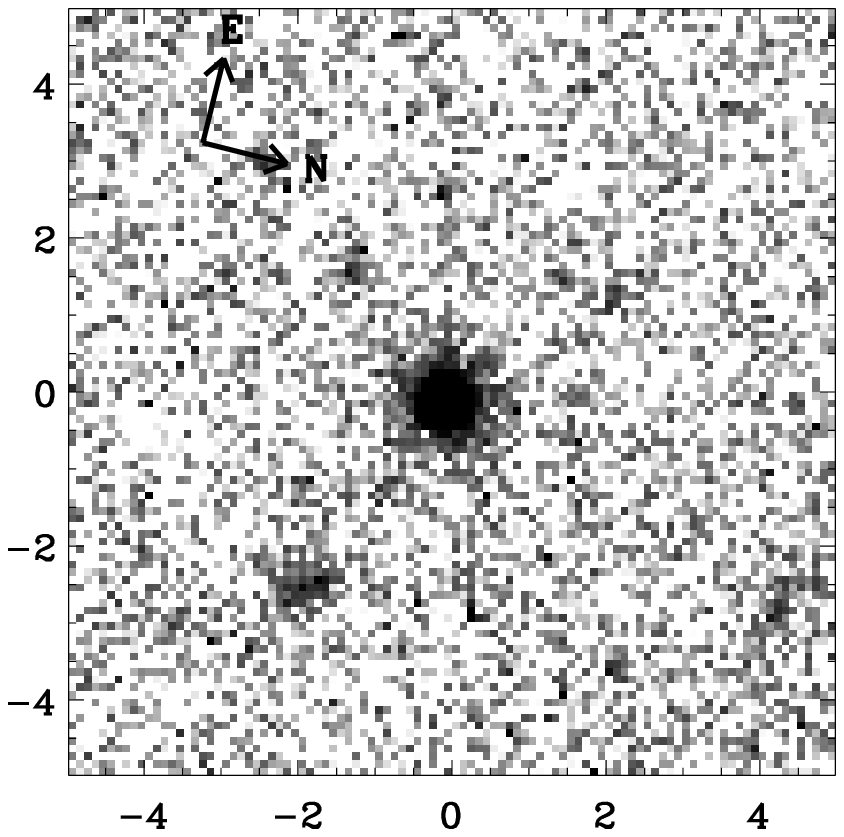}}
{\includegraphics[width=40mm]{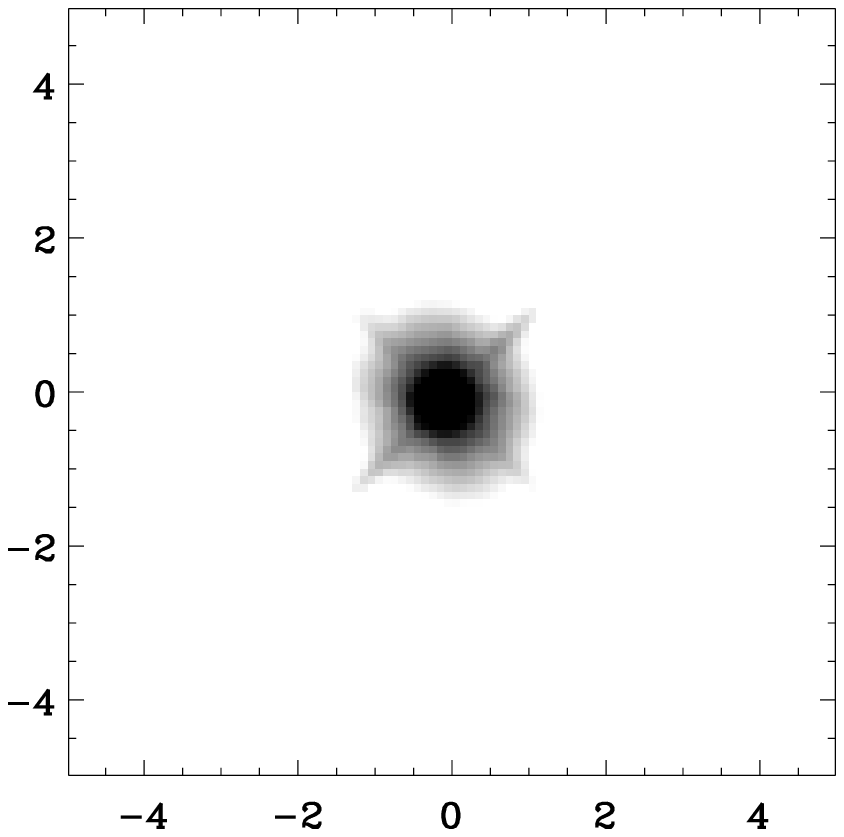}}

{\includegraphics[width=40mm]{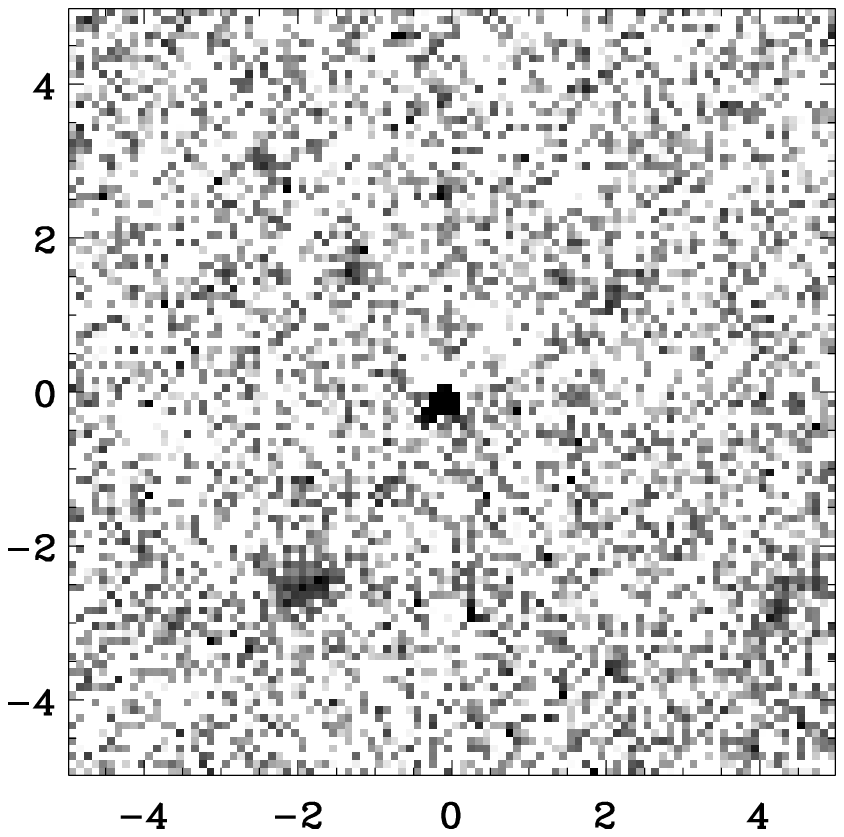}}
{\includegraphics[angle=270,origin=c,width=40mm]{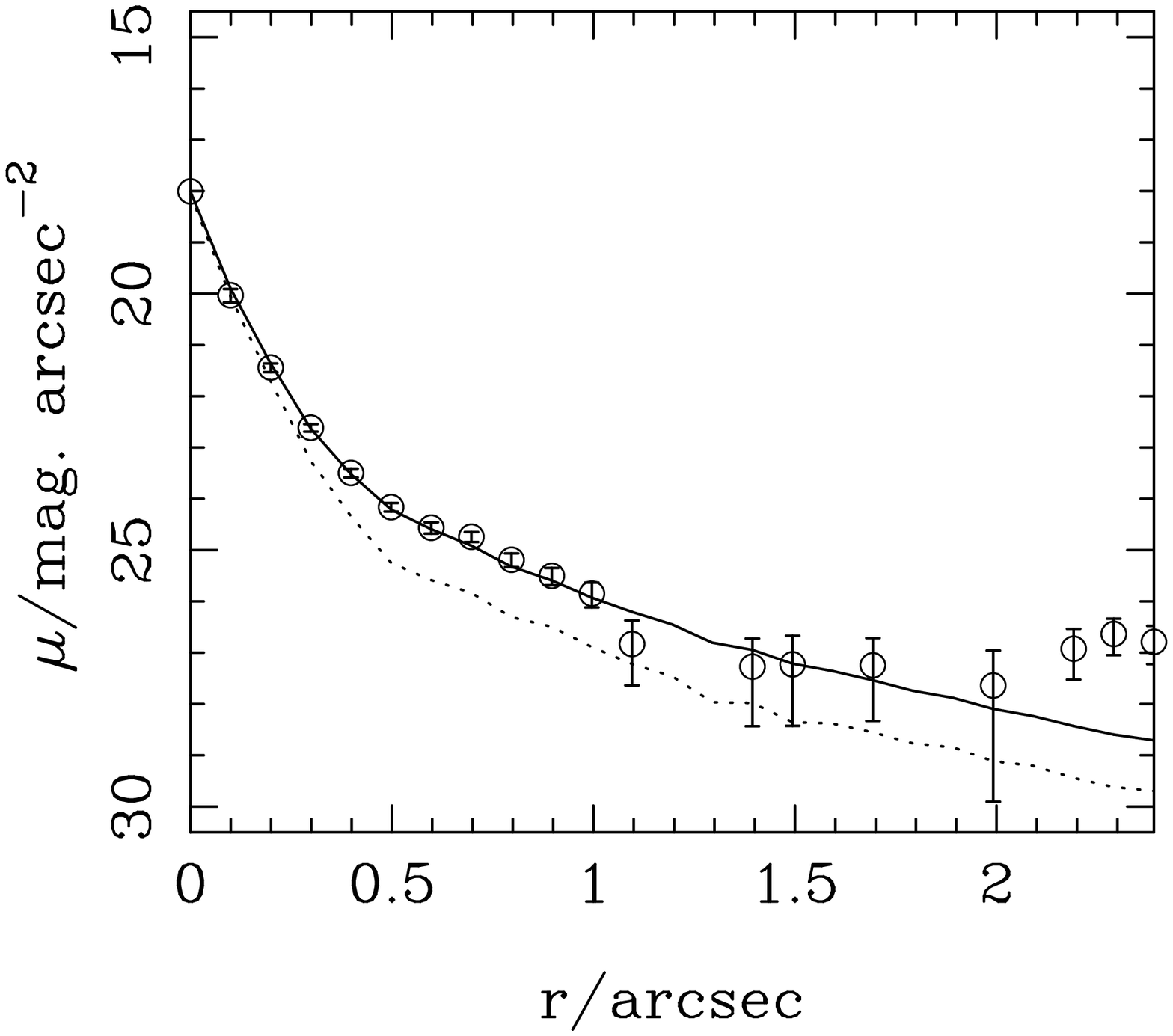}}
\caption{\label{fig-h1013} The RQQ SGP2:36 at $z=1.773$. K01 found a relatively weak nucleus in $V$: 
$[L_\mathrm{nuc}/ L_\mathrm{host}]_V= 0.8$. In $U$ we again find a relatively weak nucleus, with $[L_\mathrm{nuc}/ L_\mathrm{host}]_U= 2.5$. A large number of small objects are visible in the field.}
\end{figure}

\begin{figure}
{\includegraphics[width=40mm]{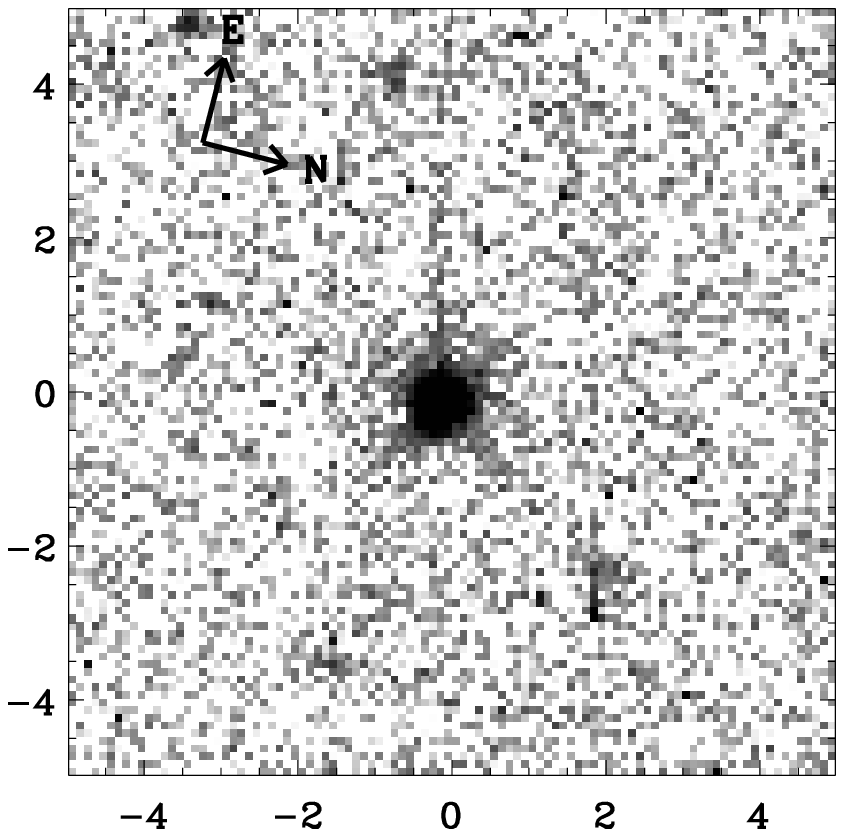}}
{\includegraphics[width=40mm]{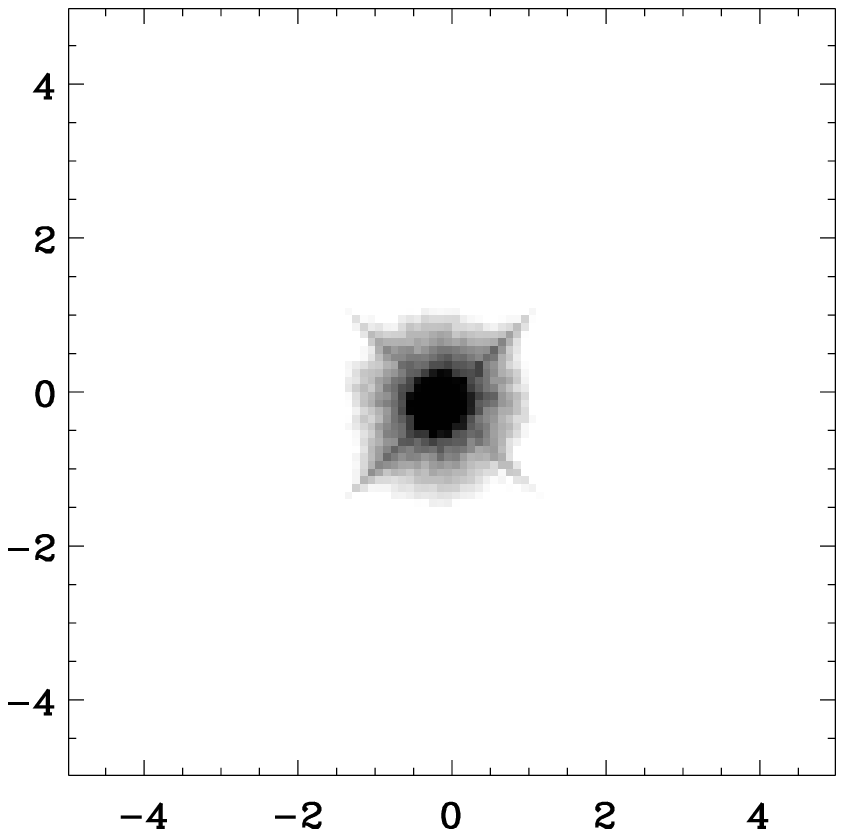}}

{\includegraphics[width=40mm]{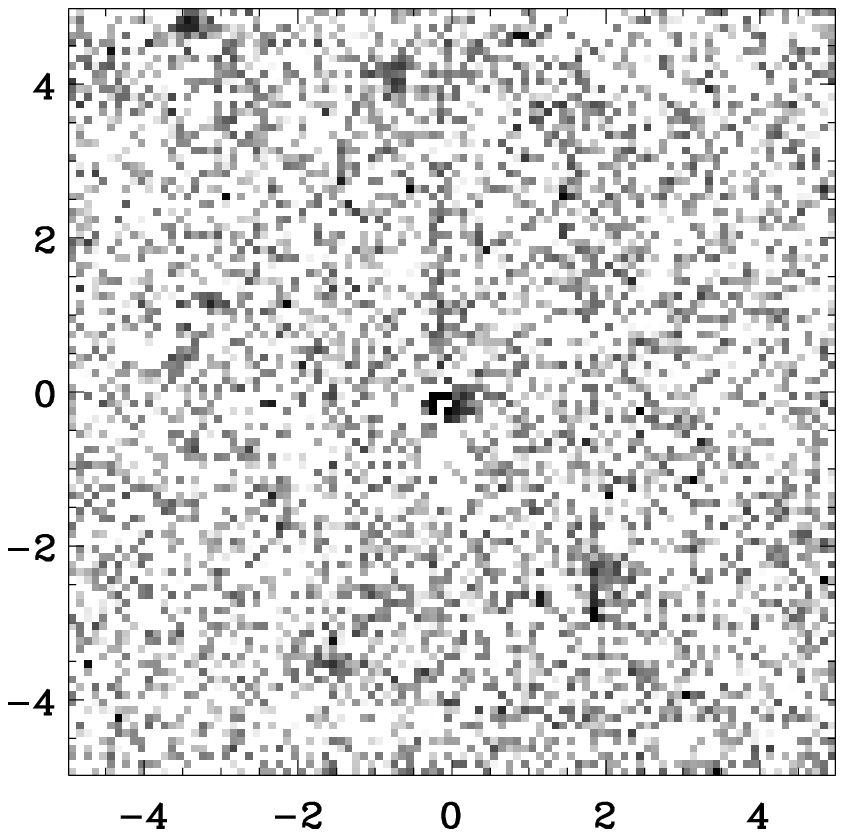}}
{\includegraphics[angle=270,origin=c,width=40mm]{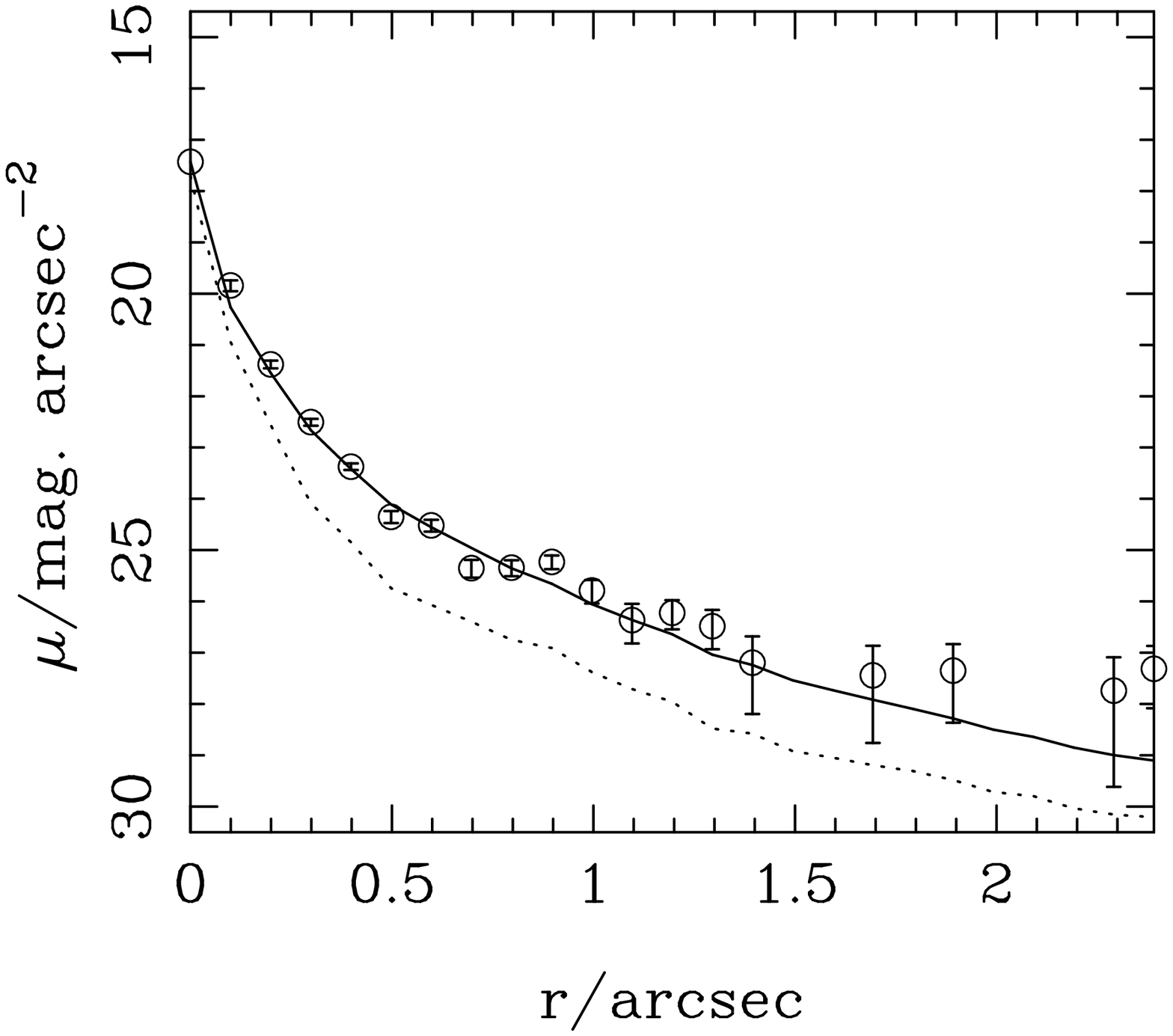}}
\caption{\label{fig-h1014} The RQQ SGP2:25 at $z=1.869$.
This quasar is quite close on the sky to the quasar SGP2:27 at $z=1.93$. K01 found a moderate nuclear component; $[L_\mathrm{nuc}/ L_\mathrm{host}]_V=1.3$. We find a much stronger nuclear component; $[L_\mathrm{nuc}/ L_\mathrm{host}]_U=8.3$. 
A large number of faint objects are observed in the field.}
\end{figure}

\begin{figure}
{\includegraphics[width=40mm]{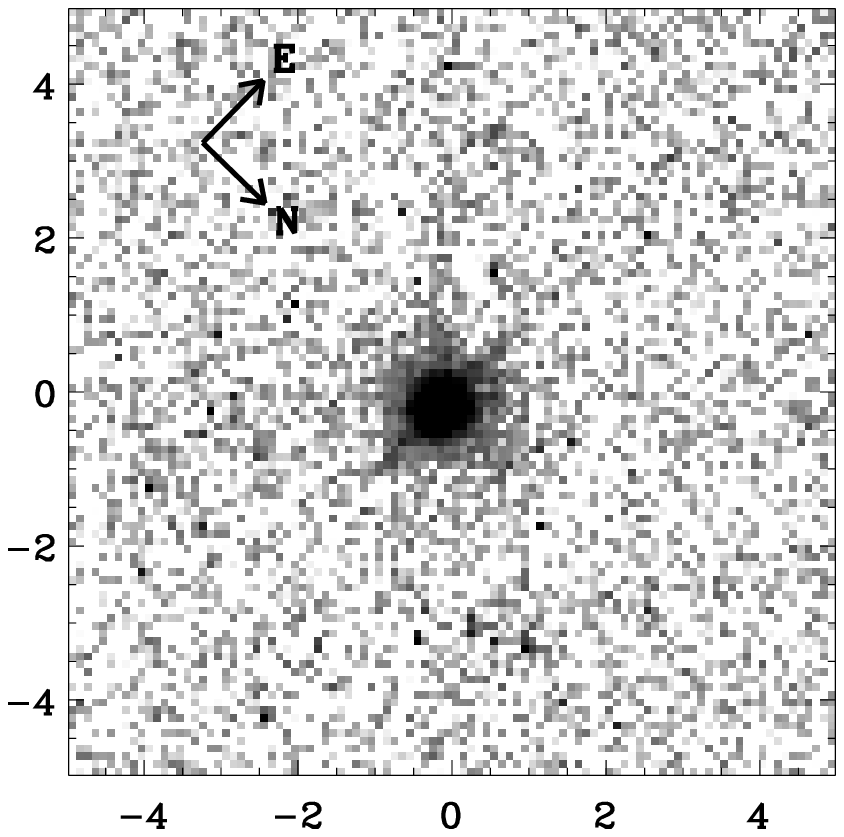}}
{\includegraphics[width=40mm]{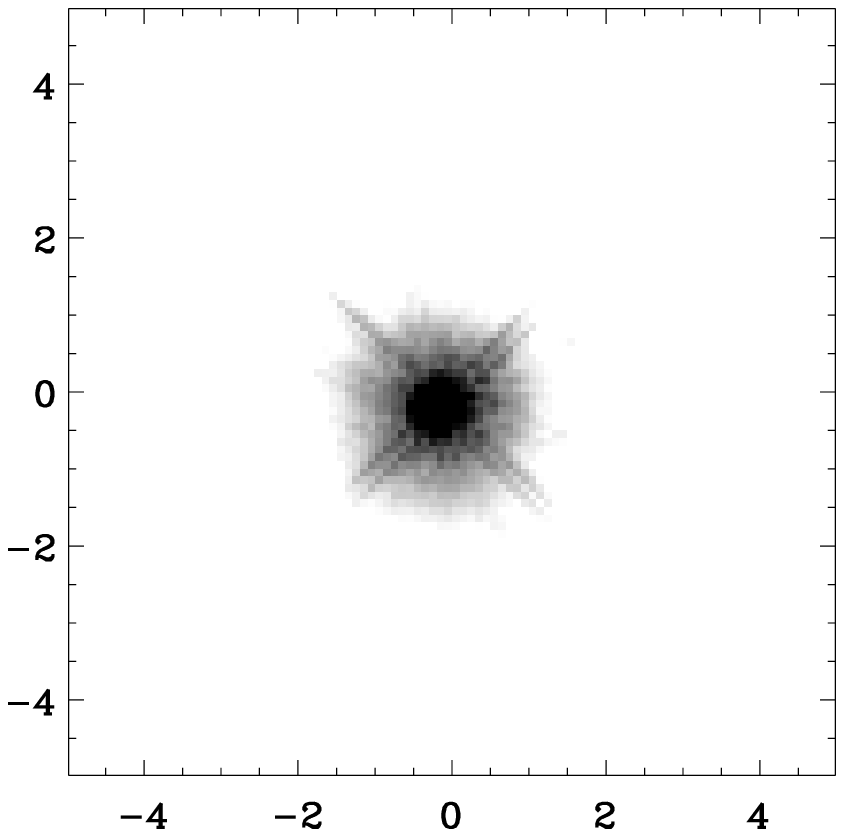}}

{\includegraphics[width=40mm]{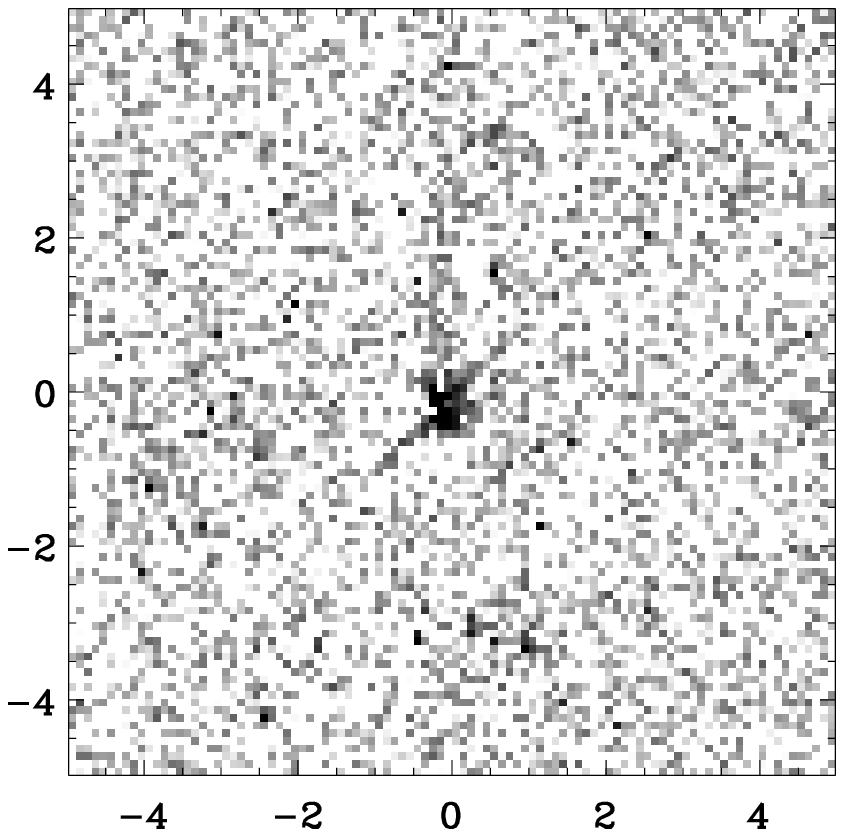}}
{\includegraphics[angle=270,origin=c,width=40mm]{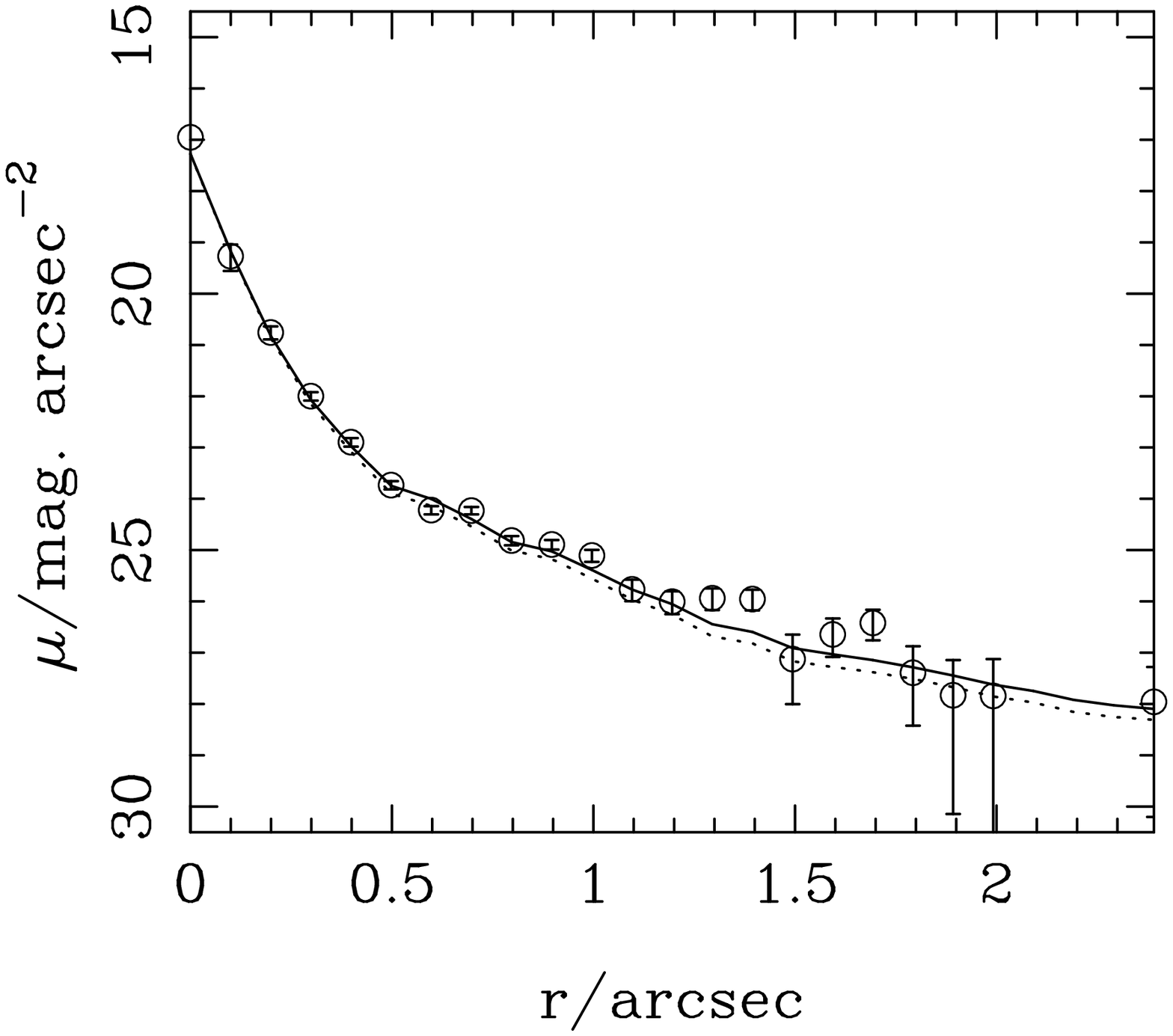}}
\caption{\label{fig-h1016} The RQQ SGP2:11 at $z=1.976$. 
We were unable to constrain the size or surface brightness of the host, but the total host-galaxy luminosity is well constrained. 
The quasar is nuclear-dominated at both wavelengths: $[L_\mathrm{nuc}/ L_\mathrm{host}]_V= 4.7$;~$L_\mathrm{nuc}/ L_\mathrm{host}]_U \sim 12.3$.
There are some circumnuclear residuals, but these are compact.}
\end{figure}

\begin{figure}
{\includegraphics[width=40mm]{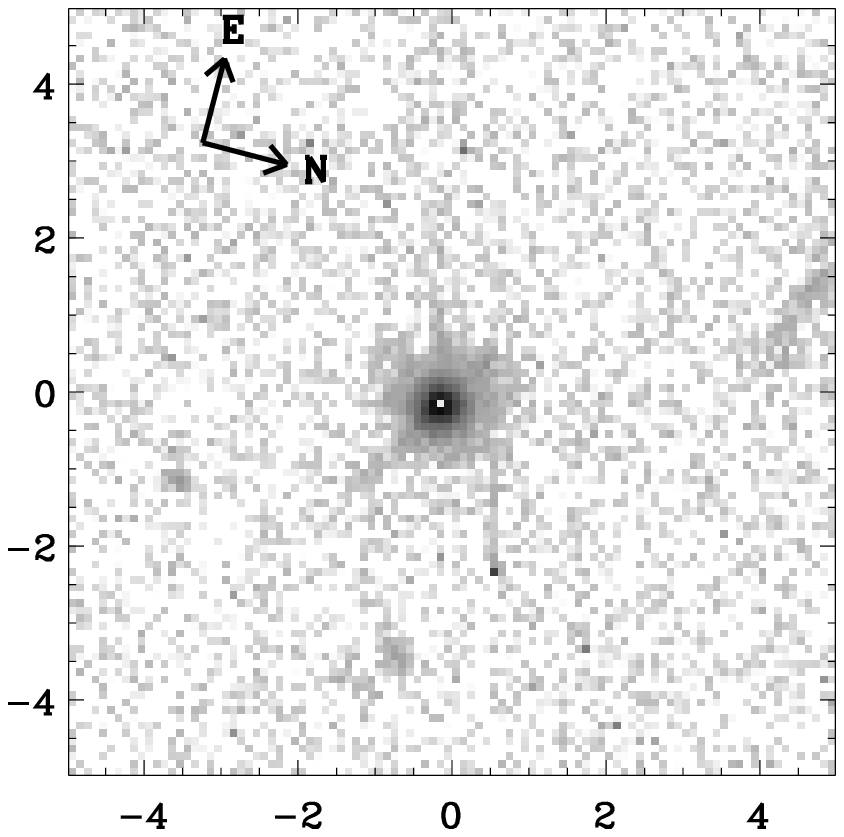}}
{\includegraphics[width=40mm]{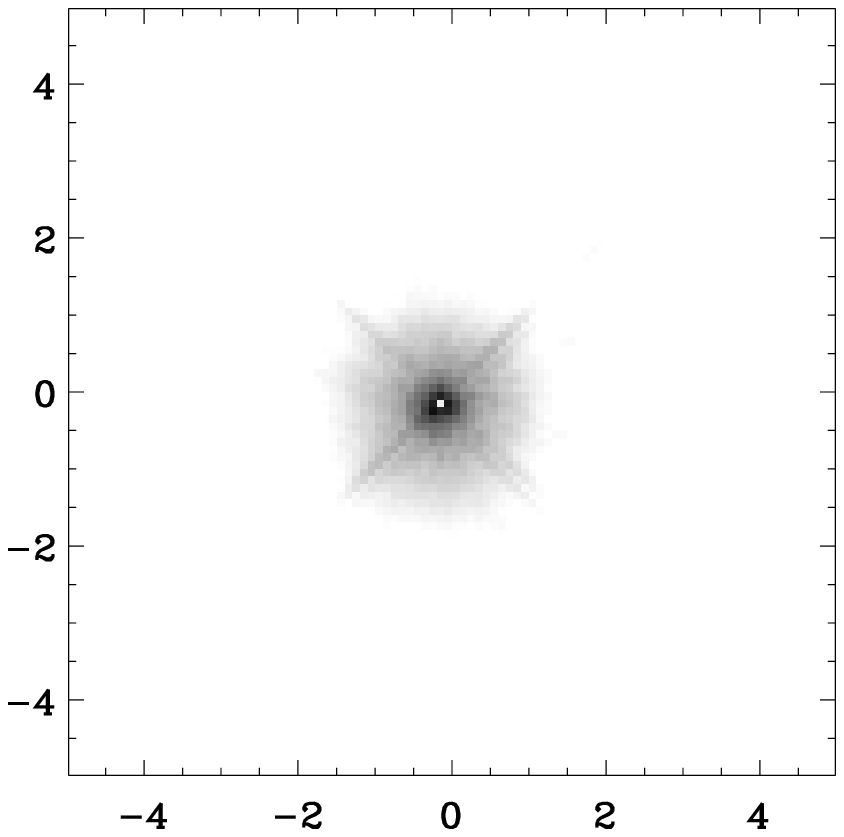}}

{\includegraphics[width=40mm]{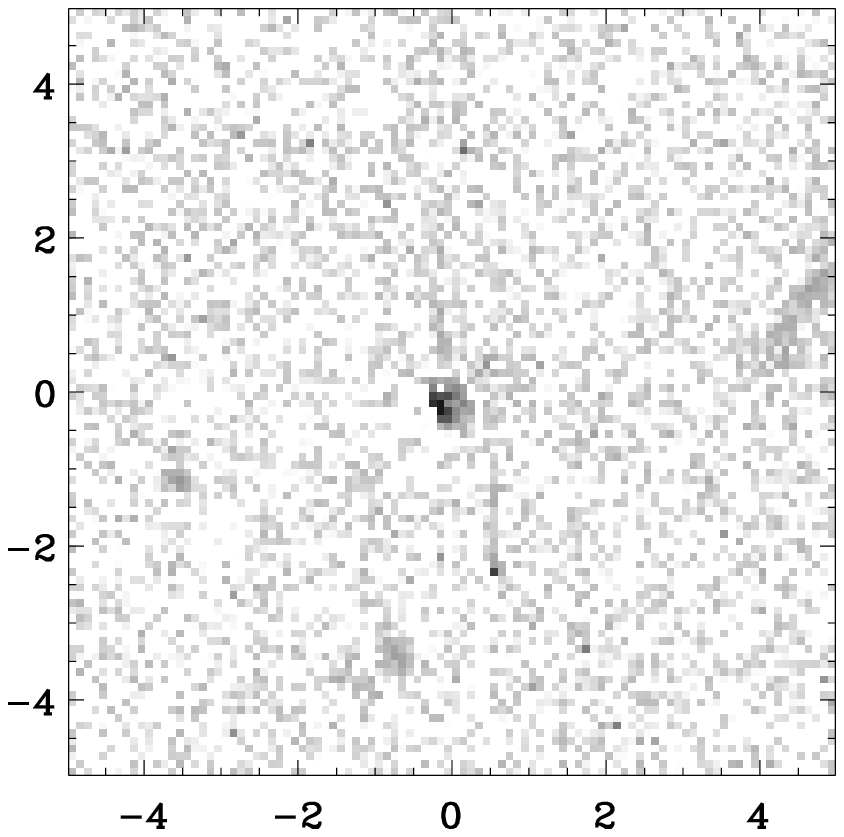}}
{\includegraphics[angle=270,origin=c,width=40mm]{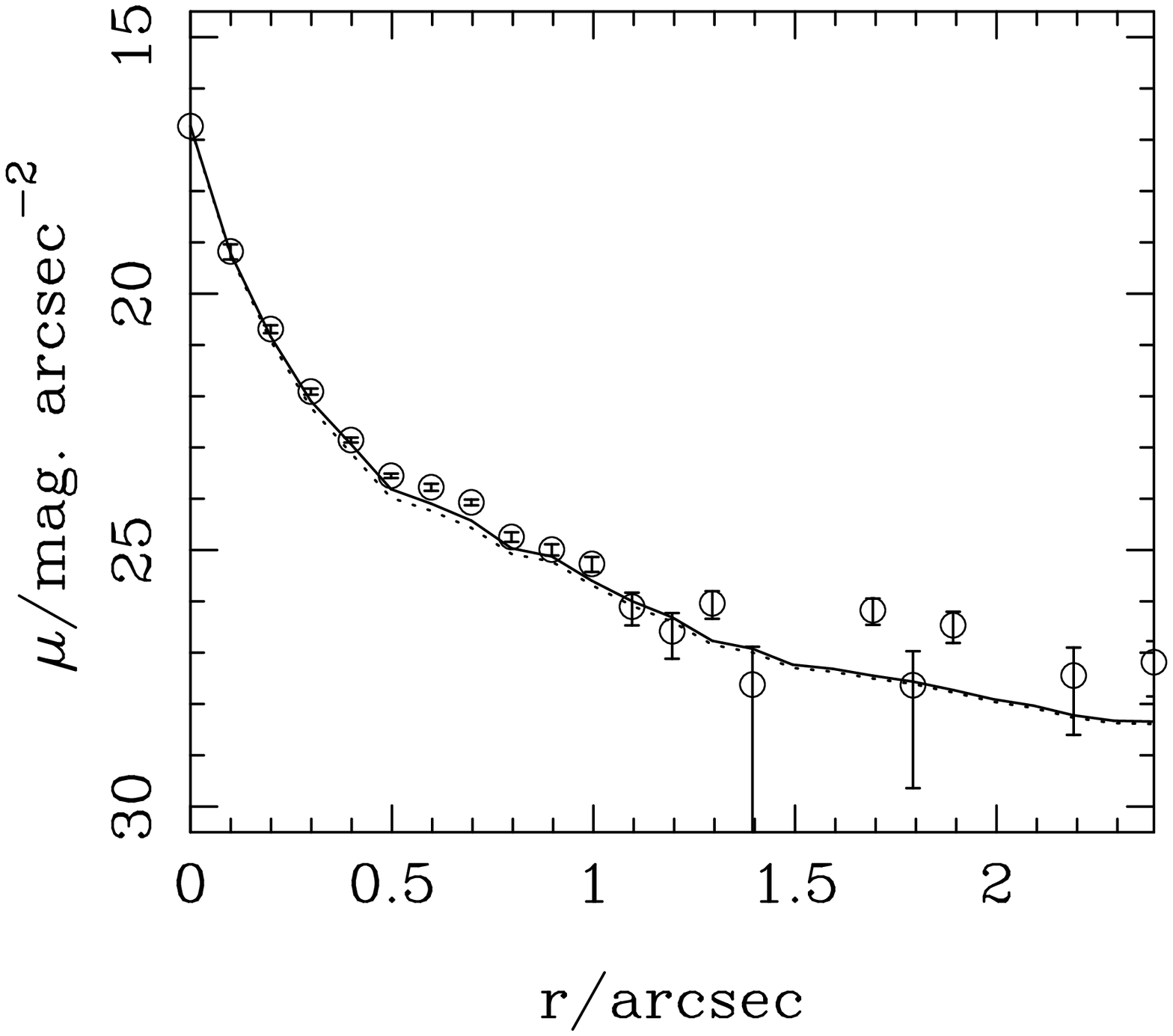}}
\caption{\label{fig-h1015} The RQQ SGP3:39 at $z=1.959$. SGP3:39 lies close on the sky to the lower redshift quasar SGP3:35 ($z=1.498$). The quasar has a fairly weak nuclear component in $V$, but much stronger in $U$: $[L_\mathrm{nuc}/ L_\mathrm{host}]_V= 1.2$;~$[L_\mathrm{nuc}/ L_\mathrm{host}]_U=5.7$.}
\end{figure}

\begin{figure}
{\includegraphics[width=40mm]{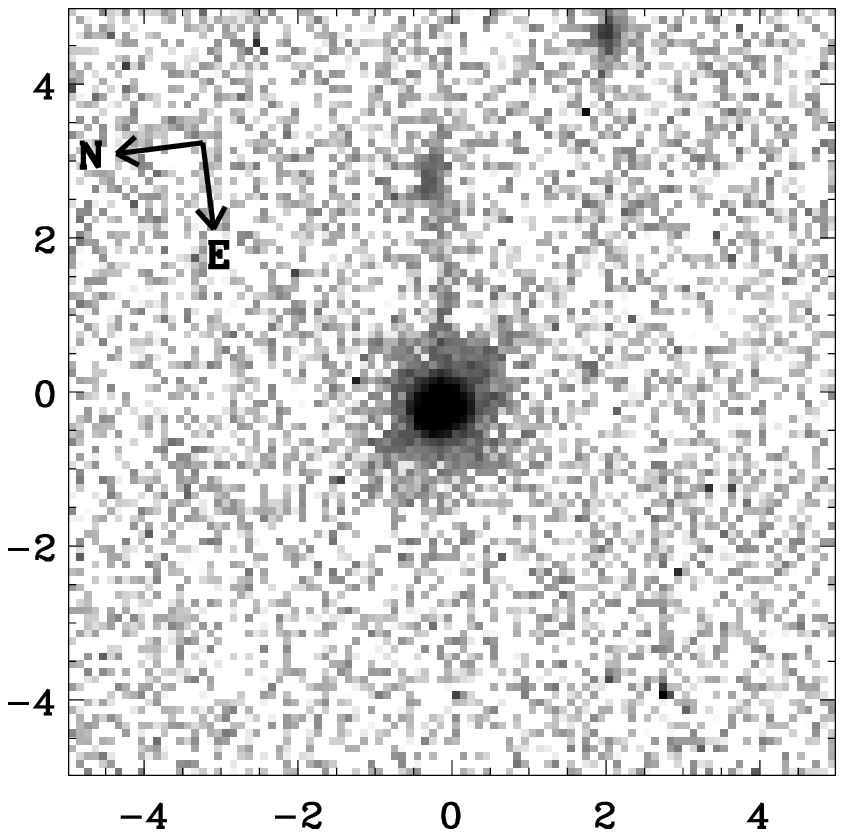}}
{\includegraphics[width=40mm]{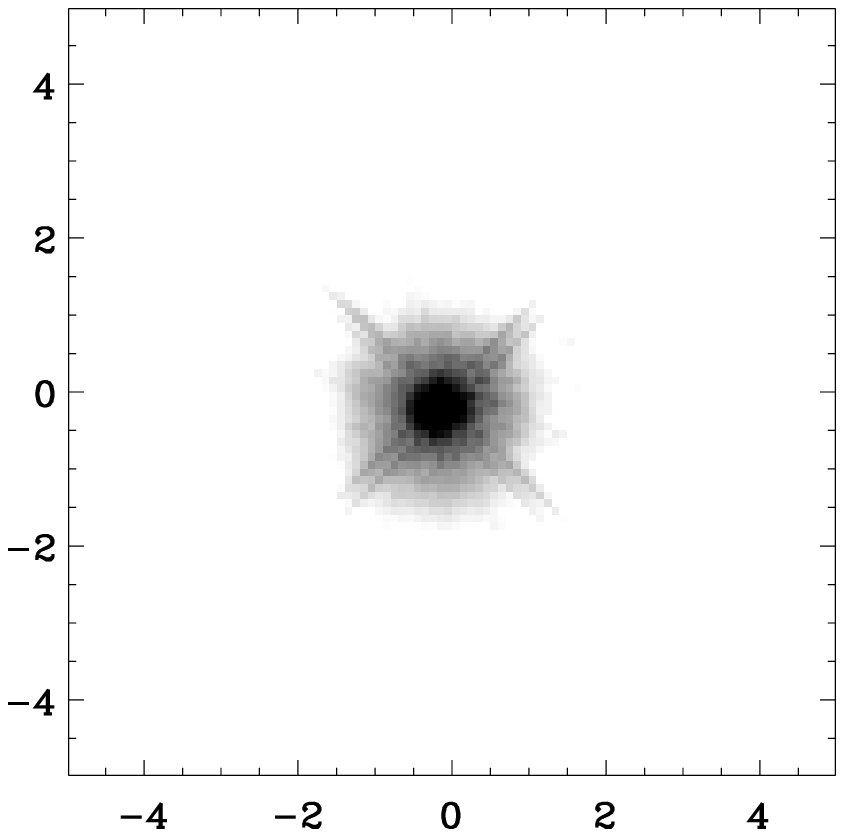}}

{\includegraphics[width=40mm]{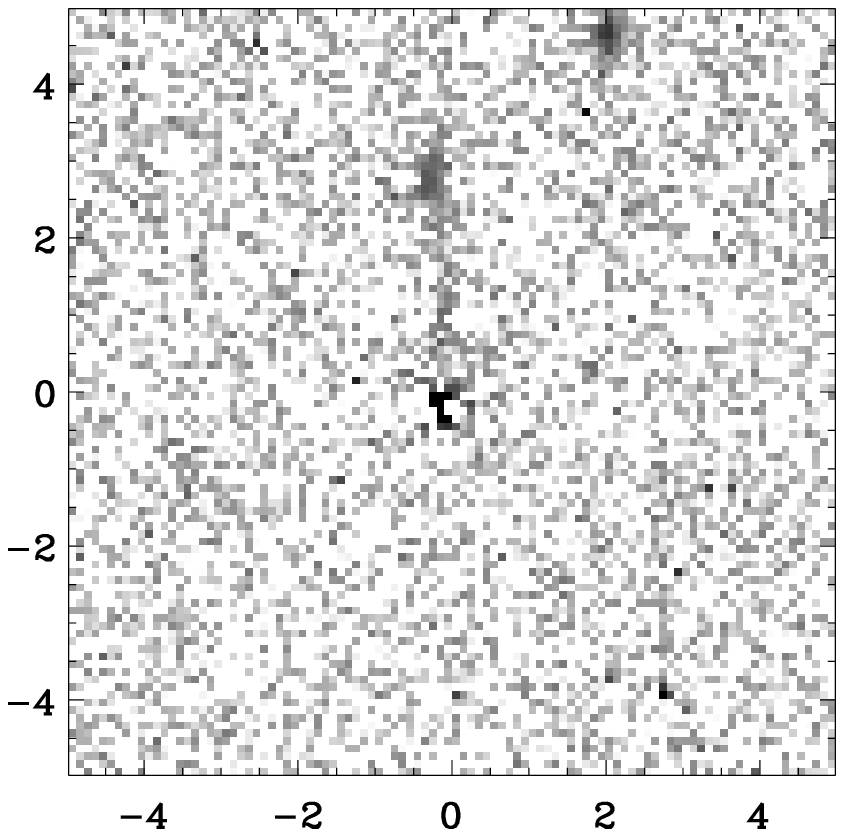}}
{\includegraphics[angle=270,origin=c,width=40mm]{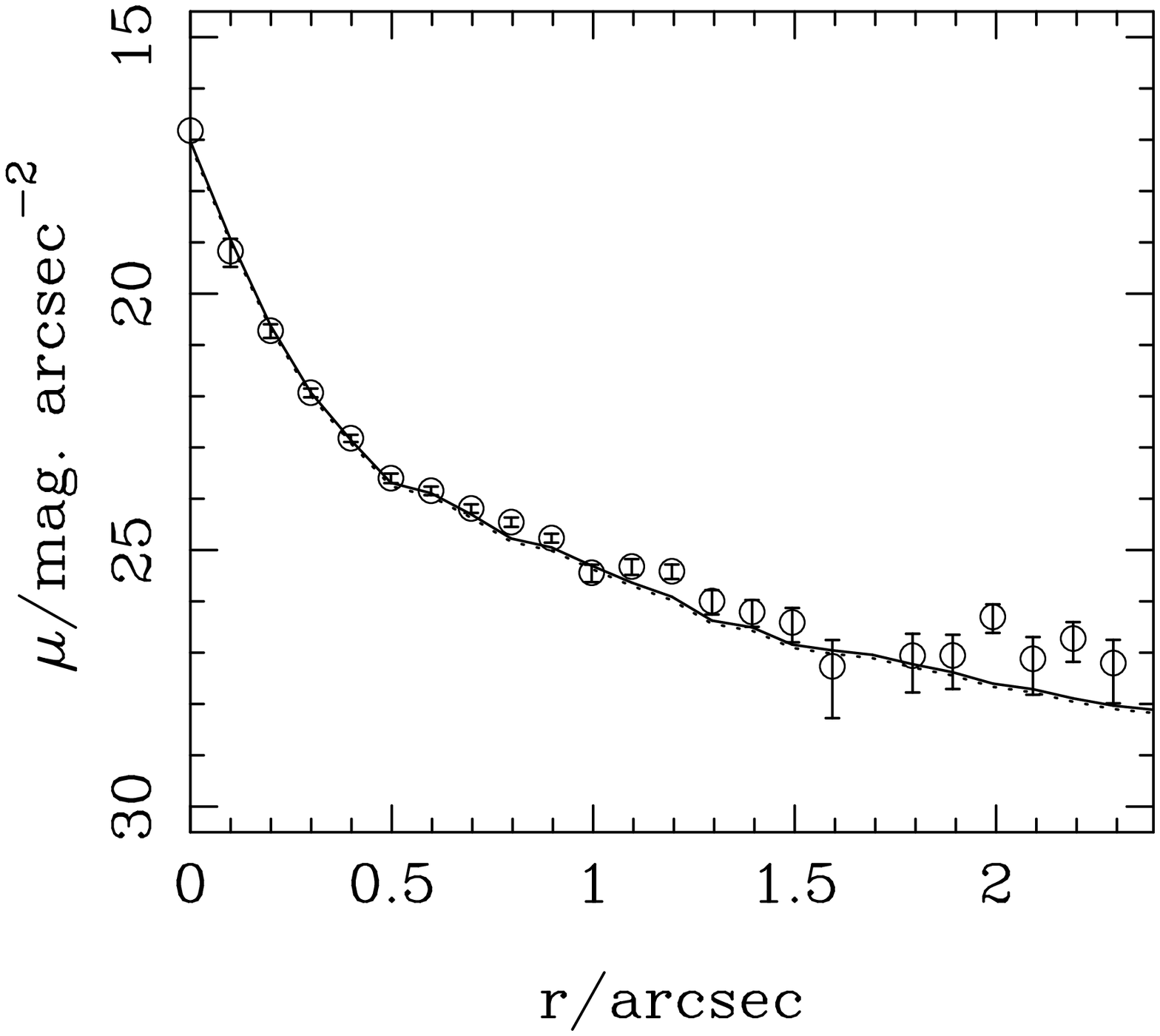}}
\caption{\label{fig-h1012} The RQQ SGP4:39 at $z=1.721$. 
K01 found it to be a heavily nuclear-dominated quasar in $V$ ($[L_\mathrm{nuc}/ L_\mathrm{host}]_V=12.4$). In $U$ the host-galaxy size and surface brightness are poorly constrained due to the strong nucleus ($[L_\mathrm{nuc}/ L_\mathrm{host}]_U\sim 5$).
However, the total host-galaxy magnitude is well constrained.
The residuals are remarkably clean. The apparent tail linking the quasar to the companion 3\arcsec west is probably diffraction spike residual that happens to coincide with the companion.}
\end{figure}


\begin{figure}
{\includegraphics[width=40mm]{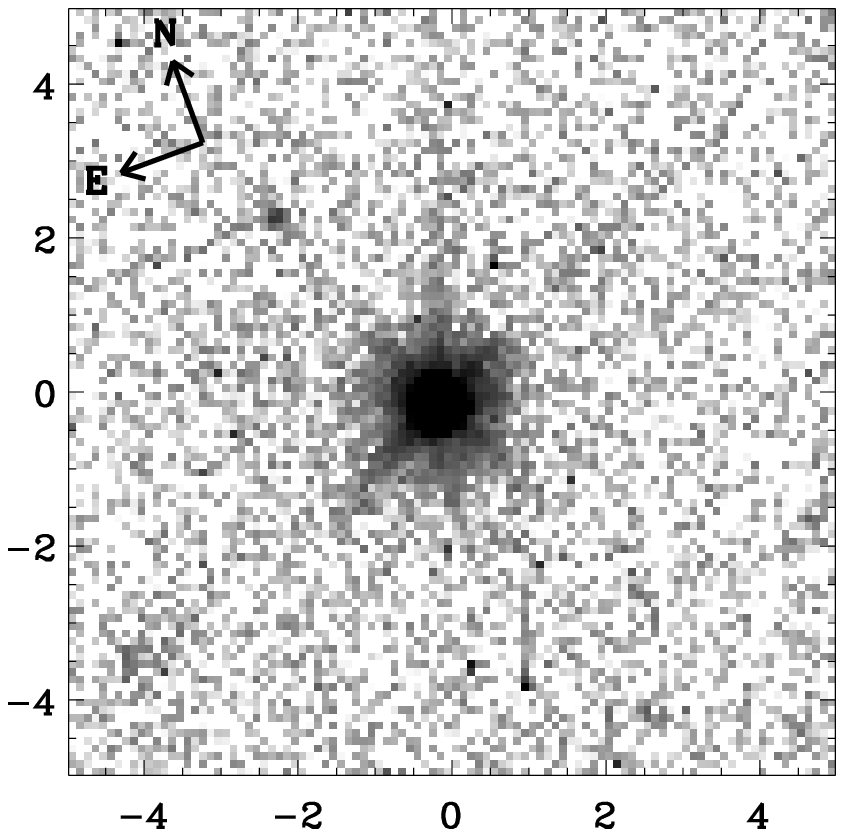}}
{\includegraphics[width=40mm]{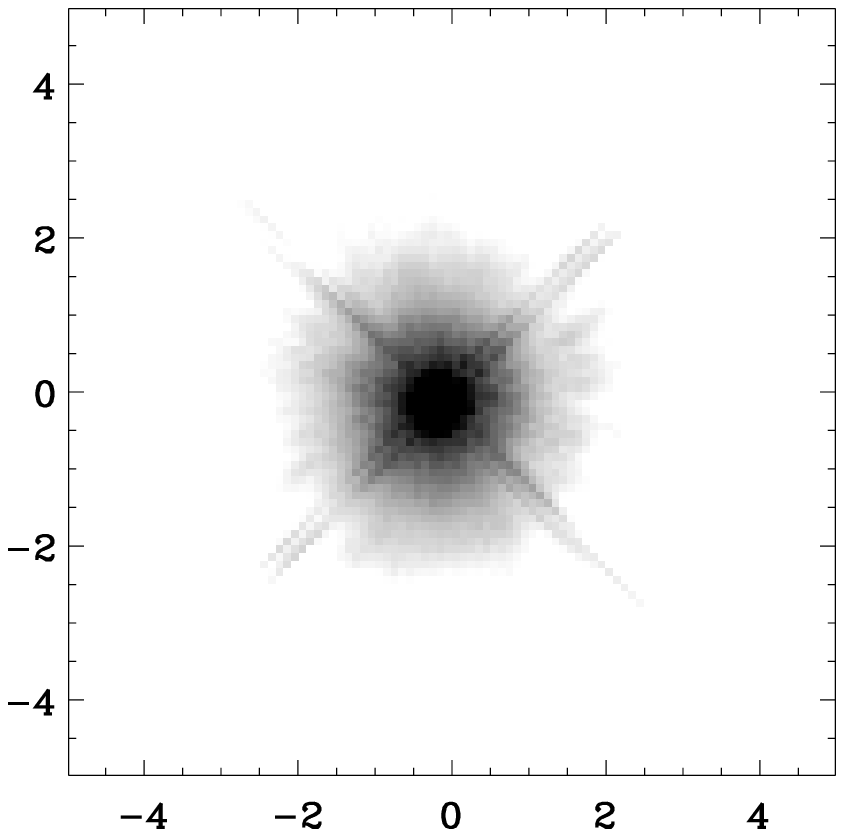}}

{\includegraphics[width=40mm]{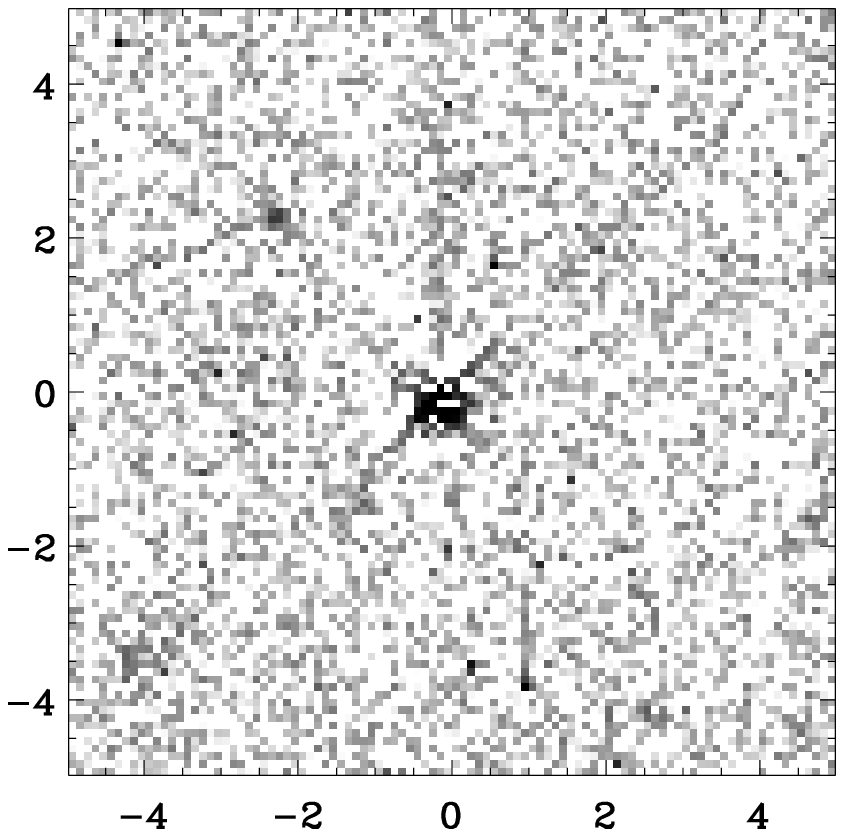}}
{\includegraphics[angle=270,origin=c,width=40mm]{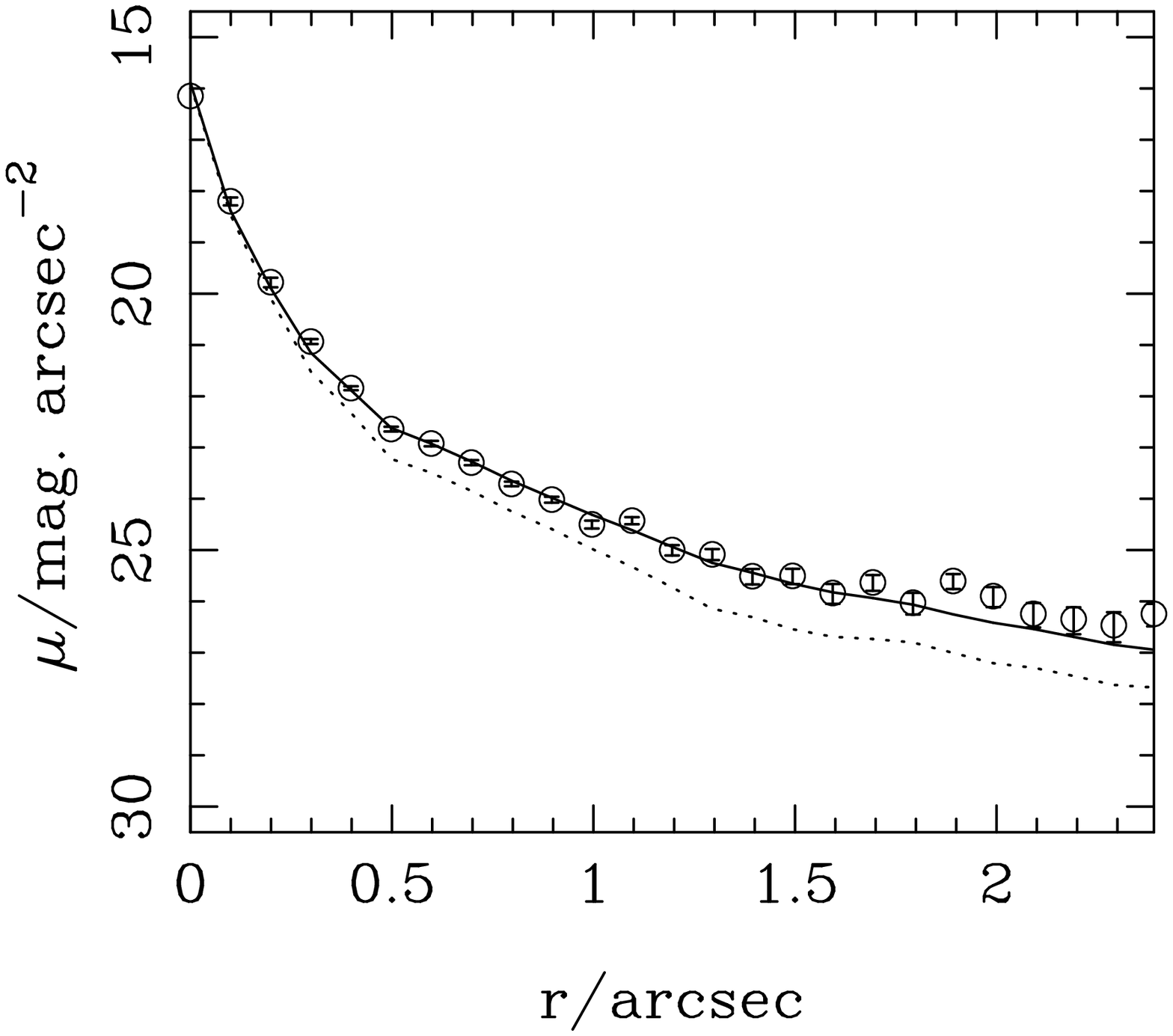}}
\caption{\label{fig-h1017} The RLQ PKS1524--13 at $z=1.687$. We see a large number of foreground / companion objects in the field, particularly to the south. The nucleus dominates at both wavelengths: $[L_\mathrm{nuc}/ L_\mathrm{host}]_V=3.2$;~$[L_\mathrm{nuc}/ L_\mathrm{host}]_U=5.5$.}
\end{figure}

\begin{figure}
{\includegraphics[width=40mm]{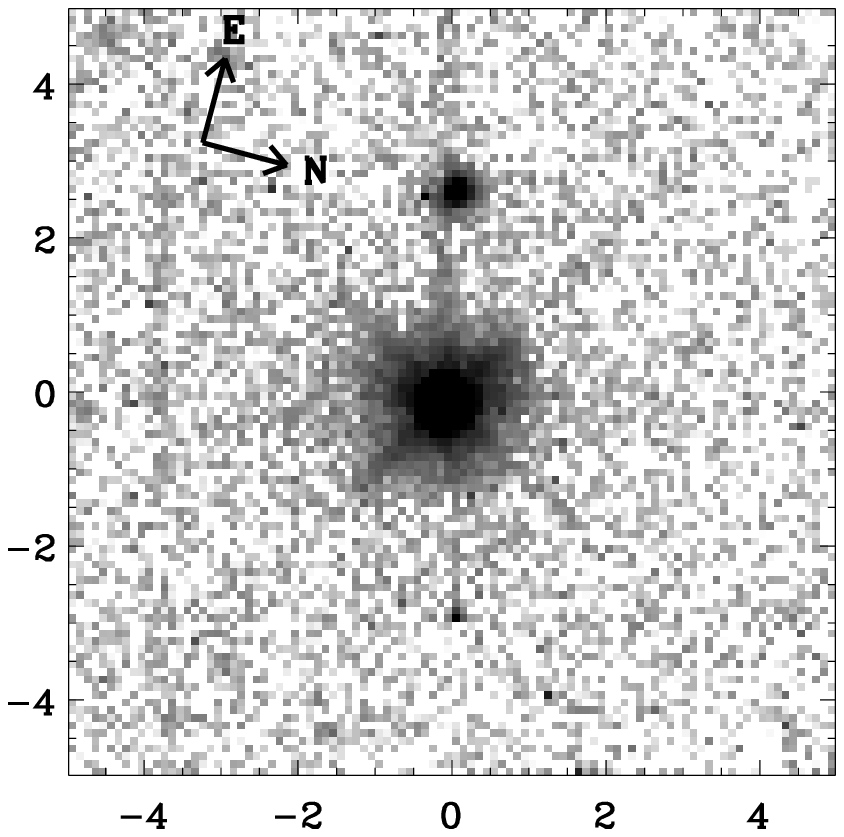}}
{\includegraphics[width=40mm]{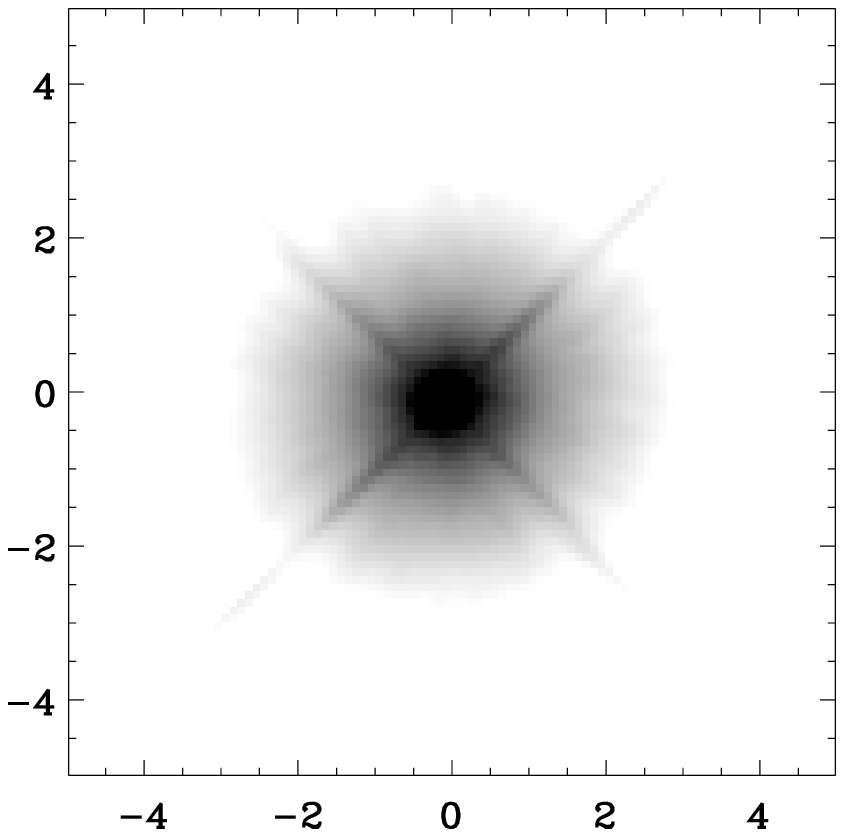}}

{\includegraphics[width=40mm]{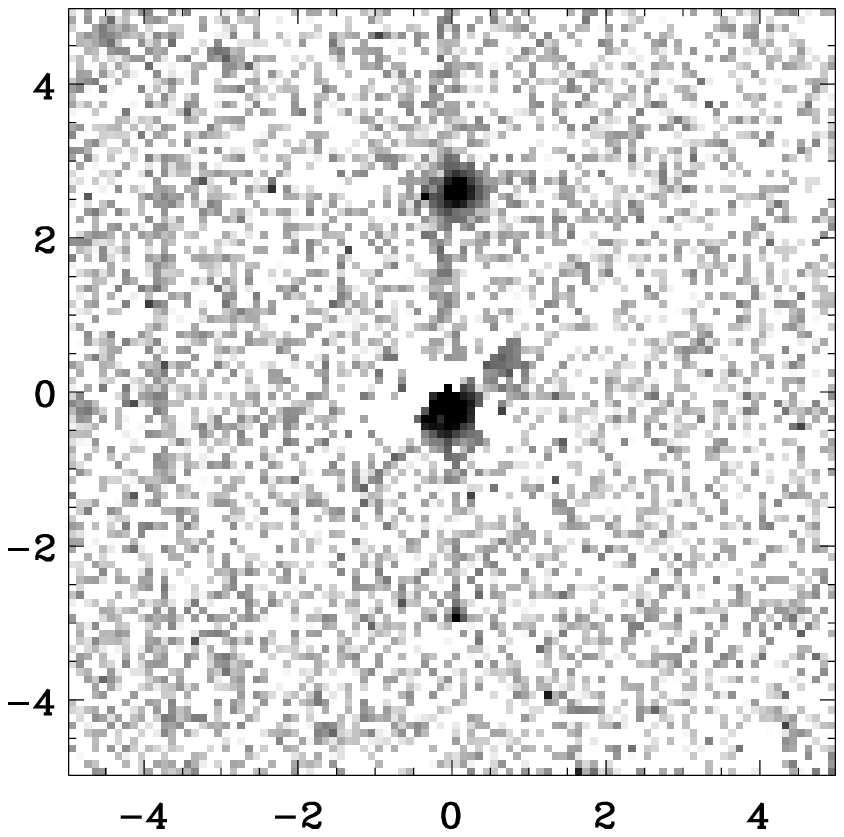}}
{\includegraphics[angle=270,origin=c,width=40mm]{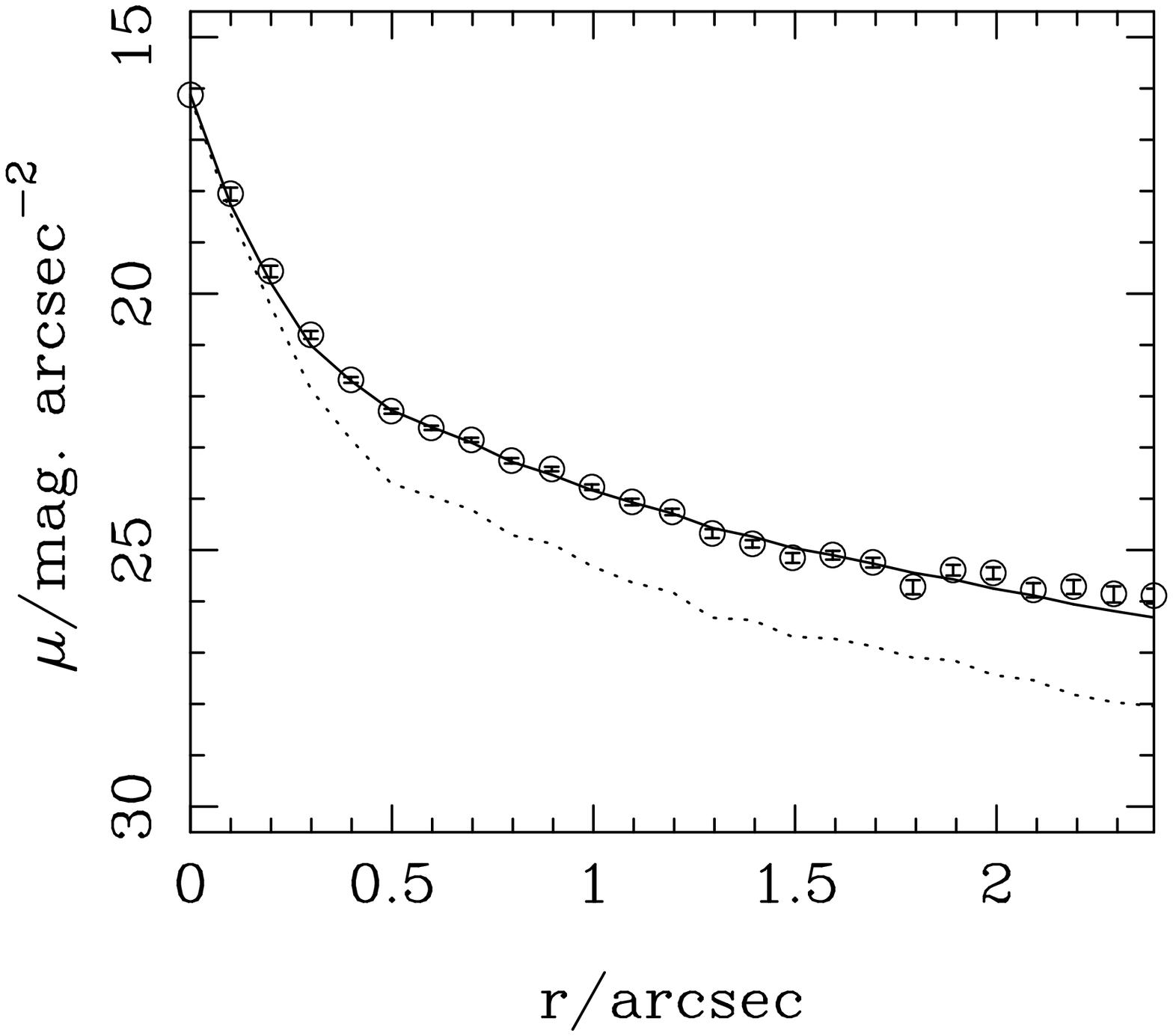}}
\caption{\label{fig-h1019} The RLQ B2~2156+29 at $z=1.753$.
A number of nearby galaxies~\citep{thomas+95} are visible in the image. The nucleus 
dominates in $U$: $[L_\mathrm{nuc}/ L_\mathrm{host}]_V= 0.9$;~$[L_\mathrm{nuc}/ L_\mathrm{host}]_U= 3.4$. There is a significant circumnuclear residual, likely a PSF artefact.}
\end{figure}

\begin{figure}
{\includegraphics[width=40mm]{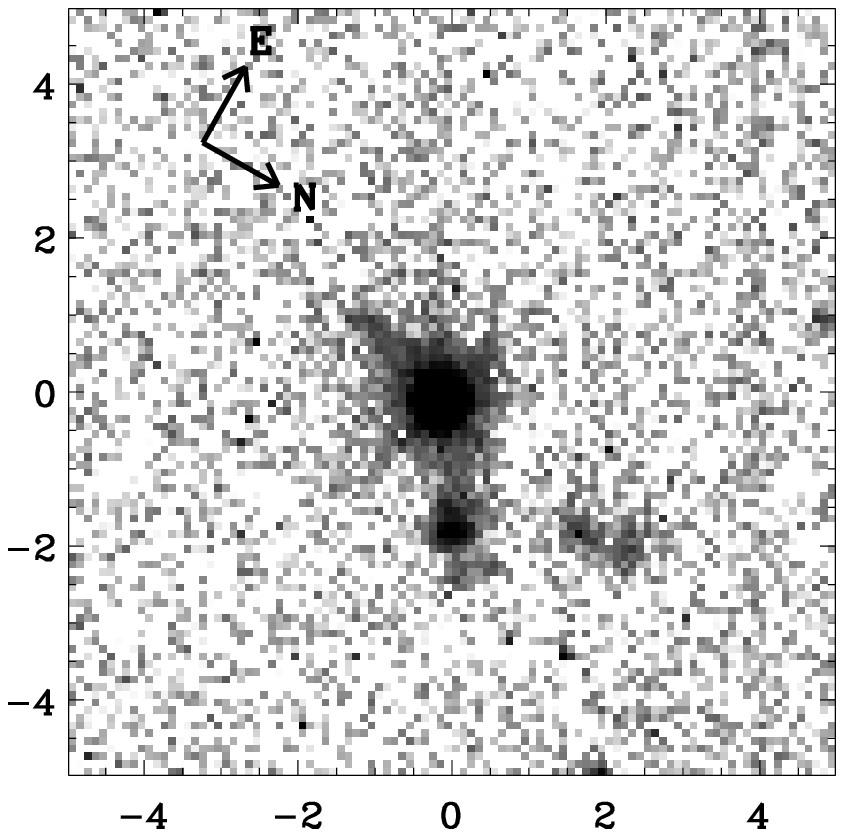}}
{\includegraphics[width=40mm]{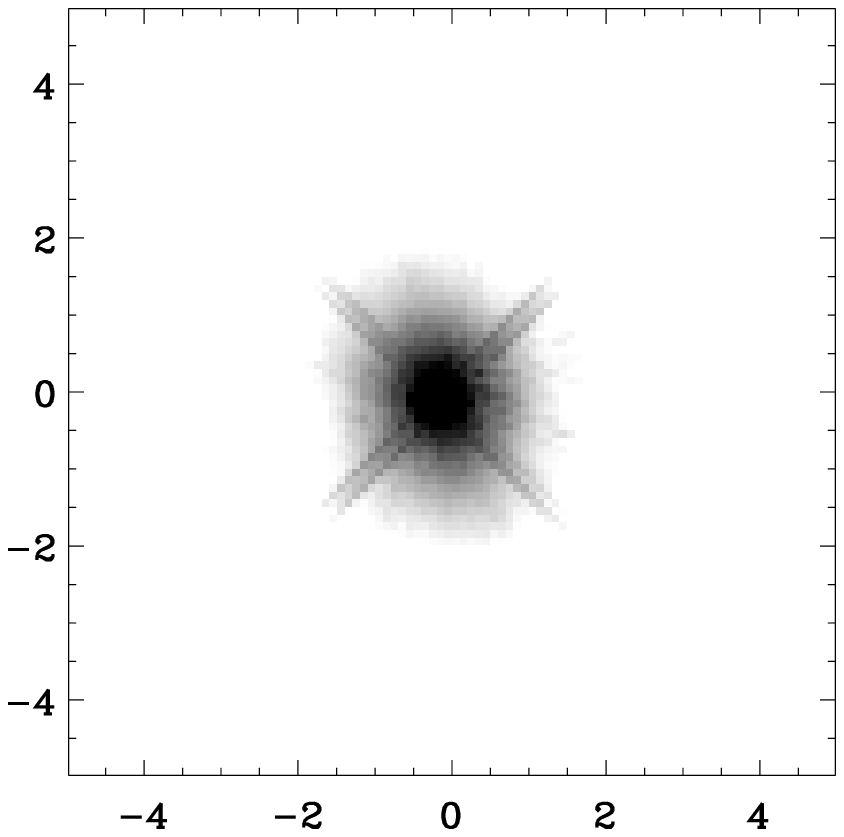}}

{\includegraphics[width=40mm]{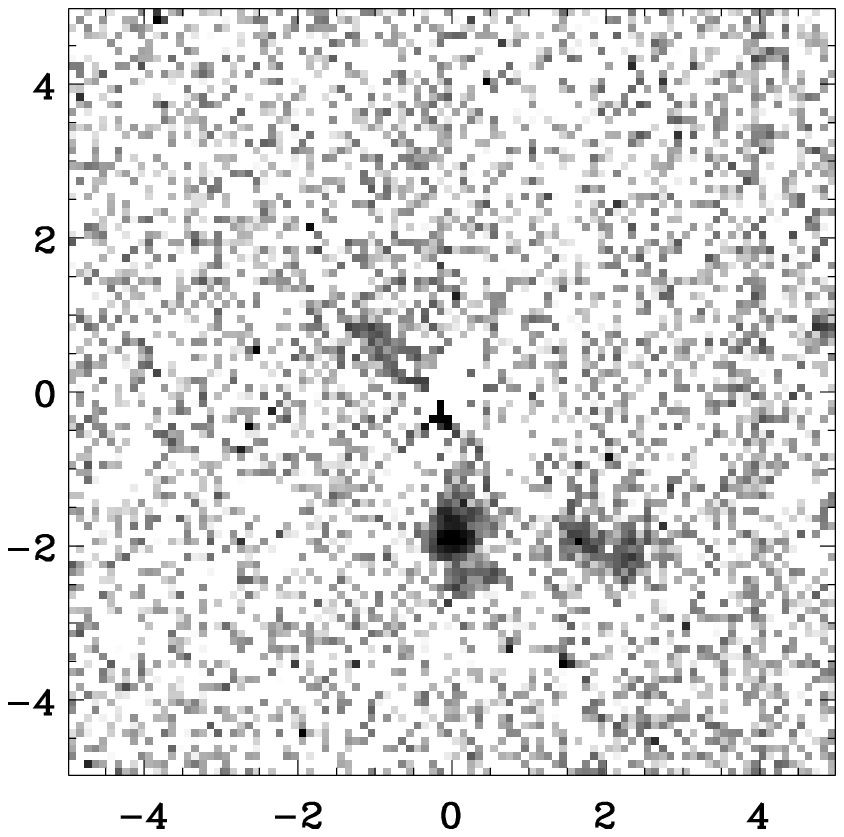}}
{\includegraphics[angle=270,origin=c,width=40mm]{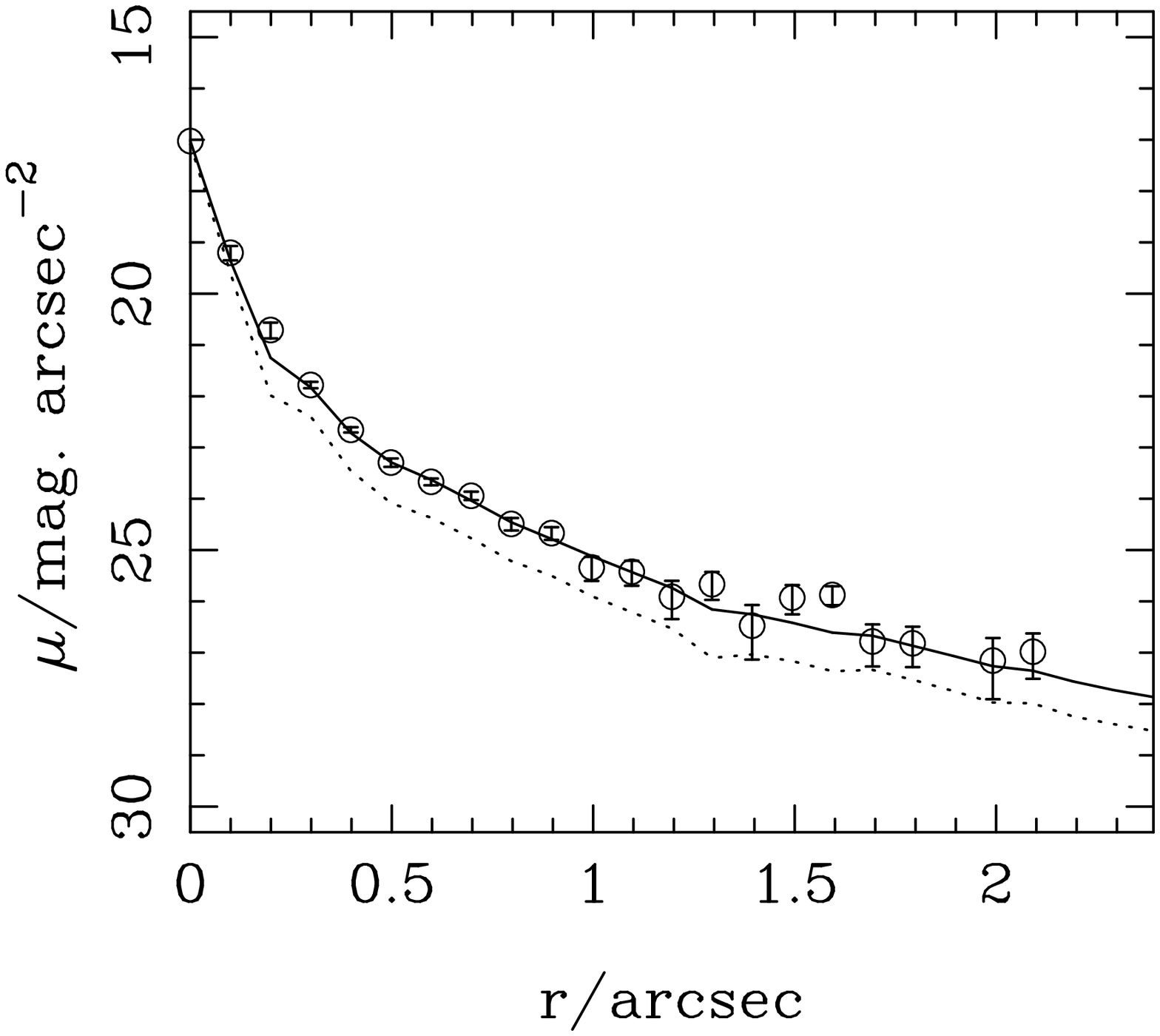}}
\caption{\label{fig-h1020} The RLQ PKS2204--20 at $z=1.923$.
There are two quasars with this PKS identifier, this one being the more distant one. The other, at $z=1.62$, is out of our field of view. 
Unusually, the nuclear domination in rest-frame $U$ appears to be less than that in  the optical: $[L_\mathrm{nuc}/ L_\mathrm{host}]_V= 3.2$;~$[L_\mathrm{nuc}/ L_\mathrm{host}]_U= 2.0$. There is a notable companion $\approx 2$\arcsec\ NW of the nucleus that appears to be interacting with the host galaxy. }
\end{figure}

\begin{figure}
{\includegraphics[width=40mm]{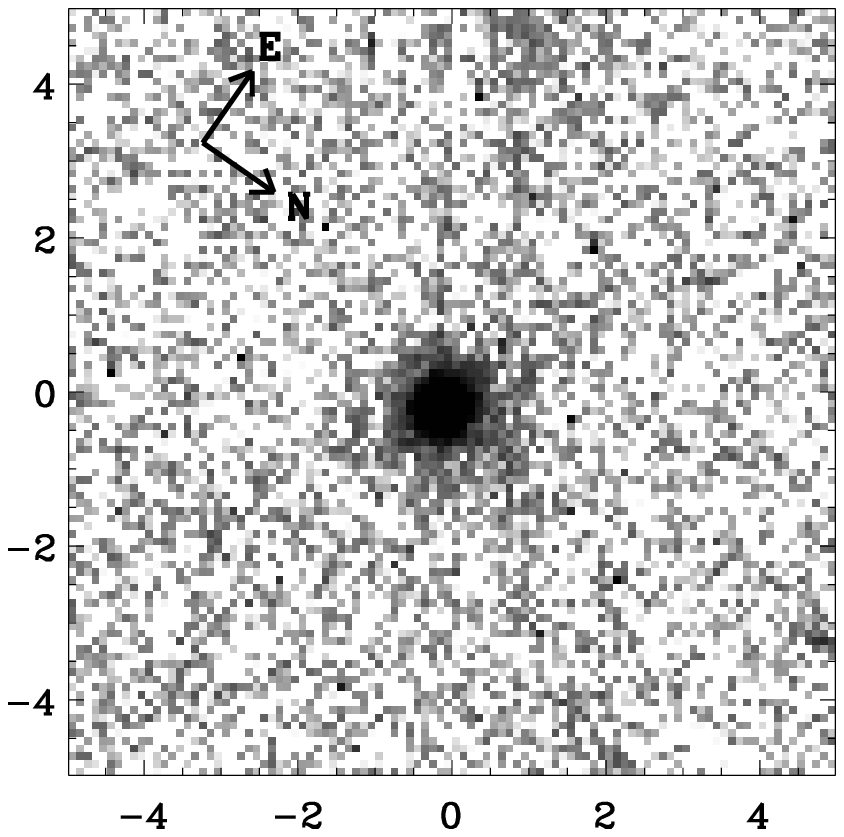}}
{\includegraphics[width=40mm]{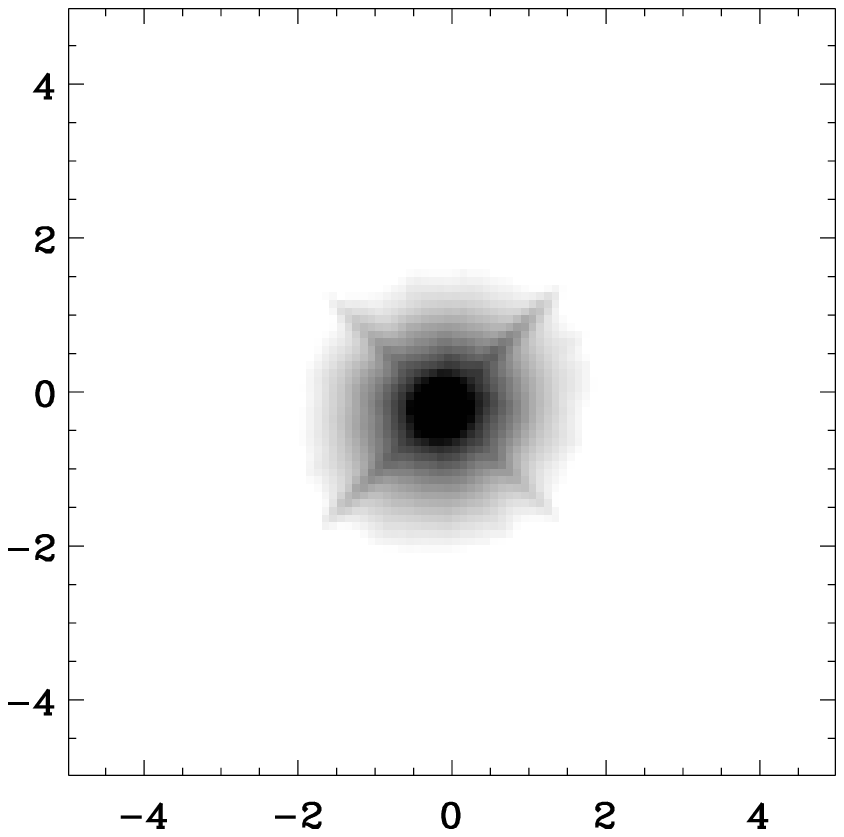}}

{\includegraphics[width=40mm]{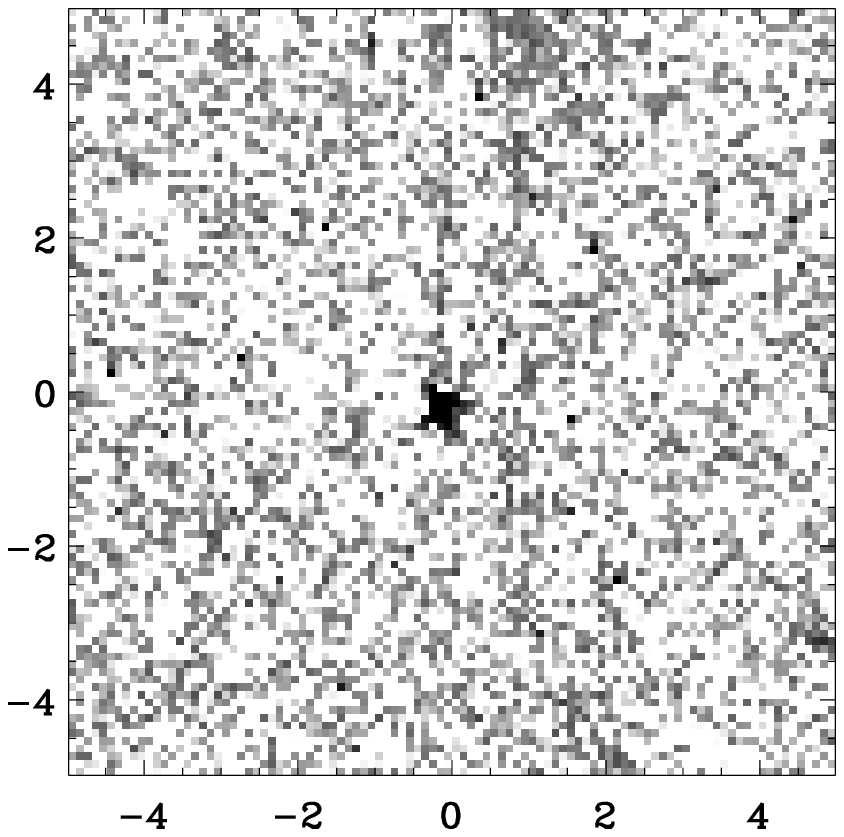}}
{\includegraphics[angle=270,origin=c,width=40mm]{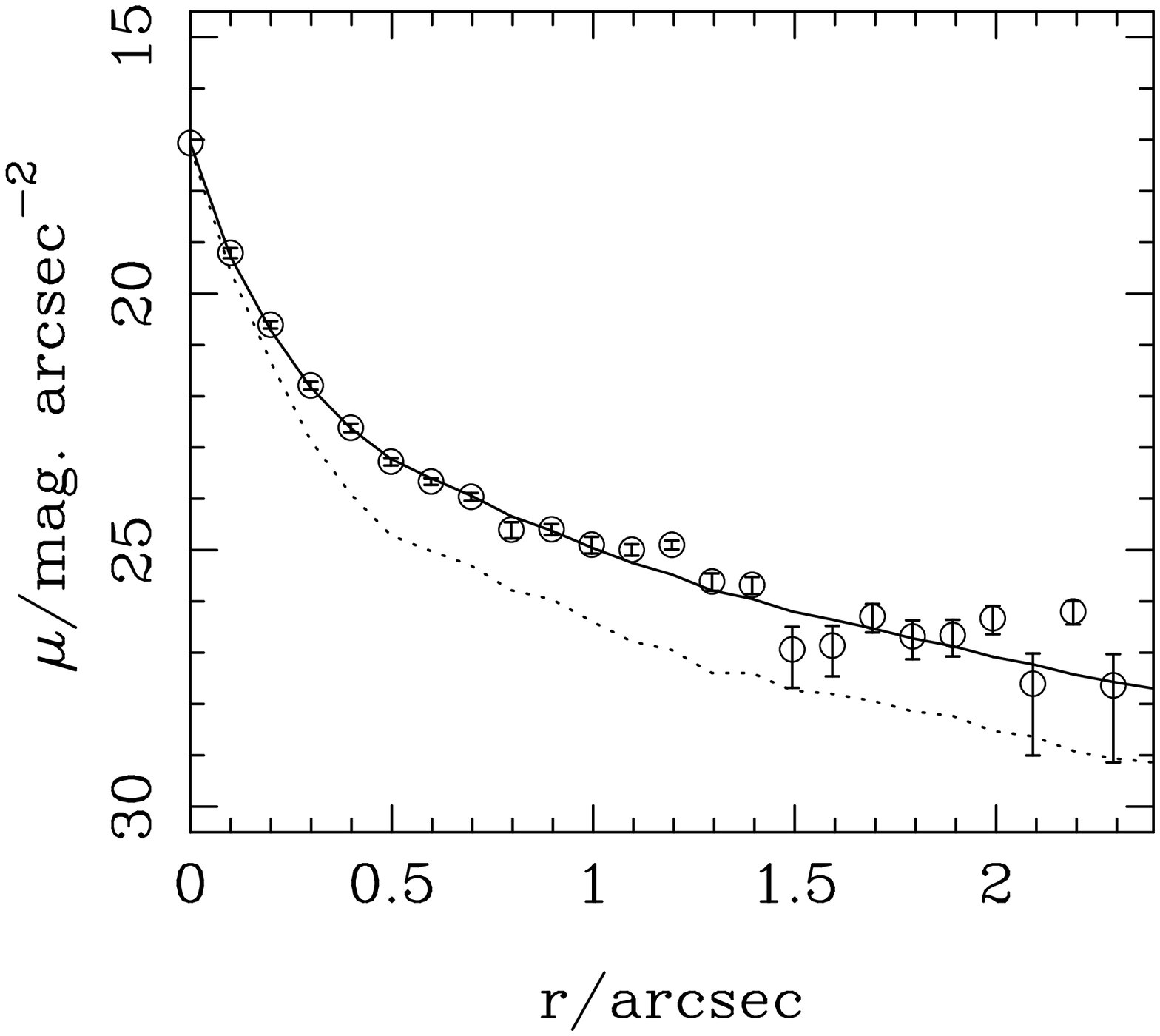}}
\caption{\label{fig-h1018} The RLQ 4C45.51 at $z=1.992$. $[L_\mathrm{nuc}/ L_\mathrm{host}]_V= 1.4$;~$[L_\mathrm{nuc}/ L_\mathrm{host}]_U= 2.2$. We see diffraction spikes from a nearby star at the top (southeast) of the image. There are four noticeable foreground / companion objects within 10\arcsec\ and the residuals show a compact (2~pixel) artefact just north of the nucleus that may be attributable to the PSF.}
\end{figure}


\bsp

\clearpage

\section{Fixed effective radius results}
Tables~\ref{tab-fixres1} and ~\ref{tab-fixres2} show the results from the fixed $R_e$ fits to our data, compared with the free radius fits and the best fits to the rest-frame optical images of K01.
\begin{table}
  \begin{center}
    \begin{small}
      \caption{\label{tab-fixres1} Fixed $R_e$ results for the \zo\ sample.
      The best-fit model from Table~\ref{tab-res1} is presented, together 
      with models at fixed $R_e$ = 2, 5 and 10~kpc, and the best fit ($J$-band) model from K01.
      Note that the K01 sizes have been updated for the current cosmology and converted to $R_e$.}
      \centering
      \begin{tabular}{crrrcccc}
        \hline
        \hline
        Object&$\chi^{2}_{red}$&$R_\mathrm{e}$&$\mu_{e}$&$V_\mathrm{AB}^\mathrm{Nuc}$&$V_\mathrm{AB}^\mathrm{Host}$&PA&$a/b$\\
        \hline
	BVF225 	& 1.122 & 5.5& 23.9 & 18.5 & 21.2 & 176 & 1.24 \\
			&1.189 & 2.0 & 21.2 & 18.5 & 20.9 & 4 & 1.20\\
			&1.134 & 5.0 & 23.5 & 18.4 & 21.2 & 178 & 1.23\\
			&1.130 & 10.0 & 25.0 & 18.4 & 21.3 & 174 & 1.29\\
	$J$-band 	&  ... 	& 5.0 & 24.4 & 18.5 & 20.7 & 33 & 3.20\\
        \hline
	BVF247 	& 1.143 & 4.5 & 24.5 & 20.4 & 22.3 & 155 & 1.25 \\
			&1.194 & 2.0 & 	22.1 & 20.7 & 21.9 & 154 & 1.26\\
			&1.189 & 5.0 & 24.3 & 20.6 & 22.0 & 156 & 1.25\\
			&1.209 & 10.0& 25.7 & 20.6 & 22.0 & 157 & 1.26\\
	$J$-band 	&  ... & 9.2 & 23.4 & 20.7 & 19.5 & 3 & 1.20\\
        \hline
	BVF262 	& 1.077 & 3.0 & 23.2 & 20.0 & 22.0 & 4 & 1.33 \\
			& 1.164 & 2.0 & 22.3 & 19.9 & 22.1 & 3 & 1.44\\
			& 1.153 & 5.0 & 24.5 & 19.9 & 22.3 & 7 & 1.52\\
			& 1.155 & 10.0& 26.0 & 19.9 & 22.3 & 10 & 1.61\\
	$J$-band 	&  ... & 3.1 & 22.2 & 19.8 & 20.4 & 5. & 1.36\\
        \hline
	PKS0440 	& 1.097 &11.1 & 25.5 & 18.9 & 21.3 & 47 & 1.12 \\
			& 1.160 & 2.0 & 21.2 & 19.2 & 20.9 & 43 & 1.05\\
			& 1.133 & 5.0 & 23.5 & 19.1 & 21.2 & 48 & 1.08\\
			& 1.137 & 10.0& 24.9 & 19.1 & 21.1 & 47 & 1.10\\
	$J$-band &  ... & 7.3 & 23.3 & 19.1 & 19.4 & 2 & 1.64\\
        \hline
	PKS0938 & 1.106 & 3.8 & 23.3 & 19.3 & 21.5 & 154 & 1.20 \\
			& 1.110 & 2.0 & 21.5 & 19.3 & 21.3 & 135 & 1.22\\
			& 1.108 & 5.0 & 23.8 & 19.3 & 21.6 & 140 & 1.25\\
			& 1.118 & 10.0& 25.3 & 19.3 & 21.6 & 142 & 1.28\\
	$J$-band &  ... & 3.0 & 21.7 & 20.4 & 20.1 & 136 & 1.33\\
        \hline
	3C422 	& 1.073 & 2.7 & 22.2 & 19.2 & 21.2 &87 & 1.57 \\
			& 1.076 & 2.0 & 21.2 & 19.2 & 21.0 & 88 & 1.58\\
			& 1.079 & 5.0 & 23.8 & 19.1 & 21.5 & 89 & 1.73\\
			& 1.090 & 10.0 & 25.6 & 19.0 & 21.8 & 91 & 2.05\\
	$J$-band 	&  ... & 11.9 & 23.5 & 18.5 & 18.9 & 139 & 1.33\\
        \hline
	MC2112 	& 1.040 & 1.8 & 20.5 & 19.9 & 20.2 & 102 & 1.74 \\
			& 1.040 & 2.0 & 20.4 & 20.2 & 20.2 & 101 & 1.65\\
			& 1.148 & 5.0 & 22.9 & 20.0 & 20.7 & 102 & 1.69\\
			& 1.265 & 10.0& 24.6 & 20.0 & 20.8 & 102 & 1.81\\
	$J$-band 	&  ... & 15.8 & 23.7 & 19.6 & 18.8 & 168. & 1.02\\
        \hline
	4C02.54 	& 1.070 & 6.3 & 24.2 & 18.8 & 21.4 & 111 & 1.42 \\
			& 1.102 & 2.0 & 21.1 & 19.0 & 20.9 & 112 & 1.26\\
			& 1.071 & 5.0 & 23.5 & 18.9 & 21.3 & 112 & 1.36\\
			& 1.072 & 10.0 & 25.1 & 18.8 & 21.4 & 111 & 1.50\\
	$J$-band 	&  ... & 2.4 & 22.2 & 18.2 & 19.9 & 16. & 4.08\\
        \hline
        \hline
      \end{tabular}
    \end{small}
  \end{center}
\end{table}
\begin{table}
  \begin{center}
    \begin{small}
      \caption{\label{tab-fixres2} Fixed $R_e$ results for the \zt\ sample.
      The best-fit model from Table~\ref{tab-res2} is presented, together 
      with models at fixed $R_e$ = 2, 5 and 10~kpc, and the best fit ($H$-band) model from K01.
      Note that the K01 sizes have been updated for the current cosmology and converted to $R_e$.}
      \centering
      \begin{tabular}{crrrcccc}
        \hline
        \hline
        Object&$\chi^{2}_{red}$&$R_\mathrm{e}$&$\mu_{e}$&$I_\mathrm{AB}^\mathrm{nuc}$&$I_\mathrm{AB}^\mathrm{host}$&PA&$a/b$\\
	\hline
	SGP2:36 	& 1.244 & 1.6 & 22.3 & 21.5 & 22.5 & 136 & 1.29\\
			& 1.246 & 2.0 & 22.9 & 21.4 & 22.5 & 156 & 1.28\\
			& 1.272 & 5.1 & 25.0 & 21.3 & 22.7 & 152 & 1.32\\
			& 1.300 & 10.2 & 26.5 & 21.2 & 22.7 & 150 & 1.39\\
	$H$-band	& ... 		&$\sim5$	& ... & 21.3 & 21.0 & ... & ... \\
        \hline
	SGP2:25 	& 1.295 & 1.8 & 23.2 & 20.8 & 23.1 & 98 & 1.21\\
			& 1.617 & 2.0 & 23.2 & 20.9 & 22.9 & 120 & 1.11\\
			& 1.635 & 5.0 & 25.7 & 20.8 & 23.4 & 156 & 1.24\\
			& 1.643 & 10.0 & 27.3 & 20.7 & 23.5 & 166 & 1.80\\
	$H$-band	& ... 	&$<10$ & ... & 20.9 & 21.2 & ... & ... \\
        \hline
        	SGP2:11 	& 1.169 & 4.0 & 25.2 & 20.6 & 23.4 & 168 & 1.15\\
			& 1.264 & 2.0 & 23.4 & 20.7 & 23.0 & 35 & 1.14\\
			& 1.263 & 5.0 & 25.6 & 20.6 & 23.3 & 34 & 1.12\\
			& 1.264 & 10.0 & 27.1 & 20.6 & 23.3 & 31 & 1.11\\
	$H$-band	& ... &$<10$ & ... & 20.3 & 21.9 & ... & ... \\
        \hline
	SGP3:39 	& 1.324  & 2.2 & 22.8 & 20.6 & 22.5 & 107 & 1.13\\
			 & 1.325 & 2.0 & 22.8 & 20.6 & 22.5 & 141 & 1.14\\
			 & 1.344 & 5.0 & 25.2 & 20.5 & 22.9 & 113 & 1.25\\
			 & 1.356 & 10.0 & 26.8 & 20.4 & 23.0 & 108 & 1.47\\
        $H$-band	& ... &$<10$ & ... & 20.8 & 21.1 & ... & ... \\
        \hline
	SGP4:39 	& 1.031 & 1.5 & 22.1 & 20.7 & 22.5 & 81 & 1.09\\
			& 1.031 & 2.0 & 22.9 & 20.6 & 22.6 & 113 & 1.11\\
			& 1.036 & 5.1 & 25.3 & 20.5 & 23.0 & 115 & 1.18\\
			& 1.041 & 10.2 & 27.0 & 20.4 & 23.1 & 115 & 1.25\\
	$H$-band	& ... &$<10$ & ... & 20.2 & 22.9 & ... & ... \\
        \hline
	PKS1524 	& 1.198 & 2.2 & 22.0 & 19.6 & 21.5 & 163 & 1.31\\
			& 1.256 & 2.0 & 21.7 & 19.6 & 21.4 & 163 & 1.30\\
			& 1.266 & 5.0 & 24.1 & 19.5 & 21.8 & 162 & 1.46\\
			& 1.281 & 10.0 & 25.7 & 19.4 & 21.9 & 162 & 1.69\\
	$H$-band	& ... &$\sim5$ & ... & 19.4 & 20.7 & ... & ... \\
        \hline
	B2~2156 	& 1.335 & 3.6 & 22.3 & 19.4 & 20.7 & 39 & 1.10\\
			& 1.393 & 2.0 & 20.7 & 19.4 & 20.4 & 28 & 1.11\\
			& 1.347 & 5.0 & 23.1 & 19.3 & 20.8 & 24 & 1.12\\
			& 1.422 & 10.0 & 24.6 & 19.1 & 20.9 & 20 & 1.15\\
	$H$-band	& ... 	&12.1 & ... & 19.3 & 19.2 & ... & 1.77 \\
        \hline
	PKS2204 	& 1.300 & 1.6 & 21.3 & 20.8 & 21.6 & 137 & 1.48\\
			& 1.364 & 2.0 & 22.0 & 20.8 & 21.7 & 140 & 1.42\\
			& 1.400 & 5.0 & 24.2 & 20.5 & 21.9 & 138 & 1.78\\
			& 1.431 & 10.0 & 25.8 & 20.3 & 22.0 & 136 & 2.31\\
	$H$-band	& ... &$\sim5$ & ... & 19.9 & 22.0 & ... & ... \\
        \hline
	4C45.51 	& 1.767 & 2.2 & 22.0 & 20.6 & 21.5 & 78 & 1.13\\
			& 1.829 & 2.0 & 21.8 & 20.7 & 21.5 & 76 & 1.10\\
			& 1.859 & 5.0 & 24.0 & 20.5 & 21.7 & 86 & 1.10\\
			& 1.906 & 10.0 & 25.6 & 20.4 & 21.7 & 91 & 1.12\\
	$H$-band	& ... 	      & 16.5	& ... & 18.8 & 19.2 & ... & 1.23 \\
        \hline
        \hline
      \end{tabular}
    \end{small}
  \end{center}
\end{table}
\label{lastpage}
\end{document}